\documentclass[aps,prd,twocolumn,showpacs,superscriptaddress,showpacs]{revtex4}

\usepackage{graphicx}
\usepackage{color}
\usepackage{verbatim}
\usepackage{subfigure}
\usepackage{placeins}

\usepackage{epsfig}
\usepackage{colordvi}
\usepackage{amsmath}
\usepackage{dcolumn}
\usepackage{bm}     
\usepackage{url}    

\usepackage{xspace}

\def\urltilde{\kern -.15em\lower .7ex\hbox{\~{}}\kern .04em}
\def\urldot{\kern -.10em.\kern -.10em}
\def\urlhttp{http\kern -.10em\lower -.1ex\hbox{:}\kern -.12em\lower 0ex\hbox{/}\kern -.18em\lower 0ex\hbox{/}}
\def\urltilde{\kern -.15em\lower .7ex\hbox{\~{}}\kern .04em}
\def\urldot{\kern -.10em.\kern -.10em}
\def\urlhttp{http\kern -.10em\lower -.1ex\hbox{:}\kern -.12em\lower 0ex\hbox{/}\kern -.18em\lower 0ex\hbox{/}}
\def\mevc {\ifmmode {\rm MeV}/c \else MeV$/c$\fi}
\def\mevcc {\ifmmode {\rm MeV}/c^2 \else MeV$/c^2$\fi}
\def\gevc {\ifmmode {\rm GeV}/c \else GeV$/c$\fi}
\def\gevcc {\ifmmode {\rm GeV}/c^2 \else GeV$/c^2$\fi}
\def\tevcc {\ifmmode {\rm TeV}/c^2 \else TeV$/c^2$\fi}

\def\ol   {\overline}

\def\vtd  {\ifmmode |V_{td}| \else $|V_{td}|$\fi}
\def\vtb  {\ifmmode |V_{tb}| \else $|V_{tb}|$\fi}
\def\vts  {\ifmmode |V_{ts}| \else $|V_{ts}|$\fi}
\def\vcb  {\ifmmode |V_{cb}| \else $|V_{cb}|$\fi}

\newcommand{\Ds} {\ifmmode D_{\mbox{\sl s}}^{-}
                       \else $D_{\mbox{\sl s}}^{-}$\fi}
\newcommand{\Bs} {\ifmmode B_{\mbox{\sl s}}^{0}
                       \else $B_{\mbox{\sl s}}^{0}$\fi}
\newcommand{\Bsb} {\ifmmode \ol B_{\mbox{\sl s}}^{0}
                       \else $\ol B_{\mbox{\sl s}}^{0}$\fi}
\newcommand{\Bsh} {\ifmmode B_{\mbox{\sl s}}^H
                       \else $B_{\mbox{\sl s}}^H$\fi}
\newcommand{\Bsl} {\ifmmode B_{\mbox{\sl s}}^L
                       \else $B_{\mbox{\sl s}}^L$\fi}
\newcommand{\Dsl} {\ifmmode D_{\mbox{\sl s}}^{-} \ell^+
                       \else $D_{\mbox{\sl s}}^{-} \ell^+$\fi}
\newcommand{\xs} {\ifmmode x_{\mbox{\sl s}}
                       \else $x_{\mbox{\sl s}}$\fi}
\newcommand{\xd} {\ifmmode x_d \else $x_d$\fi}
\newcommand{\lxy} {\ifmmode L_{\rm xy} \else $L_{\rm xy}$\fi}
\newcommand{\dgam} {\ifmmode \Delta\Gamma \else $\Delta\Gamma$\fi}
\newcommand{\dm} {\ifmmode \Delta m \else $\Delta m$\fi}
\newcommand{\ctau} {\ifmmode c\tau \else $c\tau$\fi}
 
\newcommand{\et}{E_T}

\newcommand{\peetee}{\mbox{${p_T}$}}
\newcommand{\ptran}{\mbox{${p_T^{\ttbar}}$}}
\newcommand{\etran}{\mbox{${E_T}$}}
\newcommand{\met}{\mbox{$\protect \raisebox{.3ex}{$\not$}\et$}}

\newcommand{\ppbar}{p\bar{p}}
\newcommand{\qqbar}{q\bar{q}} 
\newcommand{\ttbar}{t\bar{t}}
\newcommand{\tbar}{\bar{t}}
\newcommand{\mttb}{M_{t\bar{t}}}

\newcommand{\afb}{A_\mathrm{FB}}

\newcommand{\yt}{y_{t}}

\newcommand{\ytbar}{y_{\tbar}}

\newcommand{\dy}{\Delta y}

\newcommand{\ifb}{ {\rm fb}^{-1} }

\newcommand{\pythia}{\textsc{pythia}\xspace}
\newcommand{\powheg}{\textsc{powheg}\xspace}
\newcommand{\mcnlo}{\textsc{mc@nlo}\xspace}
\newcommand{\mcfm}{\textsc{mcfm}\xspace}
\newcommand{\absdely}{\ensuremath{\left\lvert \Delta y \right\rvert}\xspace}

\begin{document}

\title{Measurement of the top quark forward-backward production asymmetry and its dependence on event kinematic properties}

\affiliation{Institute of Physics, Academia Sinica, Taipei, Taiwan 11529, Republic of China}
\affiliation{Argonne National Laboratory, Argonne, Illinois 60439, USA}
\affiliation{University of Athens, 157 71 Athens, Greece}
\affiliation{Institut de Fisica d'Altes Energies, ICREA, Universitat Autonoma de Barcelona, E-08193, Bellaterra (Barcelona), Spain}
\affiliation{Baylor University, Waco, Texas 76798, USA}
\affiliation{Istituto Nazionale di Fisica Nucleare Bologna, $^{ee}$University of Bologna, I-40127 Bologna, Italy}
\affiliation{University of California, Davis, Davis, California 95616, USA}
\affiliation{University of California, Los Angeles, Los Angeles, California 90024, USA}
\affiliation{Instituto de Fisica de Cantabria, CSIC-University of Cantabria, 39005 Santander, Spain}
\affiliation{Carnegie Mellon University, Pittsburgh, Pennsylvania 15213, USA}
\affiliation{Enrico Fermi Institute, University of Chicago, Chicago, Illinois 60637, USA}
\affiliation{Comenius University, 842 48 Bratislava, Slovakia; Institute of Experimental Physics, 040 01 Kosice, Slovakia}
\affiliation{Joint Institute for Nuclear Research, RU-141980 Dubna, Russia}
\affiliation{Duke University, Durham, North Carolina 27708, USA}
\affiliation{Fermi National Accelerator Laboratory, Batavia, Illinois 60510, USA}
\affiliation{University of Florida, Gainesville, Florida 32611, USA}
\affiliation{Laboratori Nazionali di Frascati, Istituto Nazionale di Fisica Nucleare, I-00044 Frascati, Italy}
\affiliation{University of Geneva, CH-1211 Geneva 4, Switzerland}
\affiliation{Glasgow University, Glasgow G12 8QQ, United Kingdom}
\affiliation{Harvard University, Cambridge, Massachusetts 02138, USA}
\affiliation{Division of High Energy Physics, Department of Physics, University of Helsinki and Helsinki Institute of Physics, FIN-00014, Helsinki, Finland}
\affiliation{University of Illinois, Urbana, Illinois 61801, USA}
\affiliation{The Johns Hopkins University, Baltimore, Maryland 21218, USA}
\affiliation{Institut f\"{u}r Experimentelle Kernphysik, Karlsruhe Institute of Technology, D-76131 Karlsruhe, Germany}
\affiliation{Center for High Energy Physics: Kyungpook National University, Daegu 702-701, Korea; Seoul National University, Seoul 151-742, Korea; Sungkyunkwan University, Suwon 440-746, Korea; Korea Institute of Science and Technology Information, Daejeon 305-806, Korea; Chonnam National University, Gwangju 500-757, Korea; Chonbuk National University, Jeonju 561-756, Korea; Ewha Womans University, Seoul, 120-750, Korea}
\affiliation{Ernest Orlando Lawrence Berkeley National Laboratory, Berkeley, California 94720, USA}
\affiliation{University of Liverpool, Liverpool L69 7ZE, United Kingdom}
\affiliation{University College London, London WC1E 6BT, United Kingdom}
\affiliation{Centro de Investigaciones Energeticas Medioambientales y Tecnologicas, E-28040 Madrid, Spain}
\affiliation{Massachusetts Institute of Technology, Cambridge, Massachusetts 02139, USA}
\affiliation{Institute of Particle Physics: McGill University, Montr\'{e}al, Qu\'{e}bec H3A~2T8, Canada; Simon Fraser University, Burnaby, British Columbia V5A~1S6, Canada; University of Toronto, Toronto, Ontario M5S~1A7, Canada; and TRIUMF, Vancouver, British Columbia V6T~2A3, Canada}
\affiliation{University of Michigan, Ann Arbor, Michigan 48109, USA}
\affiliation{Michigan State University, East Lansing, Michigan 48824, USA}
\affiliation{Institution for Theoretical and Experimental Physics, ITEP, Moscow 117259, Russia}
\affiliation{University of New Mexico, Albuquerque, New Mexico 87131, USA}
\affiliation{The Ohio State University, Columbus, Ohio 43210, USA}
\affiliation{Okayama University, Okayama 700-8530, Japan}
\affiliation{Osaka City University, Osaka 588, Japan}
\affiliation{University of Oxford, Oxford OX1 3RH, United Kingdom}
\affiliation{Istituto Nazionale di Fisica Nucleare, Sezione di Padova-Trento, $^{ff}$University of Padova, I-35131 Padova, Italy}
\affiliation{University of Pennsylvania, Philadelphia, Pennsylvania 19104, USA}
\affiliation{Istituto Nazionale di Fisica Nucleare Pisa, $^{gg}$University of Pisa, $^{hh}$University of Siena and $^{ii}$Scuola Normale Superiore, I-56127 Pisa, Italy, $^{mm}$INFN Pavia and University of Pavia, I-27100 Pavia, Italy}
\affiliation{University of Pittsburgh, Pittsburgh, Pennsylvania 15260, USA}
\affiliation{Purdue University, West Lafayette, Indiana 47907, USA}
\affiliation{University of Rochester, Rochester, New York 14627, USA}
\affiliation{The Rockefeller University, New York, New York 10065, USA}
\affiliation{Istituto Nazionale di Fisica Nucleare, Sezione di Roma 1, $^{jj}$Sapienza Universit\`{a} di Roma, I-00185 Roma, Italy}
\affiliation{Texas A\&M University, College Station, Texas 77843, USA}
\affiliation{Istituto Nazionale di Fisica Nucleare Trieste/Udine; $^{nn}$University of Trieste, I-34127 Trieste, Italy; $^{kk}$University of Udine, I-33100 Udine, Italy}
\affiliation{University of Tsukuba, Tsukuba, Ibaraki 305, Japan}
\affiliation{Tufts University, Medford, Massachusetts 02155, USA}
\affiliation{University of Virginia, Charlottesville, Virginia 22906, USA}
\affiliation{Waseda University, Tokyo 169, Japan}
\affiliation{Wayne State University, Detroit, Michigan 48201, USA}
\affiliation{University of Wisconsin, Madison, Wisconsin 53706, USA}
\affiliation{Yale University, New Haven, Connecticut 06520, USA}

\author{T.~Aaltonen}
\affiliation{Division of High Energy Physics, Department of Physics, University of Helsinki and Helsinki Institute of Physics, FIN-00014, Helsinki, Finland}
\author{S.~Amerio}
\affiliation{Istituto Nazionale di Fisica Nucleare, Sezione di Padova-Trento, $^{ff}$University of Padova, I-35131 Padova, Italy}
\author{D.~Amidei}
\affiliation{University of Michigan, Ann Arbor, Michigan 48109, USA}
\author{A.~Anastassov$^x$}
\affiliation{Fermi National Accelerator Laboratory, Batavia, Illinois 60510, USA}
\author{A.~Annovi}
\affiliation{Laboratori Nazionali di Frascati, Istituto Nazionale di Fisica Nucleare, I-00044 Frascati, Italy}
\author{J.~Antos}
\affiliation{Comenius University, 842 48 Bratislava, Slovakia; Institute of Experimental Physics, 040 01 Kosice, Slovakia}
\author{G.~Apollinari}
\affiliation{Fermi National Accelerator Laboratory, Batavia, Illinois 60510, USA}
\author{J.A.~Appel}
\affiliation{Fermi National Accelerator Laboratory, Batavia, Illinois 60510, USA}
\author{T.~Arisawa}
\affiliation{Waseda University, Tokyo 169, Japan}
\author{A.~Artikov}
\affiliation{Joint Institute for Nuclear Research, RU-141980 Dubna, Russia}
% ...

\collaboration{CDF Collaboration\footnote{With visitors from
$^a$Istituto Nazionale di Fisica Nucleare, Sezione di Cagliari, 09042 Monserrato (Cagliari), Italy,
$^b$University of California Irvine, Irvine, CA 92697, USA,
$^e$Institute of Physics, Academy of Sciences of the Czech Republic, 182~21, Czech Republic,
$^f$CERN, CH-1211 Geneva, Switzerland,
$^g$Cornell University, Ithaca, NY 14853, USA,
$^h$University of Cyprus, Nicosia CY-1678, Cyprus,
$^i$Office of Science, U.S. Department of Energy, Washington, DC 20585, USA,
$^j$University College Dublin, Dublin 4, Ireland,
$^k$ETH, 8092 Z\"{u}rich, Switzerland,
$^l$University of Fukui, Fukui City, Fukui Prefecture, Japan 910-0017,
$^m$Universidad Iberoamericana, Lomas de Santa Fe, M\'{e}xico, C.P. 01219, Distrito Federal,
$^n$University of Iowa, Iowa City, IA 52242, USA,
$^o$Kinki University, Higashi-Osaka City, Japan 577-8502,
$^p$Kansas State University, Manhattan, KS 66506, USA,
$^q$Brookhaven National Laboratory, Upton, NY 11973, USA,
$^r$University of Manchester, Manchester M13 9PL, United Kingdom,
$^s$Queen Mary, University of London, London, E1 4NS, United Kingdom,
$^t$University of Melbourne, Victoria 3010, Australia,
$^u$Muons, Inc., Batavia, IL 60510, USA,
$^v$Nagasaki Institute of Applied Science, Nagasaki 851-0193, Japan,
$^w$National Research Nuclear University, Moscow 115409, Russia,
$^x$Northwestern University, Evanston, IL 60208, USA,
$^y$University of Notre Dame, Notre Dame, IN 46556, USA,
$^z$Universidad de Oviedo, E-33007 Oviedo, Spain,
$^{aa}$CNRS-IN2P3, Paris, F-75205 France,
$^{cc}$Universidad Tecnica Federico Santa Maria, 110v Valparaiso, Chile,
$^{dd}$Yarmouk University, Irbid 211-63, Jordan,
$^{ll}$Universite catholique de Louvain, 1348 Louvain-La-Neuve, Belgium,
$^{oo}$University of Z\"{u}rich, 8006 Z\"{u}rich, Switzerland,
$^{pp}$Massachusetts General Hospital and Harvard Medical School, Boston, MA 02114 USA,
$^{qq}$Hampton University, Hampton, VA 23668, USA,
$^{rr}$Los Alamos National Laboratory, Los Alamos, NM 87544, USA
}}
\noaffiliation

\begin{abstract}

We present new measurements of the inclusive forward-backward $\ttbar$ production asymmetry, $\afb$, and its dependence on several properties of the $\ttbar$ system.  The measurements are performed with the full Tevatron data set recorded with the CDF~II detector during $\ppbar$ collisions at $\sqrt{s} = 1.96$~TeV, corresponding to an integrated luminosity of 9.4 fb$^{-1}$.  We measure the asymmetry using the rapidity difference $\dy=y_{t}-y_{\bar{t}}$. Parton-level results are derived, yielding an inclusive asymmetry of $0.164\pm0.047$ (stat + syst).  We establish an approximately linear dependence of $\afb$ on the top-quark pair mass $\mttb$ and the rapidity difference $|\dy|$ at detector and parton levels.  Assuming the standard model, the probabilities to observe the measured values or larger for the detector-level dependencies are $7.4 \times 10^{-3}$ and $2.2 \times 10^{-3}$ for $\mttb$ and $|\dy|$ respectively.  Lastly, we study the dependence of the asymmetry on the transverse momentum of the $\ttbar$ system at the detector level.  These results are consistent with previous lower-precision measurements and provide additional quantification of the functional dependencies of the asymmetry.
\end{abstract}


\maketitle

\section{Introduction}\label{intro}

The creation of top quarks in $\ppbar$ collisions offers a unique test of pair-production in quantum chromodynamics (QCD) at very large momentum transfer as well as a promising potential avenue for the observation of new physical phenomena. Given the very large mass of the top quark, exotic processes may couple more strongly to top quarks than to the other known fundamental particles, and possible hints of new interactions could be first observed in top-quark production.  In particular, asymmetries in $\ttbar$ production could provide the first evidence of new interactions, such as $\ttbar$ production via a heavy axial color octet or a flavor-changing $Z^{\prime}$ boson, that might not be easily observed as excesses in the top quark production rate or as resonances in the $\ttbar$ invariant mass distribution.

The CDF and D0 collaborations have previously reported on forward-backward asymmetries ($\afb$) in $\ppbar\rightarrow \ttbar$ production at $\sqrt{s} = 1.96$~TeV at the Fermilab Tevatron. In the standard model (SM), the $\ttbar$ production process is approximately symmetric in production angle, with a $\mathcal{O}$(7\%) charge asymmetry arising at next-to-leading order (NLO) and beyond~\cite{nlotheory}.  Using a sample corresponding to $5.3~\ifb$ of integrated luminosity, CDF measured a parton-level asymmetry $\afb = 0.158 \pm 0.074$~\cite{cdfafb} in the lepton+jets decay channel ($\ttbar \rightarrow (W^{+}b)(W^{-}b) \rightarrow (l^{+}\nu)(q\bar{q^{\prime}})b\bar{b}$~\cite{cc}), and very good agreement was found by the D0 measurement $\afb = 0.196\pm 0.065$~\cite{d0afb} in a lepton+jets sample corresponding to $5.4~\ifb$. CDF and D0 have also performed simple differential measurements using two bins each in the top-antitop rapidity difference $|\dy|$ and the top-antitop invariant mass $\mttb$. The two experiments agreed on a large $|\dy|$ dependence.  CDF also saw a large $\mttb$ dependence, and while that observed at D0 was smaller, the CDF and D0 results were statistically consistent.  One of the aims of this paper is to clarify the $|\dy|$ and $\mttb$ dependence of the asymmetry using the full CDF data set.

The $5~\ifb$ results have stimulated new theoretical work, both within and outside the context of the SM.  The SM calculation has been improved by calculations of electroweak processes that contribute to the asymmetry, studies of the choice of renormalization scale, and progress on a next-to-next-to-leading order (NNLO) calculation of the asymmetry~\cite{hollikpagani,kuhnrodrigo,manohartrott,brodsky,nnlo_xsec}.  The new calculations result in a small increase in the expected asymmetry, but not enough to resolve the tension with observation. Other work has focused on the dependence of the asymmetry on the transverse momentum of the $\ttbar$ system~\cite{ttpt}, on which we report here.  

A number of speculative papers invoke new interactions in the top sector~\cite{np} to explain the large asymmetry. In one class of models, $\ttbar$ pairs can be produced via new axial $s$-channel particles arising from extended gauge symmetries or extra dimensions. For these models, the asymmetry is caused by interference between the new $s$-channel mediator and the SM gluon.  In other models, light $t$-channel particles with flavor-violating couplings create an asymmetry via a $u$, $d \to t$ flavor change into the forward Rutherford-scattering peak.  All potential models of new interactions must accommodate the apparent consistency of the measured cross section and $\mttb$ spectrum with the SM predictions. Tevatron and LHC searches for related phenomena, such as di-jet resonances, same-sign tops, and other exotic processes, can provide additional experimental limits on potential models.  Measurements by the LHC experiments of the top-quark charge asymmetry $A_{\rm C}$, an observable that is distinct from $\afb$ but correlated with it, have found no significant disagreement with the SM~\cite{aclhc}; however, any observable effect at the LHC is expected to be small, and the nature of the relationship between $\afb$ and $A_{\rm C}$ is model-dependent~\cite{afbvsac}.  A more precise measurement of the Tevatron forward-backward asymmetry and its mass and rapidity dependence may help untangle the potential new physics sources for $\afb$ from the standard model and from each other.

This paper reports on a study of the asymmetry in the lepton+jets topology, with several new features compared to the previous CDF analysis in this channel~\cite{cdfafb}.  We use the complete Tevatron Run~II data set with an integrated luminosity of $9.4~\ifb$.  We additionally expand the event selection by including events triggered by large missing transverse energy and multiple hadronic jets, increasing the total data set by approximately $30\%$ beyond what is gained by the increase in luminosity.  In total, the number of candidate events in this analysis is more than twice the number of events used in Ref.~\cite{cdfafb}.  An improved NLO Monte Carlo generator is used to describe the predicted $\ttbar$ signal, and we also add small corrections reflecting new results on the electroweak contributions to the asymmetry~\cite{hollikpagani,kuhnrodrigo,manohartrott}.  Finally, parton-level shape corrections utilize an improved algorithm which yields binned parton-level measurements of the rapidity and mass dependence of the asymmetry. We also study the dependence of the asymmetry on the $\ttbar$ transverse momentum, $\ptran$, showing that the modeling of this quanity is robust, and that the excess asymmetry above the SM prediction is consistent with being independent of $\ptran$. 

\section{Expected asymmetries and Monte Carlo models}\label{sec:mc}

The asymmetry is measured using the difference of the $t$ and $\tbar$ rapidities, $\dy = \yt - \ytbar$, where the rapidity $y$ is given by

\begin{equation} \label{eq:rap}
   y   =   \frac{1}{2} \ln \left( \frac{E + p_z}{E - p_z} \right),  
\end{equation} 

\noindent with $E$ being the total top-quark energy and $p_z$ being the component of the top-quark momentum along the beam axis as measured in the detector rest frame.  $\dy$ is invariant to boosts along the beamline, and in the limit where the transverse momentum of the $\ttbar$ system is small, the forward-backward asymmetry  

\begin{equation} \label{eq:afb}
   \afb   =   \frac{N(\dy > 0) - N(\dy< 0)}{N(\dy > 0) + N(\dy < 0)}   
\end{equation} 

\noindent is identical to the asymmetry in the top-quark production angle in the experimentally well-defined $\ttbar$ rest frame.  The standard model predictions for the top-quark asymmetry referenced in this paper are based on the NLO event generator \powheg~\cite{powheg}, using the {\sc cteq6.1M} set of parton-distribution functions (PDFs), validated by comparing \powheg to the NLO generator \mcnlo~\cite{mcnlo} as well as the NLO calculation of \mcfm~\cite{mcfm}. We find good consistency overall, as shown in Table~\ref{mc:tab:predictions}~\cite{nlo_v_lo}. Sources of asymmetry from electroweak processes in the standard model that are not included in the \powheg calculations~\cite{hollikpagani,kuhnrodrigo,manohartrott} lead to an overall increase of the asymmetry by a factor of 26\% of the QCD expectation. This is included in all the predictions shown in Table~\ref{mc:tab:predictions} and in all predicted asymmetries and $\dy$ distributions in this paper. The electroweak asymmetry is assumed to have the same $\mttb$ and $\dy$ dependence as the QCD asymmetry, and we apply a simple 26\% rescaling to the {\sc powheg} predictions there as well. Following Ref.~\cite{errors}, we include a $30\%$ uncertainty on all theoretical predictions for the SM asymmetry due to the choice of renormalization scale.

\begin{table*}[ht]
\caption{Parton-level asymmetry predictions of \powheg, \mcnlo, and \mcfm after applying electroweak corrections.\label{mc:tab:predictions}}
\begin{center}
\begin{tabular}{cccc}
\hline
\hline
                           & \mcnlo    & \powheg  & \mcfm \\ \hline
Inclusive                  & \phantom{0}$0.067 \pm 0.020$\phantom{0}  & \phantom{0}$0.066 \pm 0.020$\phantom{0} & \phantom{0}$0.073 \pm 0.022$\phantom{0} \\ \hline
$\absdely < 1$             & \phantom{0}$0.047 \pm 0.014$\phantom{0}  & \phantom{0}$0.043 \pm 0.013$\phantom{0} & \phantom{0}$0.049 \pm 0.015$\phantom{0} \\
$\absdely > 1$             & \phantom{0}$0.130 \pm 0.039$\phantom{0}  & \phantom{0}$0.139 \pm 0.042$\phantom{0} & \phantom{0}$0.150 \pm 0.045$\phantom{0} \\ \hline
$\mttb < 450~\gevcc$       & \phantom{0}$0.054 \pm 0.016$\phantom{0}  & \phantom{0}$0.047 \pm 0.014$\phantom{0} & \phantom{0}$0.050 \pm 0.015$\phantom{0} \\
$\mttb > 450~\gevcc$       & \phantom{0}$0.089 \pm 0.027$\phantom{0}  & \phantom{0}$0.100 \pm 0.030$\phantom{0} & \phantom{0}$0.110 \pm 0.033$\phantom{0}\\ 
\hline
\hline
\end{tabular}
\end{center}
\end{table*}

To test the analysis methodology in the case of a large asymmetry, we study two models in which an asymmetry is generated by the interference of the gluon with massive axial color-octet particles. Each provides a reasonable approximation of the observed data in presenting a large, positive forward-backward asymmetry, while also being comparable to the Tevatron data in other important variables such as the $\ttbar$ invariant mass, $\mttb$.

The first model, Octet A, contains an axigluon with a mass of 2~\tevcc. This hypothetical particle is massive enough that the pole is observed as only a small excess in the tail of the $\mttb$ spectrum, but it creates an asymmetry via the interference between the off-shell axigluon and the SM gluon. The couplings are tuned ($g_V(q) = g_V(t) = 0$, $g_A(q) = 3$, $g_A(t) = -3$, where $q$ refers to light-quark couplings and $t$ to top-quark couplings) to produce a parton-level asymmetry consistent with the measurement in Ref.~\cite{cdfafb}.  The second model, Octet B, contains an axigluon with the same couplings, but a smaller mass of 1.8~\tevcc.  This model produces a larger excess in the tail of the $\mttb$ spectrum and an even larger asymmetry than Octet A, allowing the measurement procedure to be tested in a regime with a very large asymmetry.

Both models are simulated using the leading order (LO) \textsc{madgraph}~\cite{madgraph,tait} Monte Carlo generator and are hadronized with \pythia~\cite{pythia} before being passed to the CDF detector simulation and reconstruction software. We emphasize that these are not hypotheses - the physical applicability of these models is, in fact, quite constrained by $\ttbar$ resonance searches at the LHC~\cite{lhcresonances}.  Rather, these models are used as controlled inputs to study the performance of the analysis in the presence of large asymmetries.  Further information about these models can be found in Ref.~\cite{cdfafb}.

\section{Measurement strategy}\label{sec:strategy}

The analysis takes place in several steps. We first consider the asymmetry observed at the reconstruction level in all selected events. Next, to study the asymmetry for a pure sample of $\ttbar$ events as recorded in the detector, the calculated non-$\ttbar$ background contribution is subtracted and the appropriate systematic uncertainties related to the background prediction are applied.  Finally, to study the asymmetry at the parton level, corrections are applied for the event reconstruction and detector acceptance, along with appropriate systematic uncertainties on the signal modeling. The reconstruction- and background-subtracted-level measurements have the advantage of fewer assumptions, while the parton-level measurement allows direct comparison to theory predictions. 

After reviewing the event selection and reconstruction in Sec.~\ref{sec:data}, we describe the various steps of the correction procedure in detail and apply them to the $\dy$ distribution and the inclusive $\afb$ measurement in Sec.~\ref{sec:inclusive_afb}.  In Sec.~\ref{sec:afb_v_dy} and Sec.~\ref{sec:afb_v_mtt}, we study the dependence of the asymmetry on $|\dy|$ and $\mttb$,  $A_{\rm FB}(|\dy|)$ and $A_{\rm FB}(\mttb)$ respectively, at all three stages of correction, and Sec.~\ref{sec:sig} discusses the significance of discrepancies observed in these dependencies between the data and the SM.  Section~\ref{sec:afb_v_pt} discusses the dependence of the asymmetry on the $\ttbar$ transverse momentum.

\section{Detector, event selection, and reconstruction}\label{sec:data}

The data sample corresponds to an integrated luminosity of $9.4~\ifb$ recorded with the CDF~II detector during $\ppbar$ collisions at $1.96$~TeV. CDF~II is a general purpose, azimuthally and forward-backward symmetric magnetic spectrometer with calorimeters and muon detectors~\cite{cdf}.  Charged particle trajectories are measured with a silicon-microstrip detector surrounded by a large open-cell drift chamber, both within a 1.4 T solenoidal magnetic field.  The solenoid is surrounded by pointing-tower-geometry electromagnetic and hadronic calorimeters for the measurement of particle energies and missing energy reconstruction.  Surrounding the calorimeters, scintillators and proportional chambers provide muon identification.  We use a cylindrical coordinate system with the origin at the center of the detector and the $z$-axis along the direction of the proton beam~\cite{coords}.

This measurement selects $t\bar{t}$ candidate events in the lepton+jets topology, where one top quark decays semileptonically ($t \rightarrow Wb \rightarrow l \nu b$) and the other hadronically ($t \rightarrow Wb \rightarrow q \bar{q}^{\prime} b$). We detect the lepton and hadronization-induced jets.  The presence of missing transverse energy ($\met$)~\cite{coords} is used to infer the passage of a neutrino through the detector.  Detector readout is initiated in one of two ways: either by indications of a high-momentum lepton (electron or muon) in the central portion of the detector or by events with indications of large $\met$ and at least two energetic jets.  Events collected in the second manner, in which we require the presence of muon candidates reconstructed offline,  make up the ``loose muon'' sample, a new addition compared to the previous version of this analysis.  After offline event reconstruction, we require that all candidate events contain exactly one electron or muon with $\etran(\peetee) >20$~GeV(GeV/$c$) and $|\eta| < 1.0$, as well as four or more hadronic jets with $E_T >20$~GeV and $|\eta| < 2.0$.  Jets are reconstructed using a cone algorithm with $\Delta R = \sqrt{\Delta\phi^2+\Delta\eta^2} < 0.4$, and calorimeter signals are corrected for various detector and measurement effects as described in Ref.~\cite{jes}. We require $\met > 20$~GeV, consistent with the presence of an undetected neutrino. We finally require that $H_T$, the scalar sum of the transverse energy of the lepton, jets, and $\met$, be $H_T > 220$~GeV.  This requirement reduces the backgrounds by $17\%$ while accepting $97\%$ of signal events. The {\sc secvtx} algorithm~\cite{secvtx} is used to identify $b$ jets by searching for displaced decay vertices within the jet cones, and at least one jet in each event must contain such a ``$b$ tag''.  The coverage of the tracking detector limits the acceptance for jets with identified $b$ tags to $|\eta| < 1$. 

The sample passing this selection, including the $b$-tag requirement, contains 2653 candidate events. The estimated non-$\ttbar$ background in the data sample is $530 \pm 124$ events. The predominant background source is QCD-induced $W$+multi-parton events containing either $b$-tagged heavy-flavor jets or erroneously tagged light-flavor jets. These events are modeled with the {\sc alpgen} Monte Carlo generator~\cite{alpgen}, with the normalizations determined by tagging efficiencies, mis-tagging rates, and other measurements in the data. QCD multi-jet (``Non-$W$'') events containing mis-measured $\met$ and jets that are mis-identified as leptons are modeled using real data events with lepton candidates that are rejected by the lepton identification requirements.  This background, which is the most difficult to model properly, is also the one that is most efficiently suppressed by the $H_T$ requirement, which reduces it by approximately 30\%. Small backgrounds from electroweak processes ($WW$, $WZ$, single-top) are estimated using Monte Carlo generators.  The expected background contributions from each source are given in Table~\ref{tab:method2}.  We note that there are correlations among the various sources of uncertainty for the different background components, so that the total background uncertainty is not a simple sum in quadrature of the uncertainties on the individual background normalizations.  Further information about the background modeling and event selection can be found in Ref.~\cite{tZxsec}. 

\begin{table}[ht]
\caption{Expected contributions of the various background sources to the selected data.}\label{tab:method2}
\begin{center}
\begin{tabular}{c c}
\hline
\hline
Background source         &  Number of events  \\
\hline
$W$+HF              & \phantom{0}256 $\pm$ 83\phantom{0}           \\  
$W$+LF              & \phantom{0}102 $\pm$ 32\phantom{0}         \\   
Non-$W$             & \phantom{00}97 $\pm$ 50\phantom{0}           \\
Single top        & \phantom{00}35 $\pm$ 3\phantom{00}           \\ 
Diboson           & \phantom{00}21 $\pm$ 3\phantom{00}           \\
$Z$+jets            & \phantom{00}19 $\pm$ 3\phantom{00}           \\
\hline
Total background  & \phantom{0}530 $\pm$ 124           \\     
$\ttbar$ (7.4 pb) & \phantom{}2186 $\pm$ 314           \\
\hline
Total prediction  & \phantom{}2716 $\pm$ 339           \\
\hline
Data              & 2653              \\
\hline
\hline
\end{tabular}
\end{center}
\end{table}

The reconstruction of the $\ttbar$ kinematics employs the measured momenta of the lepton and the four leading jets in the event, along with the measured $\vec{\met}$.  The calculation of the $\ttbar$ four-vectors uses a $\chi^2$-based fit of the lepton and jet kinematic properties to the $\ttbar$ hypothesis.  Each of the possible jet-to-parton assignments is evaluated according to its consistency with resulting from the decay of a pair of top quarks.  Two of the observed jets are required to be consistent with being decay products of a $W$ boson, while the lepton and $\met$ must be consistent with another $W$ boson. Each $W$ boson, when paired with one of the remaining ($b$) jets, is checked for consistency with having resulted from a top-quark decay.  The lepton momentum, \met, and jet energies are allowed to float within their experimental uncertainties, and we apply the constraints that $M_W=80.4~\gevcc$, $M_t=172.5~\gevcc$, and any $b$-tagged jets must be associated with $b$ partons.  The jet-to-parton assignment that best matches these requirements is chosen to define the parent top quarks in each event.

This algorithm has been studied and validated in many precision top-quark-property analyses, including mass measurements~\cite{reco}, which remove the top-quark mass constraint, and property measurements that do make use of the mass constraint~\cite{alice}.  The top- and antitop-quark four-vectors determined from this procedure are used to find the rapidities of the quarks and the $\dy = y_{t} - y_{\bar{t}}$ variable used for the asymmetry analysis, with the charges of the reconstructed top quarks being fixed by the observed lepton charge.  In the Appendix, we discuss a high-precision test of the lepton-charge determination in a large control sample with the goal of verifying that the lepton charge assigment is well-modeled by the detector simulation.

The validity of the analysis is checked at all stages by comparison to a standard model prediction created using the {\sc powheg} $\ttbar$ model, the lepton+jets background model described above, and a full simulation of the CDF~II detector.  Figure~\ref{fig:yhad} shows the rapidity distribution for the hadronically-decaying top or antitop quark.  In the measurement of the asymmetry, the observed lepton charge is used to determine whether each entry in this distribution corresponds to a top quark or an antitop quark, and this rapidity is combined with the rapidity of the leptonically decaying quark to calculate $\dy$ for each event.  In Fig.~\ref{fig:yhad} and all that follow, the $\ttbar$ signal prediction is scaled such that the total signal normalization, when added to the background prediction in Table~\ref{tab:method2}, totals number of observed events.

\begin{figure}[!ht]
\begin{center}
\includegraphics[width=0.45\textwidth, clip]{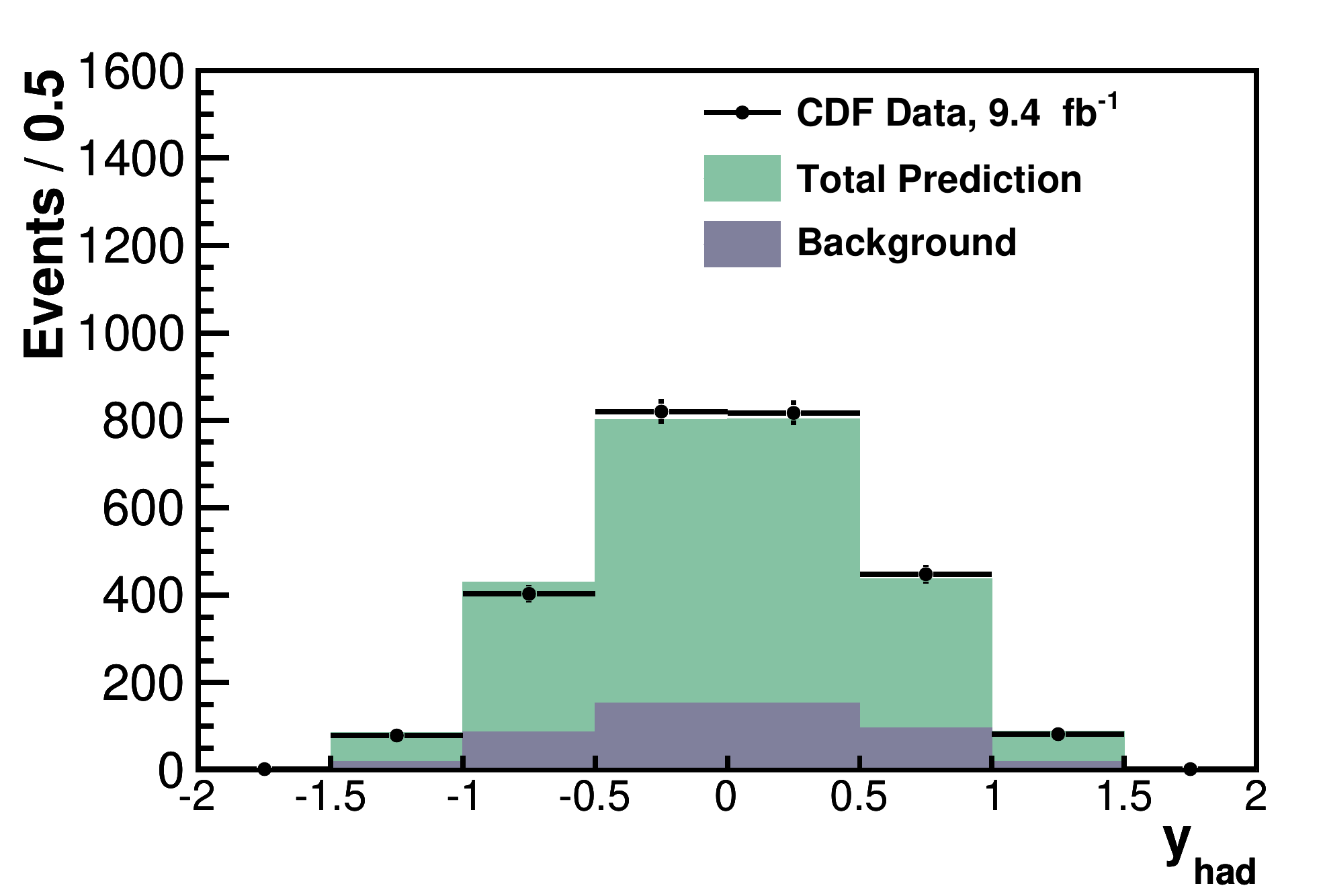}
\caption{{\small The rapidity of the hadronically-decaying top or antitop quark.} \label{fig:yhad}}
\end{center}
\end{figure}

Figure~\ref{fig:mtt} shows a comparison of the data to the prediction for the invariant mass of the $\ttbar$ system, $\mttb$; there is good agreement.  In the previous CDF analysis~\cite{cdfafb}, the forward-backward asymmetry was found to have a large dependence on this variable.   In Sec.~\ref{sec:afb_v_mtt} we report a new measurement of this dependence.  

\begin{figure}[!ht]
\begin{center}
\includegraphics[width=0.45\textwidth, clip]{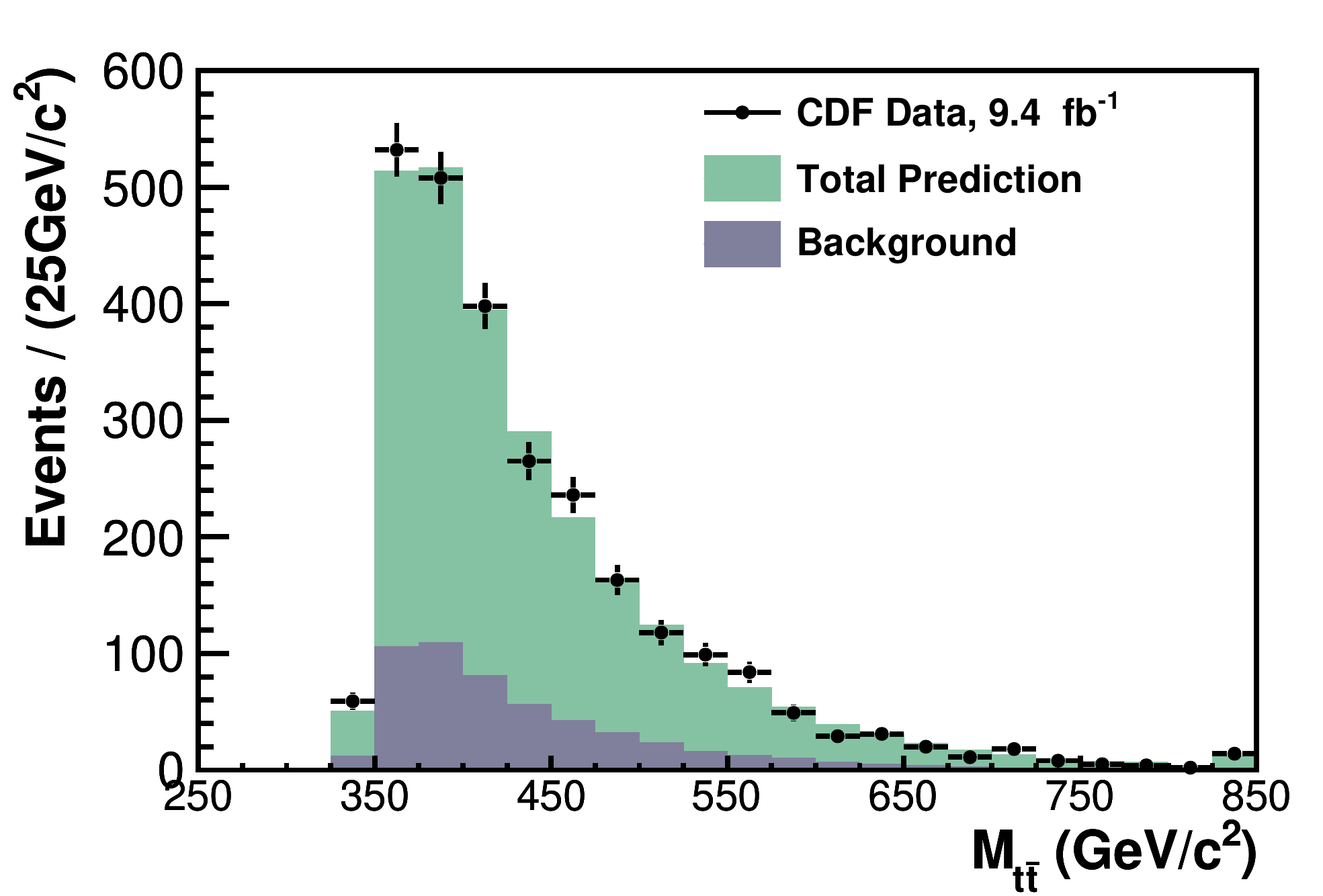}
\caption{{\small Reconstructed invariant mass of the $\ttbar$ system. The last bin contains overflow events.} \label{fig:mtt}}
\end{center}
\end{figure}

The transverse momentum of the $\ttbar$ system, $\ptran$, provides a sensitive test of the reconstruction and modeling, particularly at low momenta, where both the prediction and the reconstruction are challenged by the addition of soft gluon radiation external to the $\ttbar$ system. In Fig.~\ref{fig:ptsys_reso} we show the difference between the reconstructed and true values of the $x$-component of $\ptran$ in {\sc powheg}. The difference is centered on zero and well-fit by the sum of two gaussians with widths as shown. Most events fall in the central core with a resolution of $\sim 14~\gevc$. Doubling this in quadrature for the two transverse components gives an overall expected resolution $\delta\ptran\sim 20~\gevc$ for the bulk of the data. In Fig.~\ref{fig:ptsys} we show that the reconstructed data is in good agreement with the sum of the background prediction and the NLO $\ttbar$ model; the 10 GeV bin size here is chosen to be half the measured resolution. The $\ttbar$ forward-backward asymmetry can have a significant $\ptran$ dependence, and we discuss the expected and measured asymmetry as a function of this variable in Sec.~\ref{sec:afb_v_pt}.

\begin{figure}[!ht]
\begin{center}
\includegraphics[width=0.45\textwidth, clip]{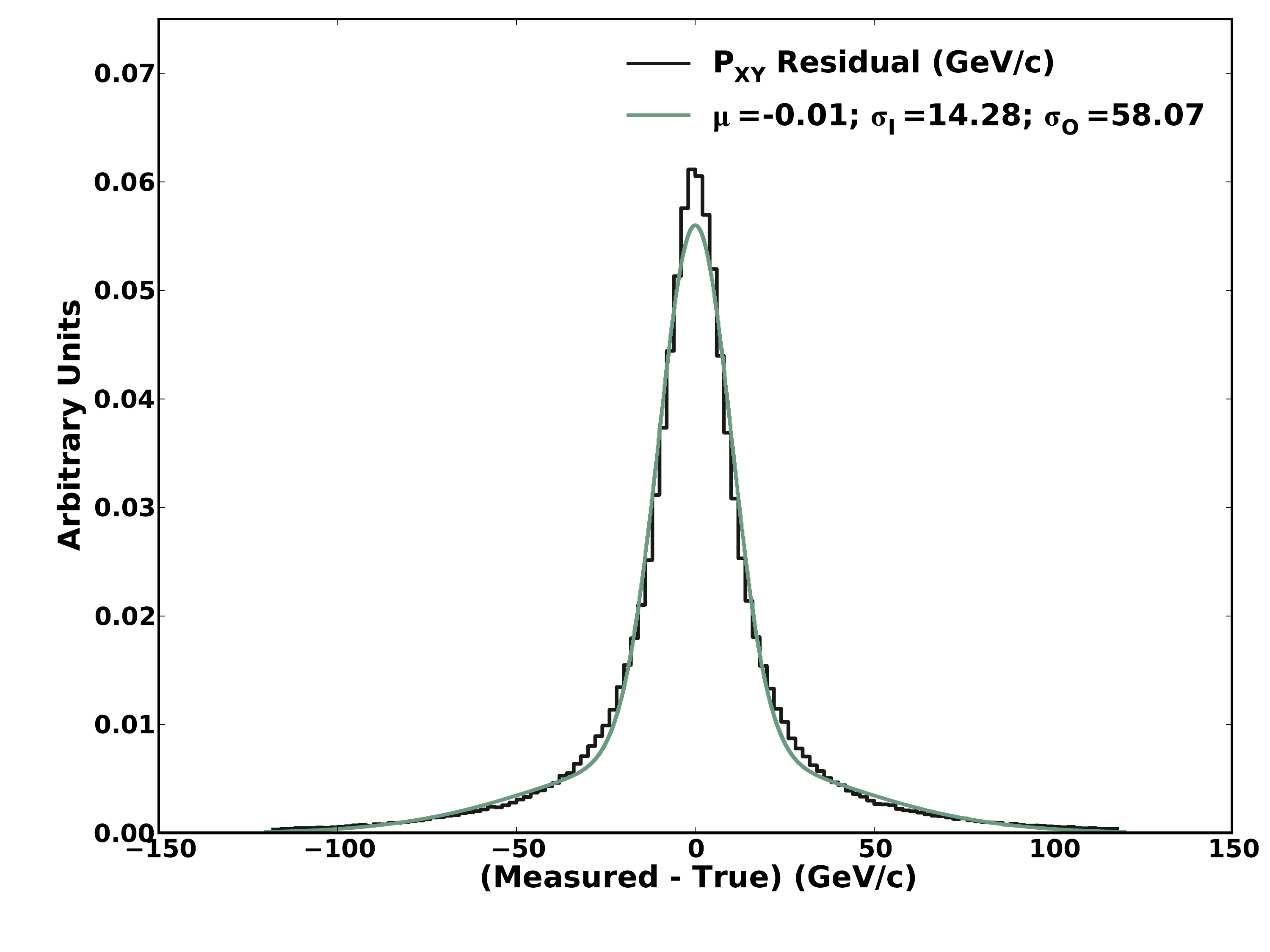}
\caption{{\small Resolution of the $x$- or $y$-component of the reconstructed $\ptran$ of the $\ttbar$ system as measured in {\sc powheg}.}\label{fig:ptsys_reso}}
\end{center}
\end{figure}

\begin{figure}[!ht]
\begin{center}
\includegraphics[width=0.45\textwidth, clip]{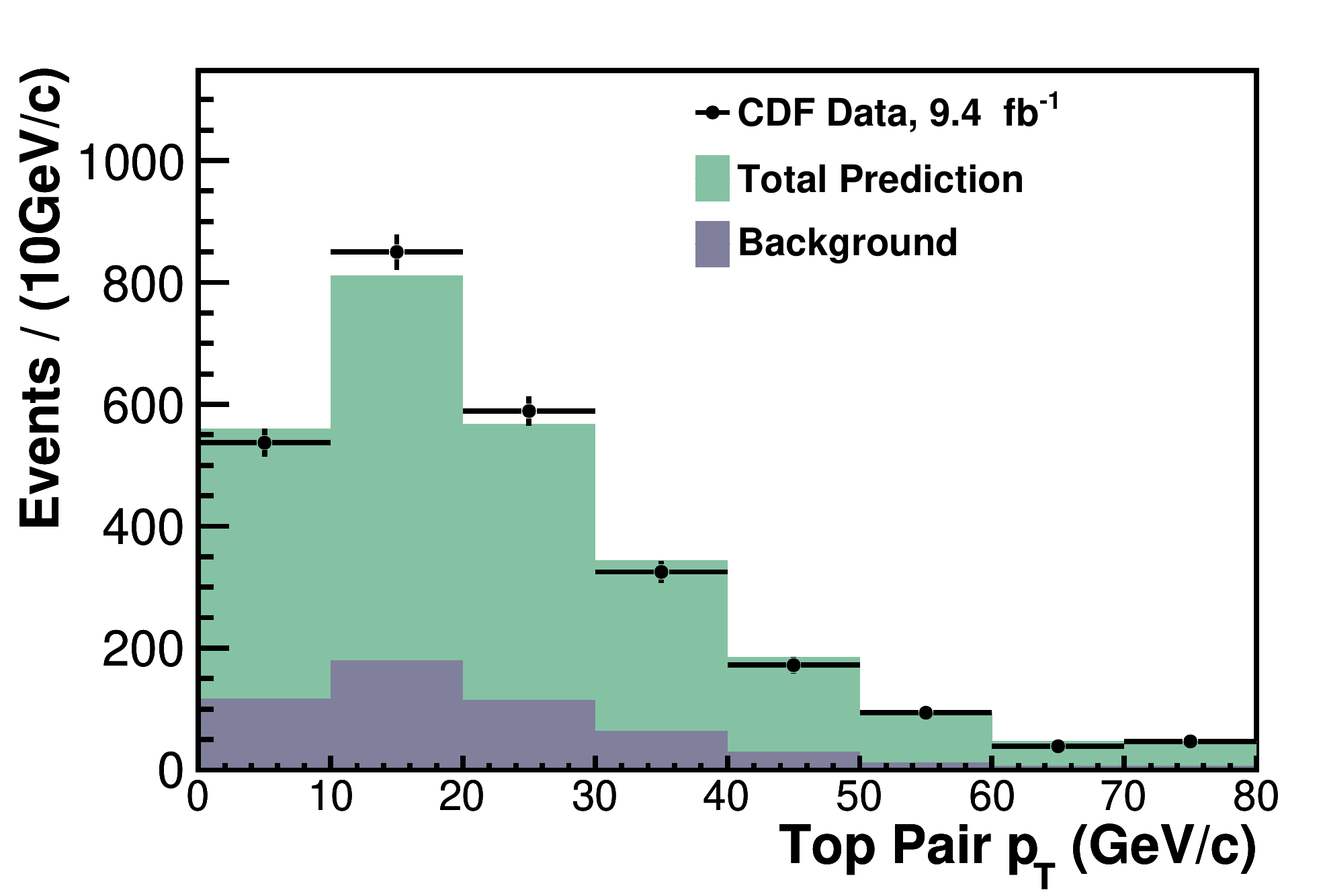}
\caption{{\small Reconstructed $\ptran$ of the $\ttbar$ system. The last bin contains overflow events.} \label{fig:ptsys}}
\end{center}
\end{figure}

\begin{figure*}[!htbp]
\begin{center}
\subfigure[]{
\includegraphics[width=0.45\textwidth, clip]{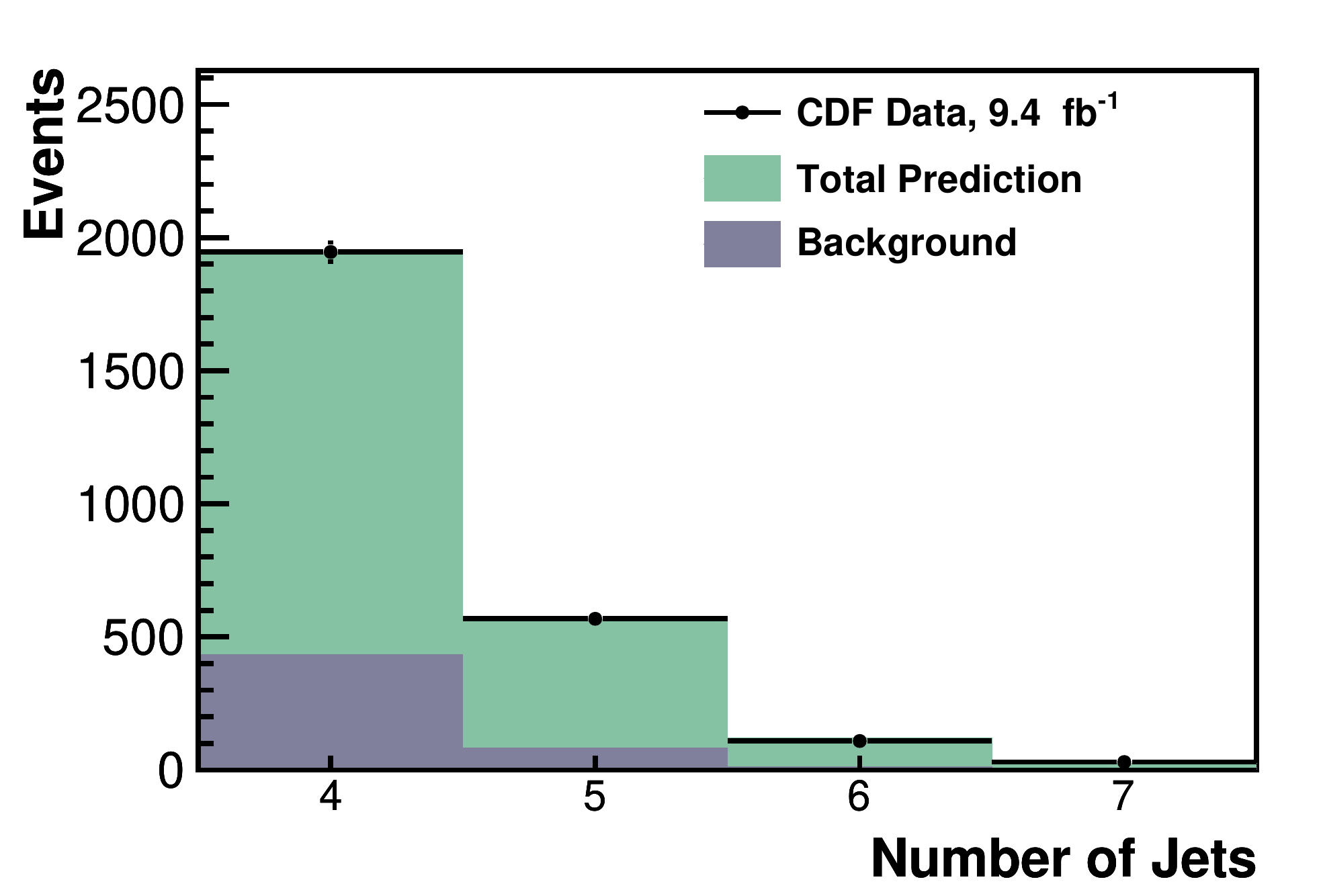}\label{fig:njet}}
\subfigure[]{
  \includegraphics[width=0.45\textwidth, clip]{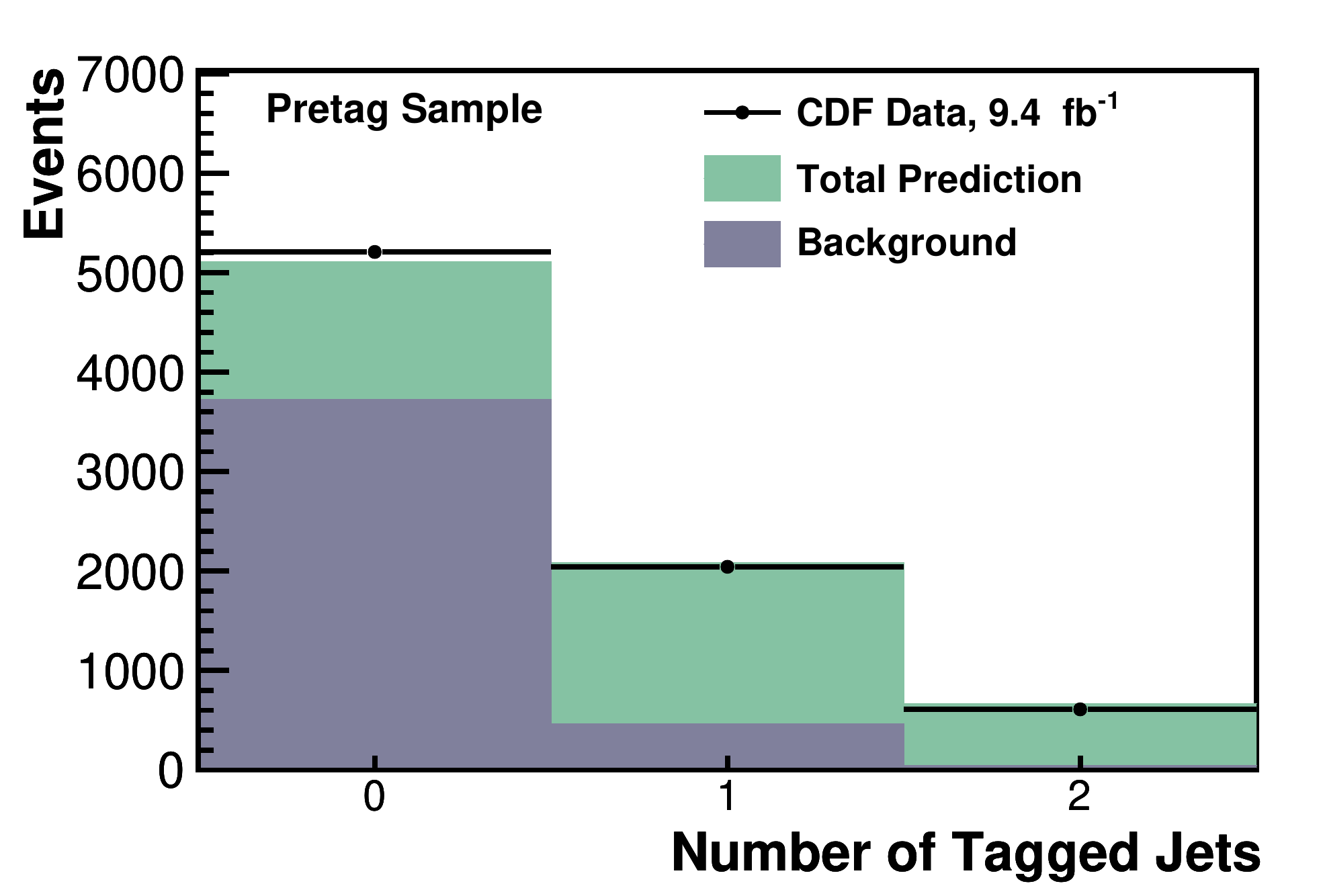}\label{fig:ntag}}
\caption{{\small \subref{fig:njet} The number of observed jets and \subref{fig:ntag} the number of jets with $b$ tags in the data compared to the signal plus background model. The last bin contains overflow events.} \label{fig:njet_ntag}}
\end{center}
\end{figure*}

We also consider a wide range of other variables, a selection of which are shown here, to validate the reconstruction algorithm and the modeling of the data set.  In Fig.~\ref{fig:njet_ntag} we show the distributions of the number of jets and number of $b$ tags in events passing the selection requirements.  Figure~\ref{fig:ntag} also includes events containing no $b$-tagged jets, which are not part of the final sample of candidate events but provide an important check on the modeling of the $b$-tagging algorithm.  Figure~\ref{fig:jetet_leppt} shows the transverse energy of the most energetic jet and the transverse momentum of the lepton, while Fig.~\ref{fig:met_ht} shows the distribution of the reconstructed $\met$ and $H_T$.   All distributions exhibit good agreement between the observed data and the model expectations.

\begin{figure*}[!htbp]
\begin{center}
\subfigure[]{
\includegraphics[width=0.45\textwidth, clip]{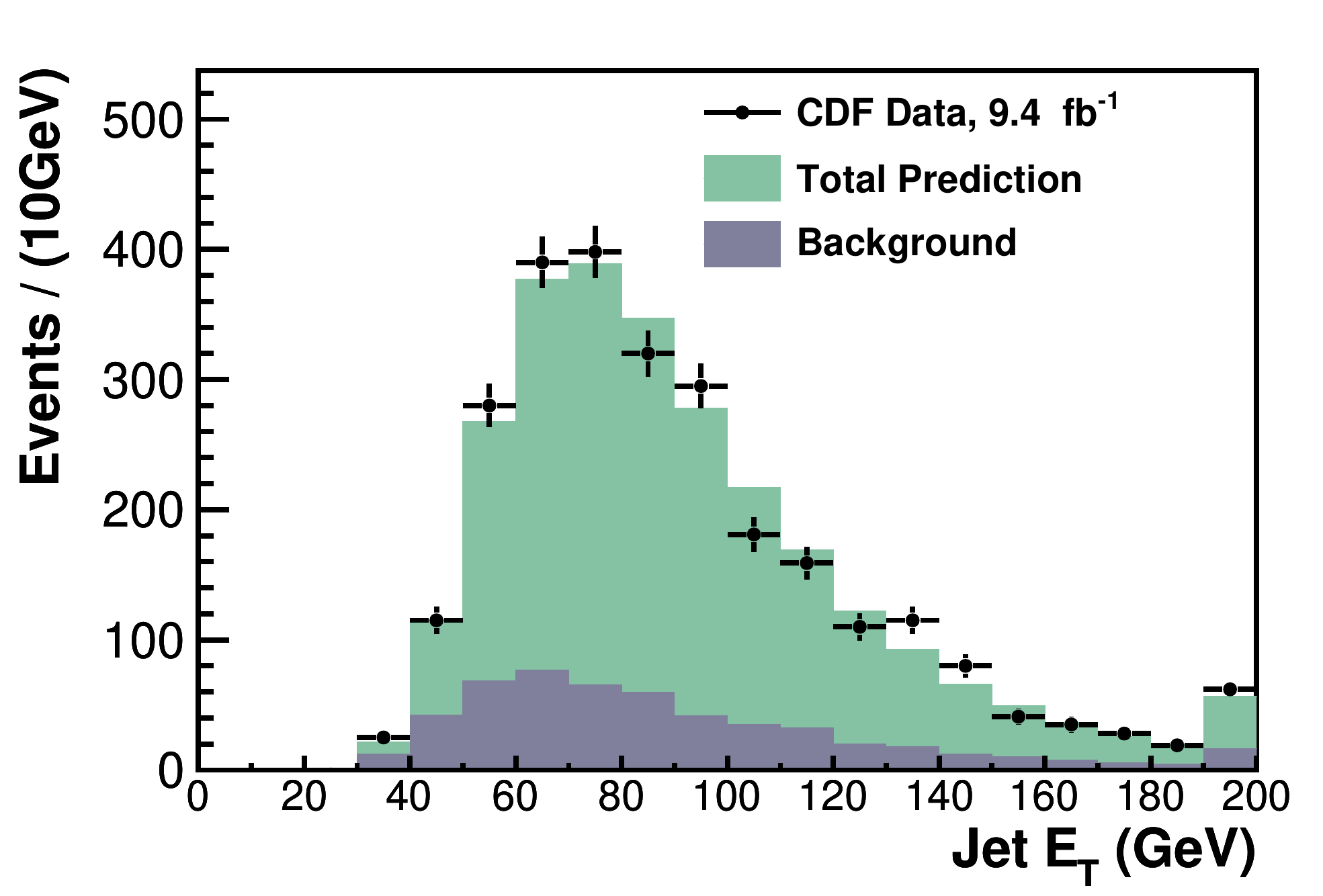}\label{fig:jetet}}
\subfigure[]{
  \includegraphics[width=0.45\textwidth, clip]{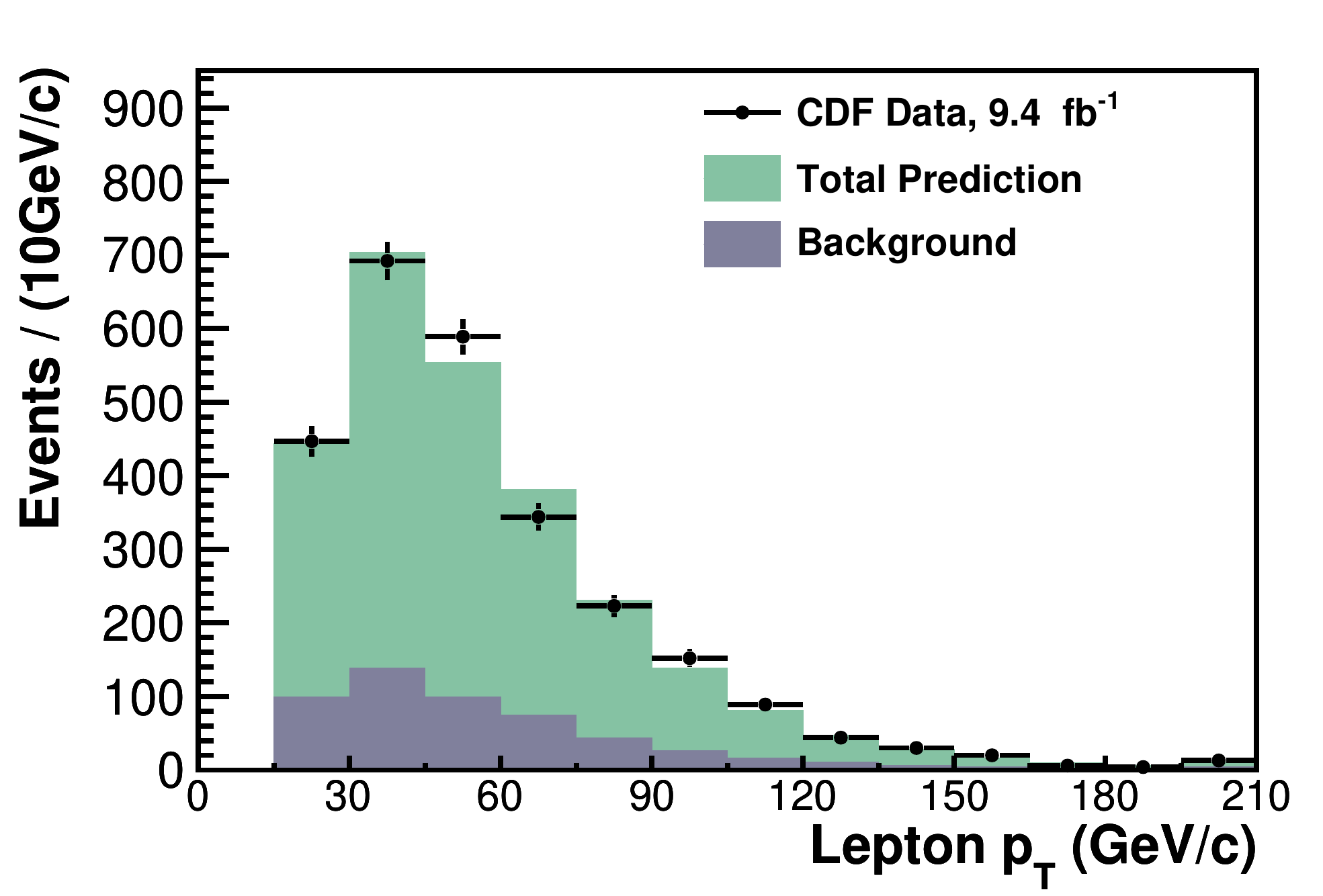}\label{fig:leppt}}
\caption{{\small \subref{fig:jetet} The $E_T$ of the most energetic jet and \subref{fig:leppt} the transverse momentum of the lepton in the data compared to the signal plus background model. The last bin contains overflow events.} \label{fig:jetet_leppt}}
\end{center}
\end{figure*}

\begin{figure*}[!htbp]
\begin{center}
\subfigure[]{
\includegraphics[width=0.45\textwidth, clip]{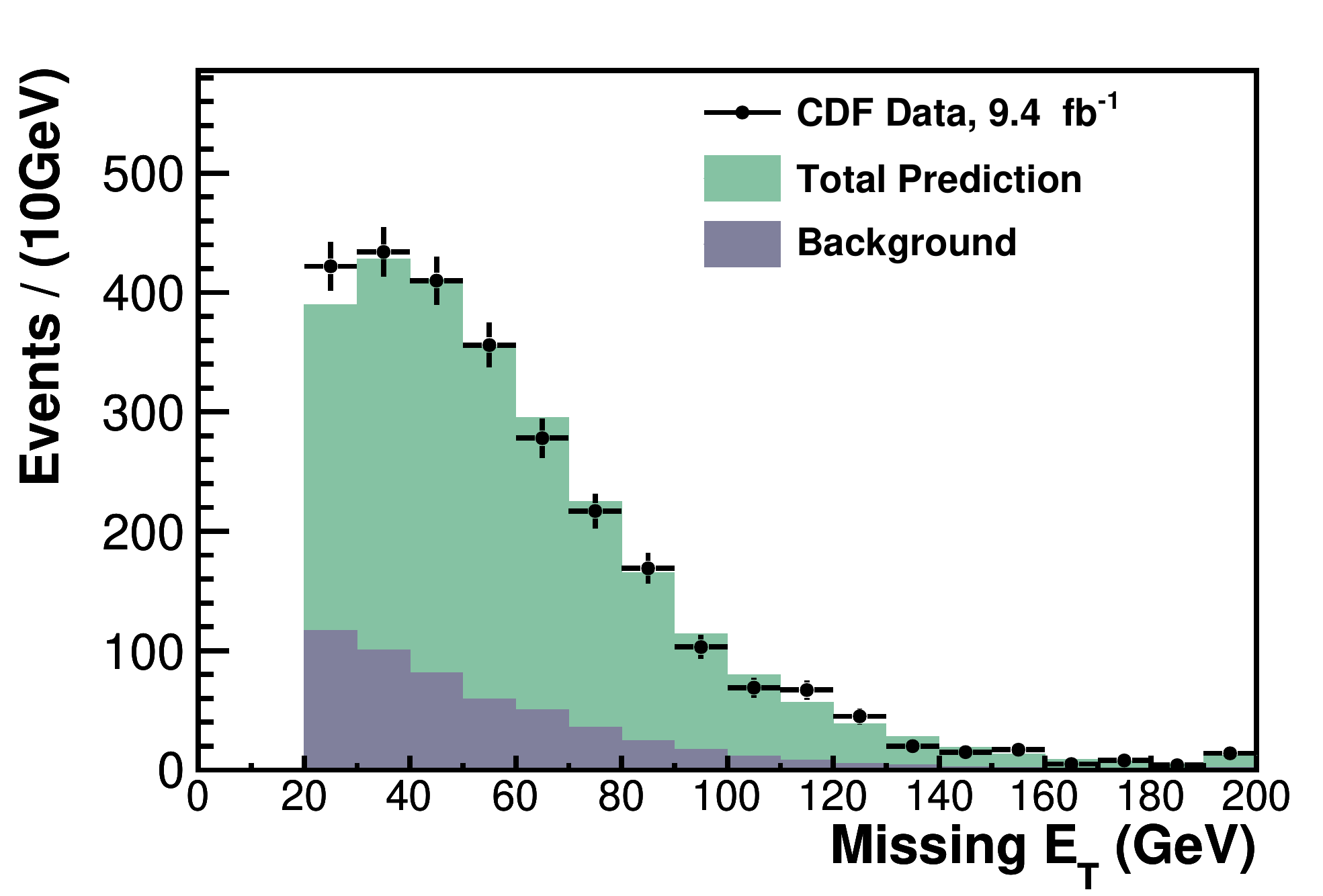}\label{fig:met}}
\subfigure[]{
  \includegraphics[width=0.45\textwidth, clip]{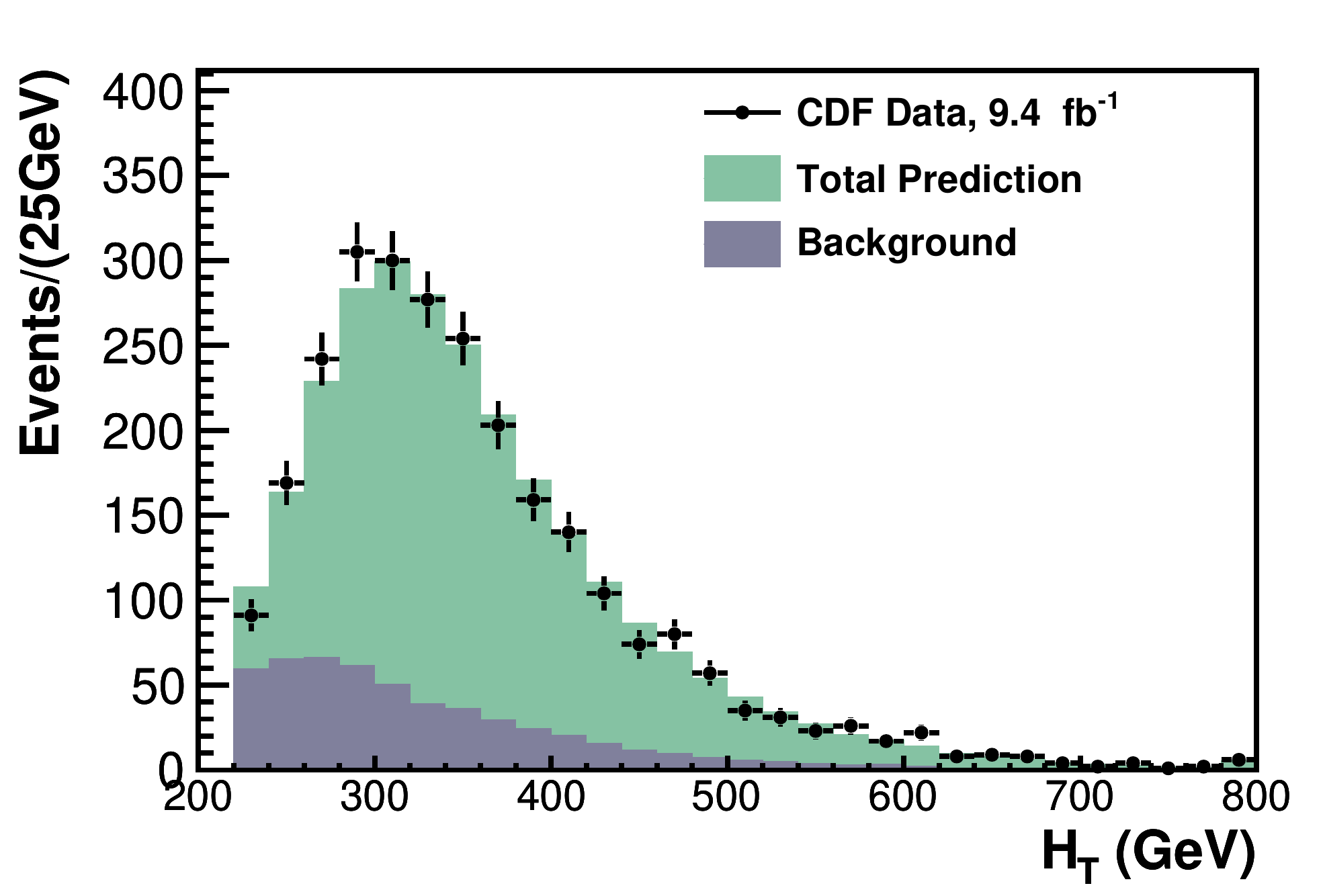}\label{fig:ht}}
\caption{{\small \subref{fig:met} The missing transverse energy and \subref{fig:ht} the scalar sum of the transverse energy of the lepton, jets, and $\met$ in the data compared to the signal plus background model. The last bin contains overflow events.} \label{fig:met_ht}}
\end{center}
\end{figure*}

\section{The inclusive asymmetry}\label{sec:inclusive_afb}

\subsection{$\dy$ in the reconstructed data}\label{sec:reco}

We first consider the reconstructed $\dy$ distribution and its asymmetry as defined in Eq.~(\ref{eq:afb}).  The $\dy$ distribution is shown in Fig.~\ref{fig:qdely}, compared to prediction for the background plus the \powheg $\ttbar$ model.  Those bins with $\dy > 0$ contain data points that are consistently higher than the prediction, while in the bins with $\dy < 0$, the data is consistently below the prediction.  This results in an inclusive reconstructed asymmetry of $\afb = 0.063 \pm 0.019$, compared to a prediction of $0.020 \pm 0.012$.  The uncertainty on the data measurement is statistical only.  Table~\ref{tab:inc_asyms_posneg} summarizes the reconstructed asymmetry values, with events split according to the charge of the identified lepton, and also reports the results of Ref.~\cite{cdfafb} for comparison.  The uncertainties scale as expected from the previous analysis according to the increase in the number of candidate events. When the sample is separated according to the charge of the lepton, the asymmetries are equal within uncertainties, as would be expected from a {\it CP}-conserving effect. 

\begin{figure}[!htbp]
\begin{center}
\includegraphics[width=0.45\textwidth, clip]{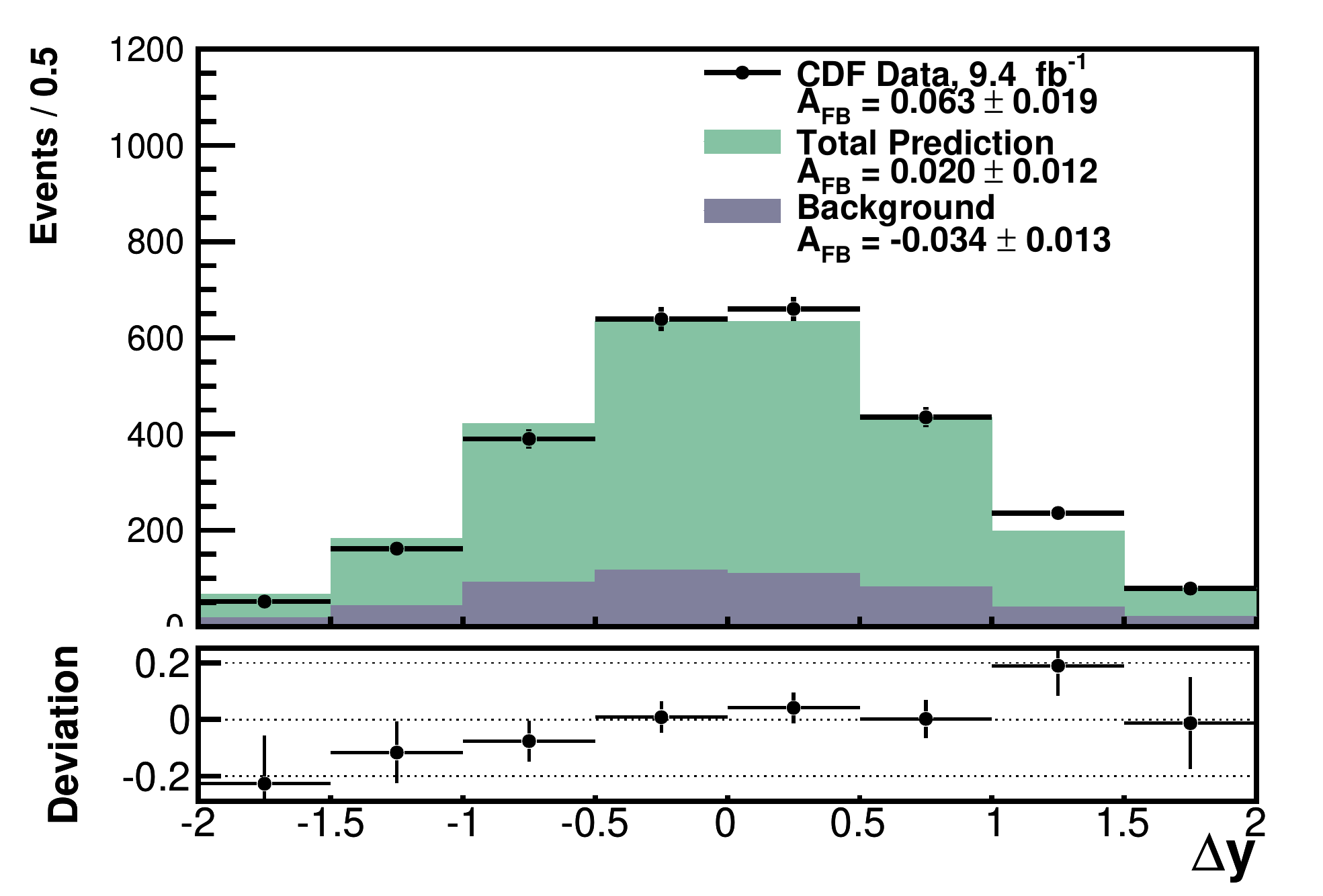}
\caption{{\small (top) The reconstructed $\dy$ distribution and the inclusive reconstruction-level asymmetry, compared to the prediction of the signal and background model.  (bottom) The difference between the data and prediction divided by the prediction.  N.B. the left-most and right-most bins are under- and over-flow bins, respectively.} \label{fig:qdely}}
\end{center}
\end{figure}

\begin{table}[!htb]
\caption{Measured reconstruction-level asymmetries in $\dy$ compared to the values measured in the previous CDF analysis~\cite{cdfafb}, as well as the predicted asymmetries for the signal and background contributions.}\label{tab:inc_asyms_posneg}
\begin{center}
\begin{tabular}{c c c}

\hline
\hline
        &  \multicolumn{2}{c}{Predicted $\phantom{0}\afb$}         \\
\hline
SM $\ttbar$      & \multicolumn{2}{c}{$\phantom{-}0.033\pm 0.011\phantom{0}$}         \\
Backgrounds                & \multicolumn{2}{c}{$          -0.034\pm 0.013\phantom{0}$}         \\
Total prediction           & \multicolumn{2}{c}{$\phantom{-}0.020\pm 0.012\phantom{0}$}        \\
\hline
                         & \multicolumn{2}{c}{ Observed  $\phantom{0}\afb \pm $ stat $ $ }            \\
                 &     $\phantom{0}9.4~\ifb$      &  $\phantom{0}5.3~\ifb$       \\

\hline
All data                   & $\phantom{-}0.063\pm 0.019\phantom{0}$   & $\phantom{-}0.057\pm 0.028\phantom{0}$     \\
Positive leptons           & $\phantom{-}0.072\pm 0.028\phantom{0}$   & $\phantom{-}0.067\pm 0.040\phantom{0}$     \\
Negative leptons           & $\phantom{-}0.055\pm 0.027\phantom{0}$   & $\phantom{-}0.048\pm 0.039\phantom{0}$     \\
\hline \hline
\end{tabular}
\end{center}
\end{table}

\subsection{Subtracting the background contributions}\label{sec:bkg_sub}

Approximately 20\% of the selected data set is composed of events originating from various background sources. We remove the effect of these events by subtracting the predicted background contribution from each bin of the reconstructed distribution.  This background-subtraction procedure introduces additional systematic uncertainty, which is added in quadrature to the statistical uncertainty for all background-subtracted results in this paper.

To derive this uncertainty, we start with a total prediction containing $n$ components ($n-1$ background sources and one signal), with each component $i$ having an asymmetry $A_i$ and contributing $N_i$ events.  This leads to a total asymmetry for the prediction of 

\begin{equation}
A_{\rm tot} = \frac{\sum\limits_{i=1}^{n} A_i N_i}{\sum\limits_{i=1}^{n} N_i} = \frac{\sum\limits_{i=1}^{n} A_i N_i}{N_{\rm tot}}.
\end{equation}

\noindent For the $i$'th component, we let $\sigma_{A_{i}}$ and $\sigma_{N_{i}}$ be the uncertainties on the asymmetry and the normalization respectively.  For $\sigma_{N_{i}}$, we use the predicted uncertainty of each background component, as listed in Table~\ref{tab:method2}.  The uncertainty due to the finite sample size of the model for a given background component is included as $\sigma_{A_{i}}$, though this is only appreciable for the non-$W$ component, which is taken from a statistically limited sideband in the data.

These uncertainties can be propagated in the usual way by calculating derivatives and adding in quadrature, leading to the term within the summation in Eq.~(\ref{eq:bkgsub_syst}).  For the uncertainty due to background subtraction, the summation runs over the $n-1$ background components.  We also include an overall uncertainty $\sigma_{A_{\rm bkg}}$ as the final term.

\begin{equation}\label{eq:bkgsub_syst}
\sigma^2_{\rm syst} = \sum\limits_{i=1}^{n-1}\left[\frac{N^2_i}{N^2_n}\sigma_{A_{i}}^2 + \frac{(A_i - A_{\rm tot})^2}{N^2_n}\sigma_{N_{i}}^2\right] + \sigma_{A_{\rm bkg}}^2.
\end{equation}

\noindent For the uncertainty $\sigma_{A_{\rm bkg}}$ on the overall background shape, we substitute an alternate model for the non-$W$ background component and determine the effect on the measured asymmetry, contributing an uncertainty of $0.002$ to the inclusive $\afb$ result.  The summation term in Eq.~(\ref{eq:bkgsub_syst}) results in a total uncertainty of $0.008$.  In total, the sum of the systematic contributions to the uncertainty is small compared to the statistical uncertainty.

The $\dy$ distribution after background subtraction is shown in Fig.~\ref{fig:qdely_signal}.  Because the total background prediction is nearly symmetric, the removal of the backgrounds increases the asymmetry attributable to the $\ttbar$ signal.  The resulting observed asymmetry in the background-subtracted sample is $0.087 \pm 0.026$ (stat$+$syst), compared to the \powheg prediction of $0.033 \pm 0.011$.

\begin{figure}
\begin{center}
\includegraphics[width=0.45\textwidth, clip]{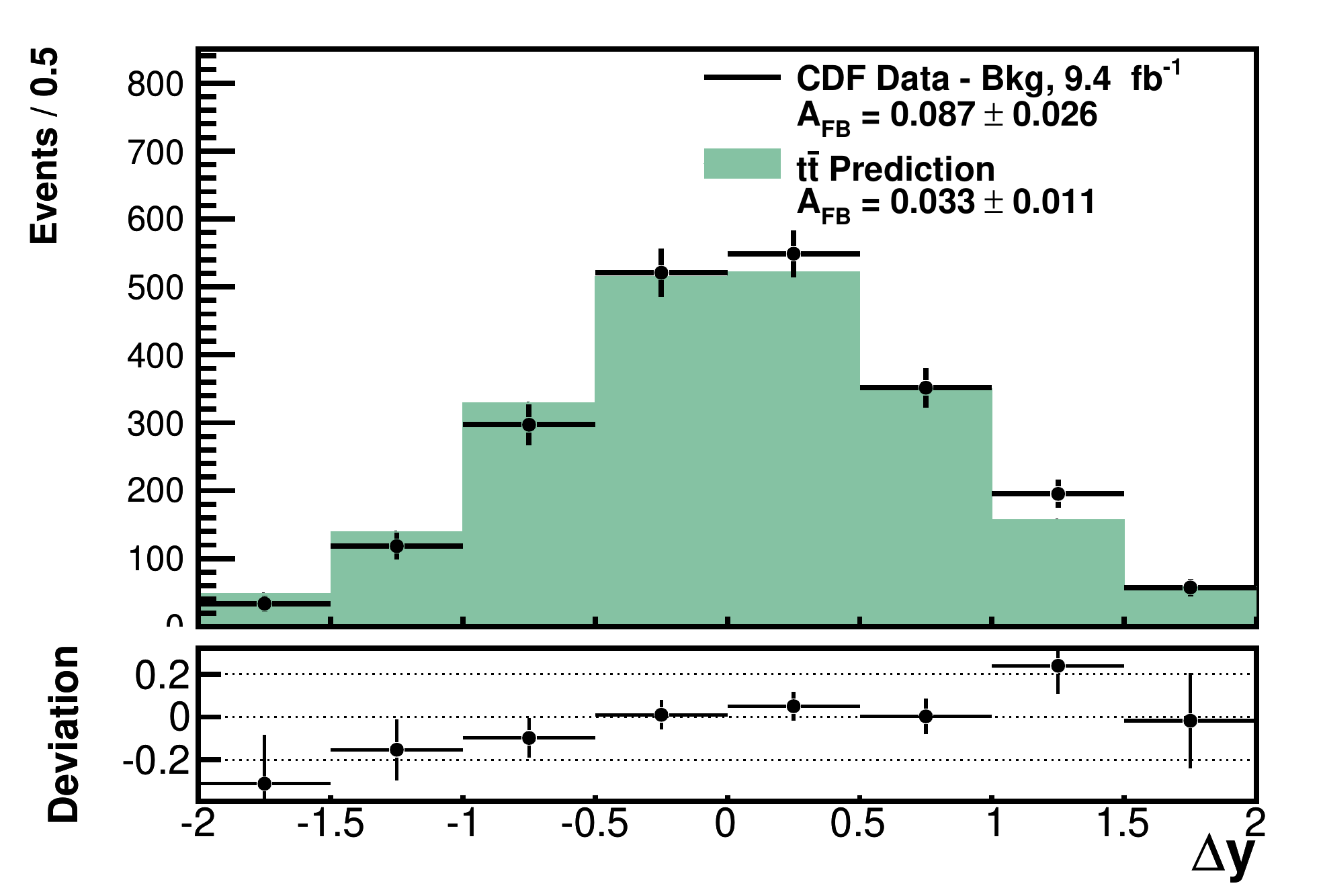}
\caption{{\small (top) The observed background-subtracted $\Delta y$ distribution compared to the SM prediction.  Error bars include both statistical and background-related systematic uncertainties.  (bottom) The difference between the data and prediction divided by the prediction.} \label{fig:qdely_signal}}
\end{center}
\end{figure}

\subsection{Correction to the parton level}\label{sec:parton}

The background-subtracted results provide a measurement of the asymmetry due to $\ttbar$ events.  However, these results are not directly comparable to theoretical predictions because they include the effects of the limited acceptance and resolution of the detector.  We correct for these effects so as to provide parton-level results, in the $\ttbar$ rest frame after radiation, that can be directly compared to theoretical predictions.

If the true parton-level binned distribution of a particular variable is given by $\vec{n}_{\text{parton}}$, then, after background subtraction, we will observe $\vec{n}_{\text{bkg.sub.}}=\mathbf{S}\mathbf{A}\vec{n}_{\text{parton}}$, where the diagonal matrix $\mathbf{A}$ encodes the effect of the detector acceptance and selection requirements, while the response matrix $\mathbf{S}$ describes the bin-to-bin migration that occurs in events passing the selection due to the limited resolution of the detector and $\ttbar$ reconstruction algorithm.  To recover the parton-level distribution, the effects of $\mathbf{S}$ and $\mathbf{A}$ must be reversed.

The $5.3~\ifb$ CDF analysis~\cite{cdfafb} used simple matrix inversion (``unfolding'') to perform the correction to the parton level.  While effective, this technique was limited in its application because unfolding via matrix inversion tends to enhance statistical fluctuations (due to small eigenvalues in the migration matrix), which makes it reliable only in densely populated distributions.  This limited the previous analysis to the extent that the determination the functional dependencies of the asymmetry could only use two bins of $|\dy|$ and $\mttb$.  In this paper, we employ a new algorithm, also based on matrix inversion but more sophisticated in application, to measure more finely-binned parton-level distributions, resulting in a more robust measurement of the functional dependence of $\afb$ on $|\dy|$ and $\mttb$ at the parton level.

We first consider $\mathbf{S}$, correcting for the finite resolution of the detector using a regularized unfolding algorithm based on Singular Value Decomposition (SVD)~\cite{svd,roounfold}.  We model the bin-to-bin migration caused by the detector and reconstruction using \powheg.  The matrix $\mathbf{S}$ in $\dy$ from \powheg is represented graphically in Fig.~\ref{fig:delyresponse}. Along each row, the box area is proportional to the probability that each possible measured value $\dy_{\text{meas}}$ is observed in events with a given true rapidity difference $\dy_{\text{true}}$. The matrix population clusters along the diagonal where $\dy_{\text{meas}} = \dy_{\text{true}}$ and is approximately symmetric, showing no large biases in the $\dy$ reconstruction.  Before inverting the matrix $\mathbf{S}$ and applying it to the background-subtracted data, a regularization term is introduced to prevent statistical fluctuations from dominating the correction procedure.  It is this smoothing via regularization that allows an increase in the number of bins in the parton-level distributions compared to the previous analysis.  Details regarding how the regularization term is included are given in Ref.~\cite{svd}, but in essence, a term $\sqrt{\tau}\mathbf{C}$, where $\mathbf{C}$ is the second-derivative matrix,

\begin{equation}\label{eq:secondderivatives}
\mathbf{C} = \left( \begin{array}{ccccc} 
-1 &  1 &  0 & 0 & ... \\ 
 1 & -2 &  1 & 0 & ... \\ 
 0 &  1 & -2 & 1 & ... \\ 
... & ... & ... & ... & ... \\ 
... & 0 & 1 & -2 & 1 \\ 
... & 0 & 0 &  1 & -1 
 \end{array} \right),
\end{equation}

\noindent is added to the matrix equation relating $\vec{n}_{\text{parton}}$ to $\vec{n}_{\text{bkg.sub.}}$.  This term imposes the {\it a priori} condition that the parton-level solution should be smooth (more precisely, the regularization assumes that the ratio between the data distribution after acceptance cuts and the model distribution after acceptance cuts is smooth, but given a smooth acceptance function and a model that is smooth at the parton level, this is equivalent to a condition that the data be smooth at parton level).  The value of $\tau$ defines how strongly the regularization condition affects the result and is determined using the methods recommended in Ref.~\cite{svd}.

\begin{figure}
\begin{center}
\includegraphics[width=0.45\textwidth, clip]{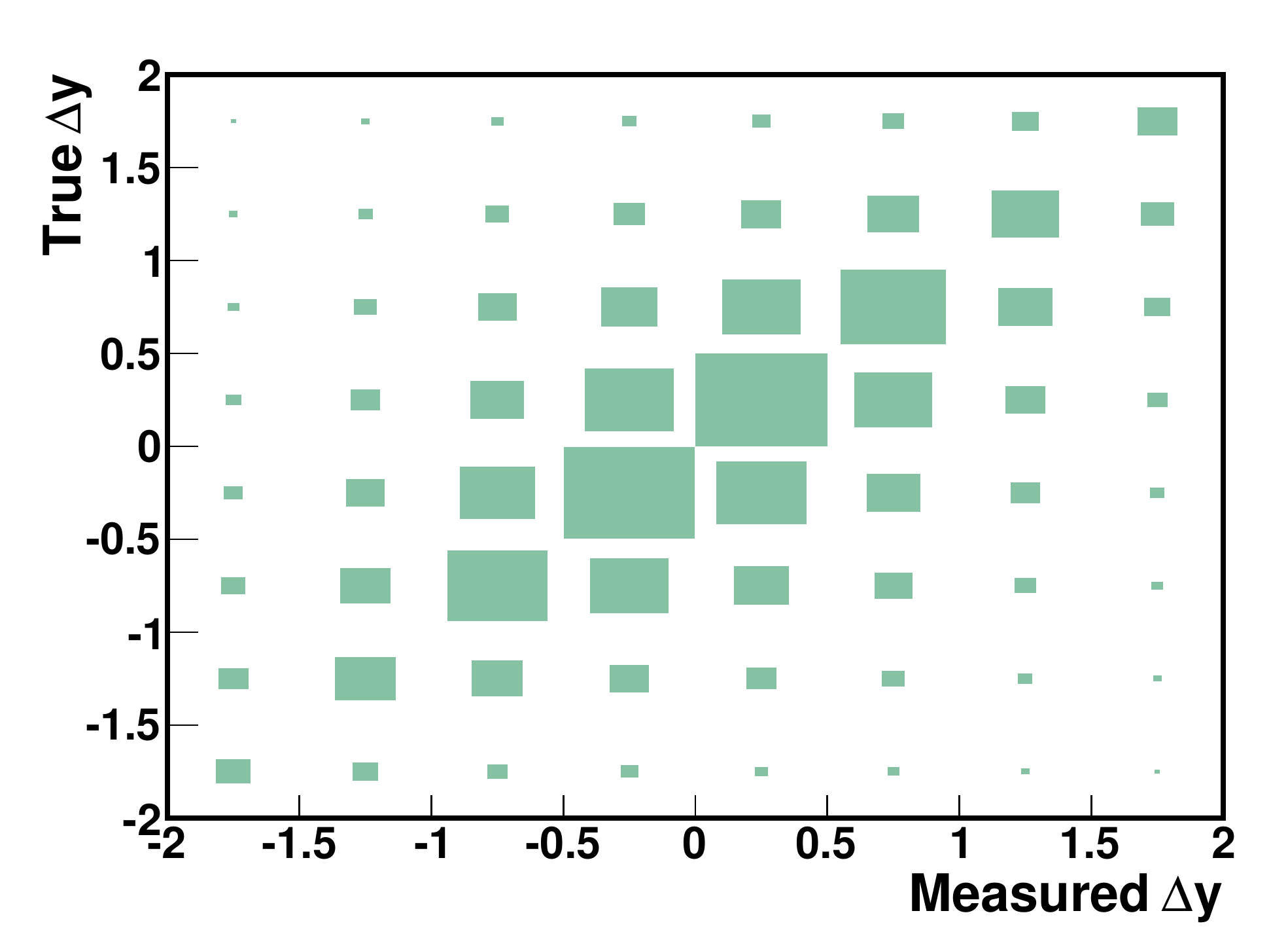}
\caption{{\small Detector response in $\dy$ as modeled by \powheg, showing the true value of $\dy$ as a function of the measured value for all events passing the selection criteria. The size of each rectangle is proportional to the number of entries in that bin.} \label{fig:delyresponse}}
\end{center}
\end{figure}

In the second step of the parton-level correction procedure, we account for events that are unobserved due to limited acceptance.  The acceptance in each bin is derived from the \powheg model, as shown in Fig.~\ref{fig:delyacceptance}, and these acceptances are applied to the data as an inverse-multiplicative correction to each bin.  The acceptance is asymmetric in $\dy$, with backwards events passing the selection requirements more often than forward events.  This effect is related to the $\ptran$ dependence of the asymmetry that is discussed in Sec.~\ref{sec:afb_v_pt}.  Large $\ptran$ in a given event leads to $\ttbar$ decay products that also have large $\peetee$, and thus events with large $\ptran$ pass the selection requirements more often than events with small $\ptran$.  As is shown in Sec.~\ref{sec:afb_v_pt}, high-$\ptran$ events are also predicted by \powheg (and various other SM calculations) to have a negative asymmetry.  The result is that events with a negative asymmetry are more likely to fulfill the selection requirements, leading to the asymmetric acceptance distribution in Fig.~\ref{fig:delyacceptance}.

The SVD unsmearing and bin-by-bin acceptance correction have similarly-sized impact on the final result.  Both of the corrections lead to an increase in the asymmetry.  The population of poorly reconstructed events tends to have zero asymmetry, and thus dilutes the true asymmetry.  One effect of the unsmearing is to remove this dilution.  The acceptance correction also increases $\afb$ because of the asymmetric acceptance shown in Fig.~\ref{fig:delyacceptance}.

\begin{figure}
\begin{center}
\includegraphics[width=0.45\textwidth, clip]{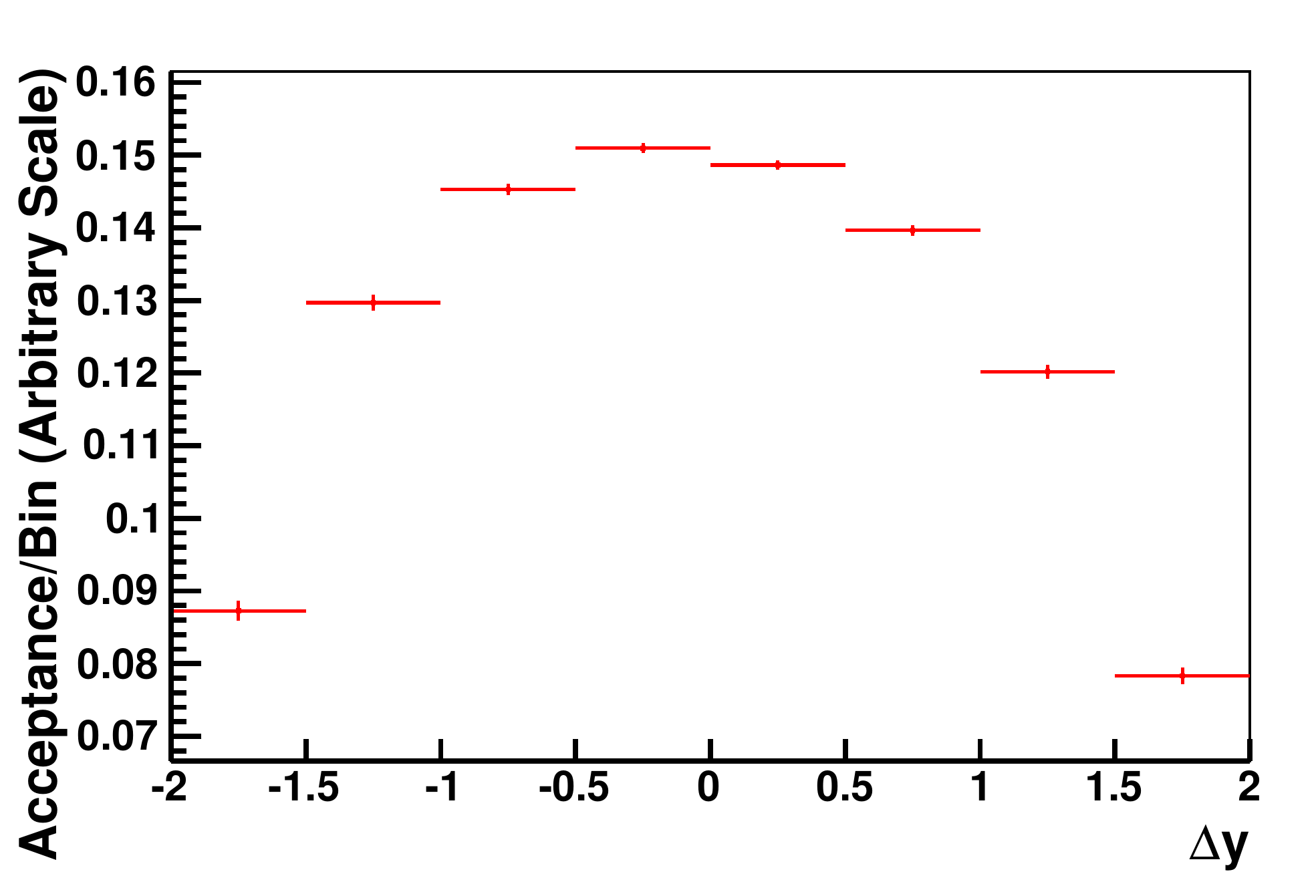}
\caption{{\small Acceptance as a function of $\dy$ as modeled by \powheg.} \label{fig:delyacceptance}}
\end{center}
\end{figure}

\begin{figure}
\begin{center}
\includegraphics[width=0.45\textwidth, clip]{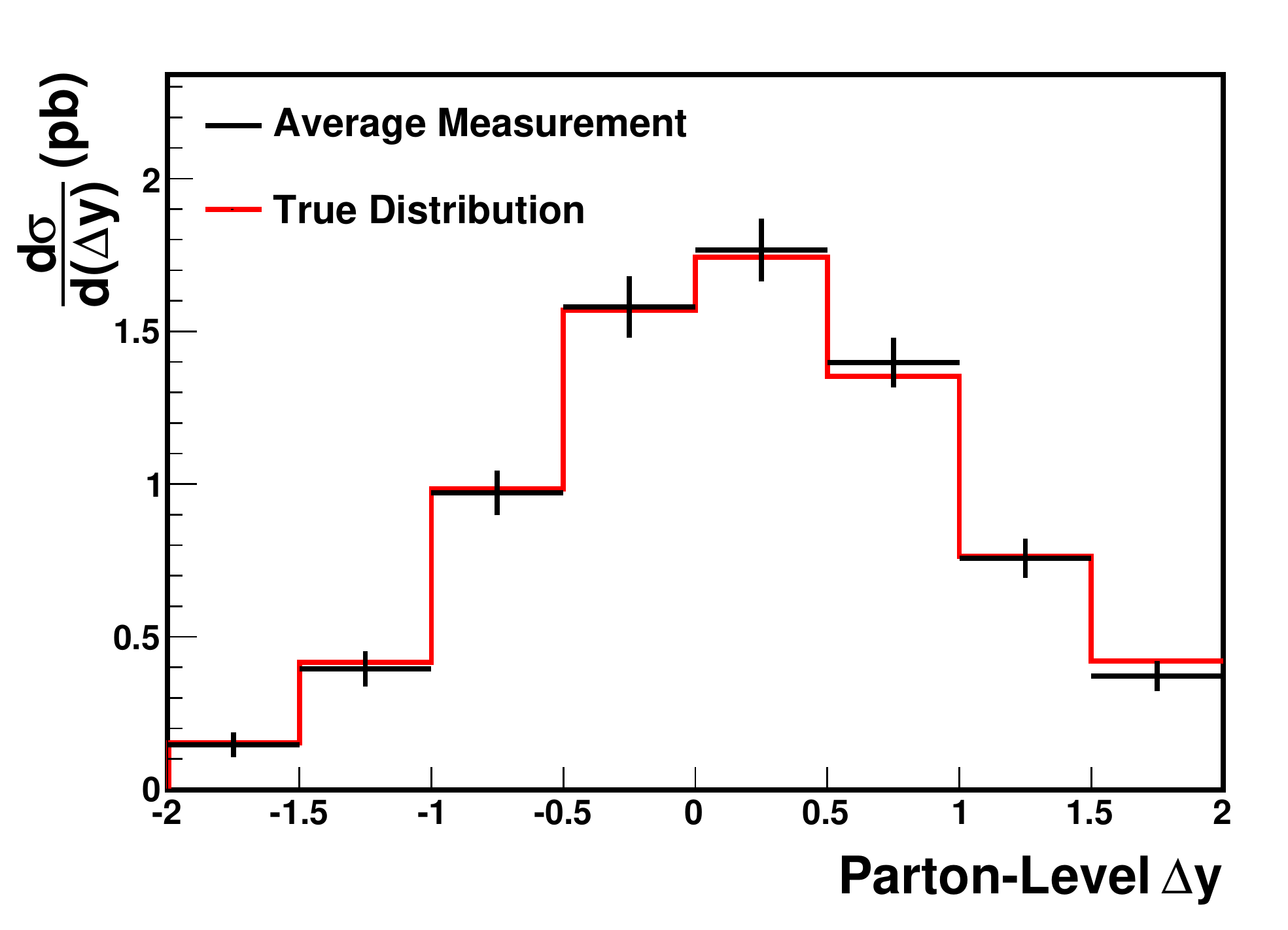}
\caption{{\small Results from simulated parton-level $\dy$ measurements based on Octet A.  The data points show the central values for the simulated results, with the error bars representing the $1\sigma$ spread of the results. } \label{fig:pe_octetA}}
\end{center}
\end{figure}

\begin{table*}[!htb]
\caption{Average parton-level asymmetry values in 10~000 simulated experiments with Octet A.}\label{tab:pe_octetA}
\begin{center}
\begin{tabular}{c c c c }
\hline
\hline
$|\dy|$        & \phantom{0}Average measured $\afb$\phantom{0} & \phantom{0}      Average uncertainty     \phantom{0} & \phantom{0}True $\afb$\phantom{0} \\
\hline
Inclusive               & 0.162 & 0.039  & 0.156        \\
$0.0 \leq |\dy| < 0.5$  & 0.056 & 0.035  & 0.052        \\
$0.5 \leq |\dy| < 1.0$  & 0.180 & 0.055  & 0.158        \\      
$1.0 \leq |\dy| < 1.5$  & 0.316 & 0.078  & 0.295        \\ 
$|\dy| \geq 1.5$        & 0.434 & 0.128  & 0.468        \\       
\hline \hline
\end{tabular}
\end{center}
\end{table*}

\begin{table}[!htb]
\caption{Systematic uncertainties on the parton level $\afb$ measurement.}\label{tab:svd_systematics}
\begin{center}
\begin{tabular}{l c }
\hline
\hline
Source      &     Uncertainty \\     
\hline         
Background shape         &  0.018    \\
Background normalization &  0.013    \\                   
Parton shower         &  0.010     \\
Jet energy scale         &  0.007     \\
Initial- and final-state radiation         &  0.005     \\
Correction procedure         &  0.004     \\
Color reconnection         &  0.001     \\
Parton-distribution functions         &  0.001     \\
\hline 
Total systematic uncertainty           &       0.026   \\
\hline 
Statistical uncertainty     &   0.039 \\
\hline 
Total uncertainty     &   0.047 \\
\hline 
\hline 
\end{tabular}
\end{center}
\end{table}

The combination of these two parts of the correction procedure allows the determination of the parton-level distribution of $\dy$, which is reported as a differential cross section.  This algorithm is tested in various simulated $\ttbar$ samples, including standard model $\powheg$ and the non-SM samples Octet A and Octet B.  Analyzing these samples as if they are data, we measure the bias in the comparison of derived parton-level results to the true values in the generated samples. The {\sc powheg} results are self-consistent to better than 1\%, and, because the NLO standard model is assumed {\it a priori} to be the correct description of the underlying physics and is used to model the acceptance and detector response, any biases observed in this case are included as systematic uncertainties, as described below.

In the octet models, the derived distributions track the generator truth predictions well, but small biases (generally less than $3-4\%$) are observed in some of the differential asymmetry values. An example of the average corrected distribution across a set of 10~000 simulated experiments is shown in Fig.~\ref{fig:pe_octetA} for Octet A, with the asymmetry as a function of $|\dy|$ for these simulated experiments summarized in Table~\ref{tab:pe_octetA}.  We do not attempt to correct the biases seen in the non-SM models or include them in the uncertainty because there is no reason to believe that these specific octet models actually represent the real underlying physics - these models exhibit small but significant discrepancies with the data in the $\mttb$ spectrum, a variable that has a significant effect on the $\ttbar$ reconstruction, and thus the detector response matrix.  In light of this model-dependence, we emphasize that the parton-level results need to be interpreted with some caution in relation to models that differ significantly from the NLO standard model.

Because the resolution corrections can cause migration of events across bins, the populations in the final parton-level distributions are correlated.  In all binned parton-level distributions, the error bars on a given bin correspond to the uncertainty in the contents of that bin, but they are not independent of the uncertainties corresponding to other bins in the distribution. When we calculate derived quantities such as $\afb$, we use the covariance matrix associated with the unsmearing procedure to propagate the uncertainties correctly.

Several sources of systematic uncertainty must be accounted for when applying the correction procedure.  In addition to uncertainties on the size and shape of the background prediction, there are also uncertainties related to the signal Monte Carlo sample used to model the acceptance and detector response.  These signal uncertainties include the size of the jet energy scale corrections~\cite{jes}, the amount of initial- and final-state radiation, the underlying parton-distribution functions~\cite{IFSRPDFsyst}, the modeling of color reconnection~\cite{colorreconnection}, and the modeling of parton showering and color coherence.  We evaluate these uncertainties by repeating the measurement after making reasonable variations to the assumptions that are used when modeling the detector response.  For example, to estimate the effect on our measurement of uncertainty in parton shower and color coherence models, we compare two detector response models, one using the Lund string model~\cite{pythia} and one using the Catani-Seymour dipole model~\cite{herwig}.  We also include a systematic uncertainty for the correction algorithm itself, taking the difference between the true value in \powheg and the average result from the simulated experiments based on \powheg described above as the uncertainty resulting from the correction procedure.  The systematic uncertainties on the inclusive $\afb$ measurement are shown in Table~\ref{tab:svd_systematics}, and the total systematic uncertainty is found to be small compared to the statistical uncertainty.  When adding the systematic uncertainties to the the covariance matrices that result from the unfolding procedure, the systematic uncertainties are assumed to be 100\% correlated across all bins.

\begin{figure}
\begin{center}
\includegraphics[width=0.45\textwidth, clip]{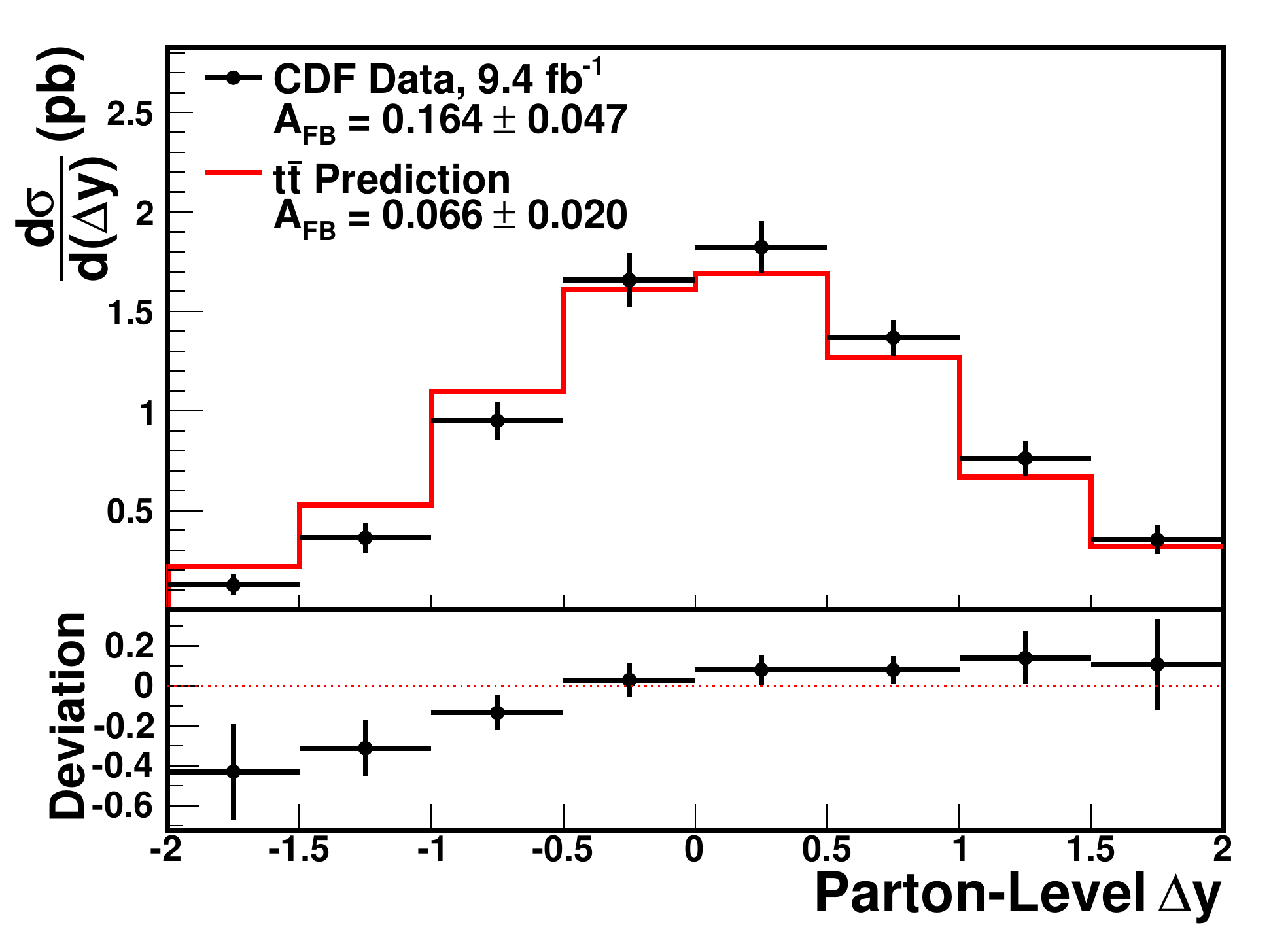}
\caption{{\small (top) The differential cross section $d\sigma/d(\dy)$ as measured in the data after correction to the parton level compared to the SM prediction.  Uncertainties include both statistical and systematic contributions and are correlated between bins.  (bottom) The difference between the data and prediction divided by the prediction.} \label{fig:dely_parton}}
\end{center}
\end{figure}

\begin{table}[!htb]
\caption{The measured differential cross section as a function of $\dy$.  The total cross section is normalized to 7.4 pb.  Errors include both statistical and systematic contributions, and are correlated across bins.}\label{tab:xsecmeas}
\begin{center}
\begin{tabular}{c c }

\hline
\hline
  $\dy$      &     $d\sigma/d(\dy)$ (pb)   \\

\hline
$\le$ $-1.5$        &   \phantom{0}0.13 $\pm$ 0.05\phantom{0}       \\
$-1.5$ to $-1.0$    &   \phantom{0}0.36 $\pm$ 0.07\phantom{0}       \\
$-1.0$ to $-0.5$    &   \phantom{0}0.95 $\pm$ 0.10\phantom{0}       \\
$-0.5$ to $0.0$     &   \phantom{0}1.66 $\pm$ 0.14\phantom{0}       \\
$0.0$ to $0.5$      &   \phantom{0}1.82 $\pm$ 0.13\phantom{0}       \\
$0.5$ to $1.0$      &   \phantom{0}1.37 $\pm$ 0.09\phantom{0}       \\
$1.0$ to $1.5$      &   \phantom{0}0.76 $\pm$ 0.09\phantom{0}       \\
$\ge$ $1.5$         &   \phantom{0}0.35 $\pm$ 0.07\phantom{0}       \\
\hline \hline
\end{tabular}
\end{center}
\end{table}

Applying the correction procedure to the data of Fig.~\ref{fig:qdely_signal} yields the distribution shown in Fig.~\ref{fig:dely_parton}, where the measured result is compared to the SM \powheg prediction.  Both the prediction and the observed data distributions are scaled to a total cross section of 7.4 pb, so that Fig.~\ref{fig:dely_parton} shows the differential cross section for $\ttbar$ production as a function of $\dy$.  The measured values are summarized in Table~\ref{tab:xsecmeas}.  We measure an inclusive parton-level asymmetry of $0.164 \pm 0.039(\rm stat)\pm 0.026(\rm syst) = 0.164 \pm 0.047$.  At the parton level, the observed inclusive asymmetry is non-zero with a significance of $3.5\sigma$ and exceeds the NLO prediction of \powheg by $1.9\sigma$, where we have included a $30\%$ uncertainty on the prediction.

\section{The dependence of the asymmetry on $|\dy|$}\label{sec:afb_v_dy}

The dependence of $\afb$ on the rapidity difference $|\dy|$ was studied in the $5~\ifb$ analyses~\cite{cdfafb,d0afb}, but with only two bins of $|\dy|$.  The CDF and D0 results were consistent and showed a rise of $\afb$ with increasing $|\dy|$.  We perform a more detailed study of the rapidity dependence of $\afb$ using the full data set and improved analysis techniques. 

The forward-backward asymmetry as a function of $|\dy|$ at the reconstruction level can be derived from the data shown in Fig.~\ref{fig:qdely} according to

\begin{equation}
\afb(|\dy|) = \frac{N_F(|\dy|) - N_B(|\dy|)}{N_F(|\dy|) + N_B(|\dy|)},
\label{afbvdy_data}
\end{equation}

\noindent where $N_F(|\dy|)$ is the number of events in a given $|\dy|$ bin with $\dy > 0$ and $N_B(|\dy|)$ is the number of events in the corresponding $|\dy|$ bin with $\dy < 0$.  One important constraint on the $\dy$ dependance of the asymmetry  may be anticipated: any theory that predicts a continuous and differentiable $\dy$ distribution must have $\afb(|\dy|=0)=0$, regardless of the size of the inclusive asymmetry.

Figure~\ref{fig:afb_v_dely} shows $\afb(|\dy|)$ in four bins of $|\dy|$, with the measured values and their uncertainties listed in Table~\ref{tab:afb_v_dy_reco}. To quantify the behavior in a simple way, we assume a linear relationship, which provides a good approximation of both the data and the \powheg prediction (see also Ref.~\cite{almeida}). From the theoretical considerations described above, we make the assumption $\afb(|\dy|=0)=0$ and fit the slope only.  The slope $\alpha_{\dy}$ of the line does not correspond to a specific parameter of any particular theory, but provides a quantitative comparison of the $|\dy|$ dependence of the asymmetry in the data and prediction.  The measurements of $\afb(|\dy|)$ in data at the reconstruction level are well-fit by a line with a $\chi^2$ per degree of freedom of $1.7/3$ and a slope $\alpha_{\dy} = (11.4 \pm 2.5) \times 10^{-2}$,  a rapidity dependence that is non-zero with significance in excess of $4 \sigma$.  The predicted slope from \powheg and the background model is $(3.6 \pm 0.9) \times 10^{-2}$.

\begin{figure}
\begin{center}
  \includegraphics[width=0.45\textwidth, clip]{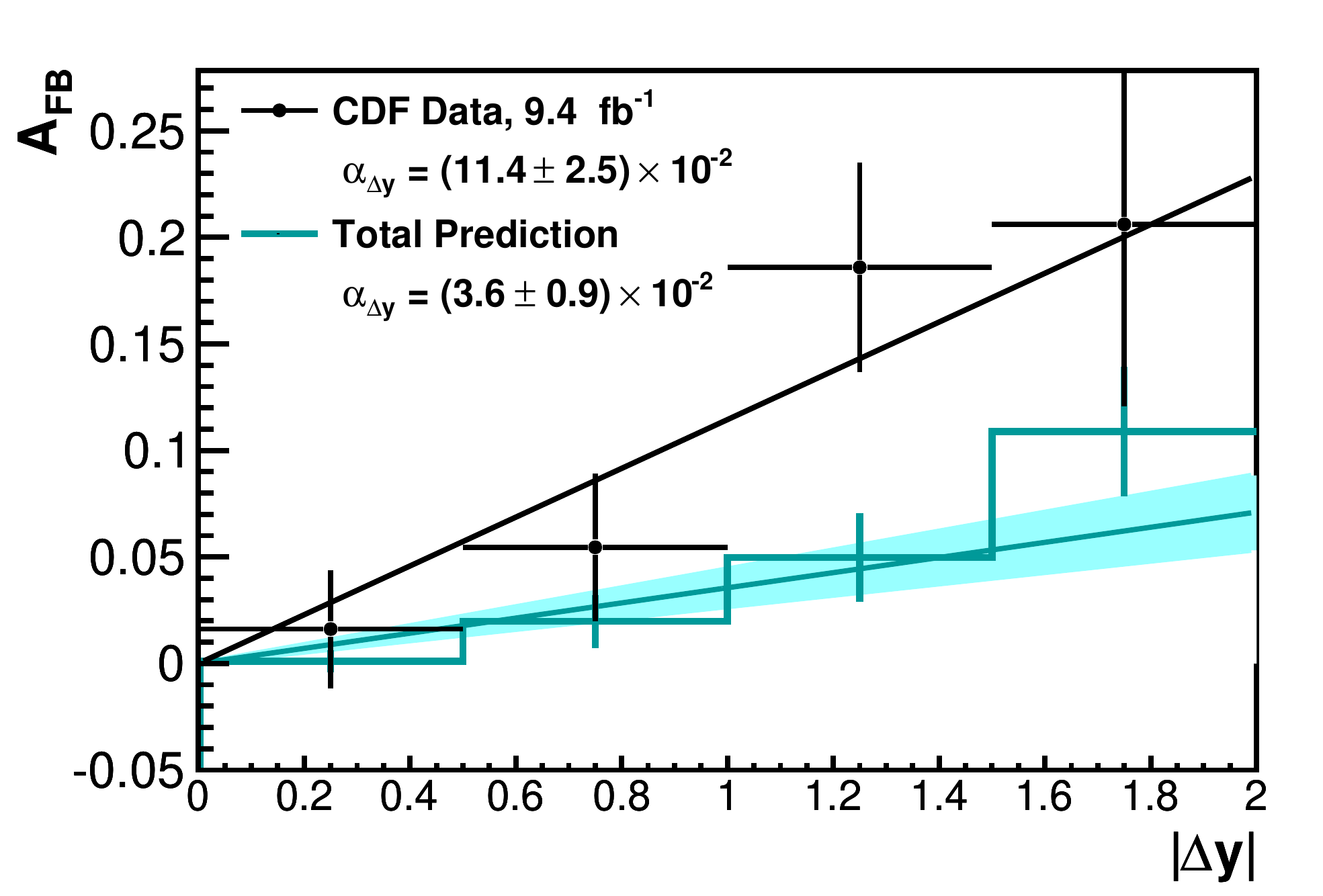}
\caption{{\small The reconstruction-level forward-backward asymmetry as a function of $|\dy|$ with a best-fit line superimposed.  The errors on the data are statistical, and the shaded region represents the uncertainty on the slope of the prediction.} \label{fig:afb_v_dely}}
\end{center}
\end{figure}

\begin{figure}
\begin{center}
  \includegraphics[width=0.45\textwidth, clip]{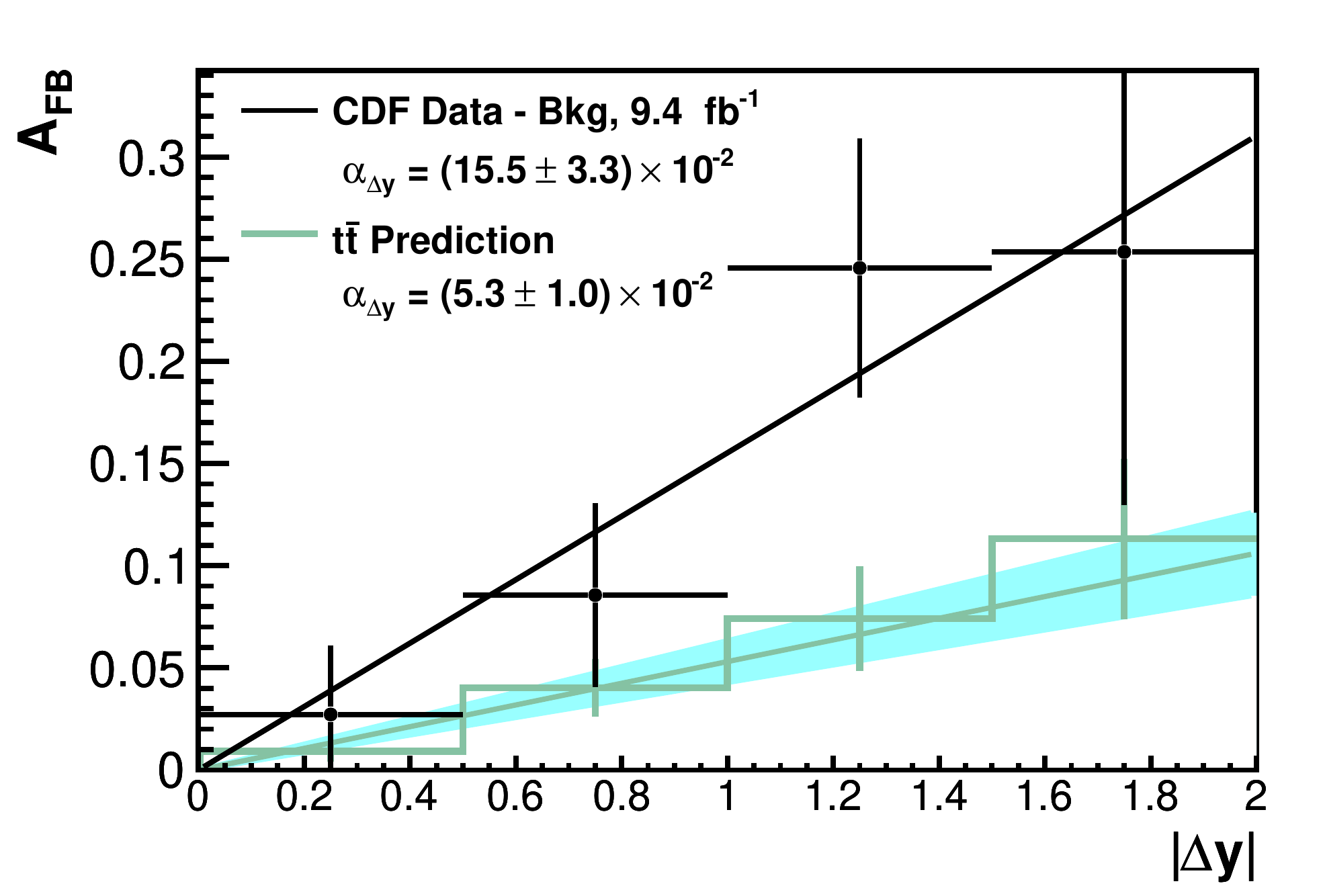}
\caption{{\small The background-subtracted asymmetry as a function of $|\dy|$ with a best-fit line superimposed.  Error bars include both statistical and background-related systematic uncertainties.  The shaded region represents the theoretical uncertainty on the slope of the prediction.} \label{fig:afb_v_dely_signal}}
\end{center}
\end{figure}

\begin{table*}[!htb]
\caption{The asymmetry at the reconstructed level as measured in the data, compared to the SM $\ttbar$ plus background expectation, as a function of $|\dy|$.}\label{tab:afb_v_dy_reco}
\begin{center}
\begin{tabular}{c c c }

\hline
\hline
               &    Data                   &  SM $\ttbar$ + Bkg.        \\
  $|\dy|$      &     $\afb$ $\pm$ stat      &  $\afb$       \\

\hline
0.0 - 0.5      &   \phantom{0}0.016 $\pm$ 0.028\phantom{0}                  &  \phantom{0}0.001 $\pm$ 0.005\phantom{0}    \\
0.5 - 1.0      &   \phantom{0}0.055 $\pm$ 0.035\phantom{0}                  &  \phantom{0}0.020 $\pm$ 0.012\phantom{0}    \\
1.0 - 1.5      &   \phantom{0}0.186 $\pm$ 0.049\phantom{0}                  &  \phantom{0}0.050 $\pm$ 0.021\phantom{0}    \\
$\ge$ 1.5      &   \phantom{0}0.206 $\pm$ 0.085\phantom{0}                  &  \phantom{0}0.109 $\pm$ 0.030\phantom{0}    \\
\hline \hline
\end{tabular}
\end{center}
\end{table*}

\begin{table*}[!htb]
\caption{The asymmetry at the background-subtracted level as measured in the data, compared to the SM $\ttbar$ expectation, as a function of $|\dy|$.}\label{tab:afb_v_dy_signal}
\begin{center}
\begin{tabular}{c c c }

\hline
\hline
        &    Data     &  SM $\ttbar$        \\
  $|\dy|$      &     $\afb$ $\pm$ (stat$+$syst)     &   $\afb$       \\
\hline
0.0 - 0.5      &    \phantom{0}0.027 $\pm$ 0.034\phantom{0}                &  \phantom{0}0.009 $\pm$ 0.005\phantom{0}   \\
0.5 - 1.0      &    \phantom{0}0.086 $\pm$ 0.045\phantom{0}                &  \phantom{0}0.040 $\pm$ 0.014\phantom{0}    \\
1.0 - 1.5      &    \phantom{0}0.246 $\pm$ 0.063\phantom{0}                &  \phantom{0}0.074 $\pm$ 0.026\phantom{0}   \\
$\ge$ 1.5      &    \phantom{0}0.254 $\pm$ 0.124\phantom{0}                &  \phantom{0}0.113 $\pm$ 0.039\phantom{0}    \\
\hline \hline
\end{tabular}
\end{center}
\end{table*}

\begin{figure}[!htbp]
\begin{center}
  \includegraphics[width=0.45\textwidth, clip]{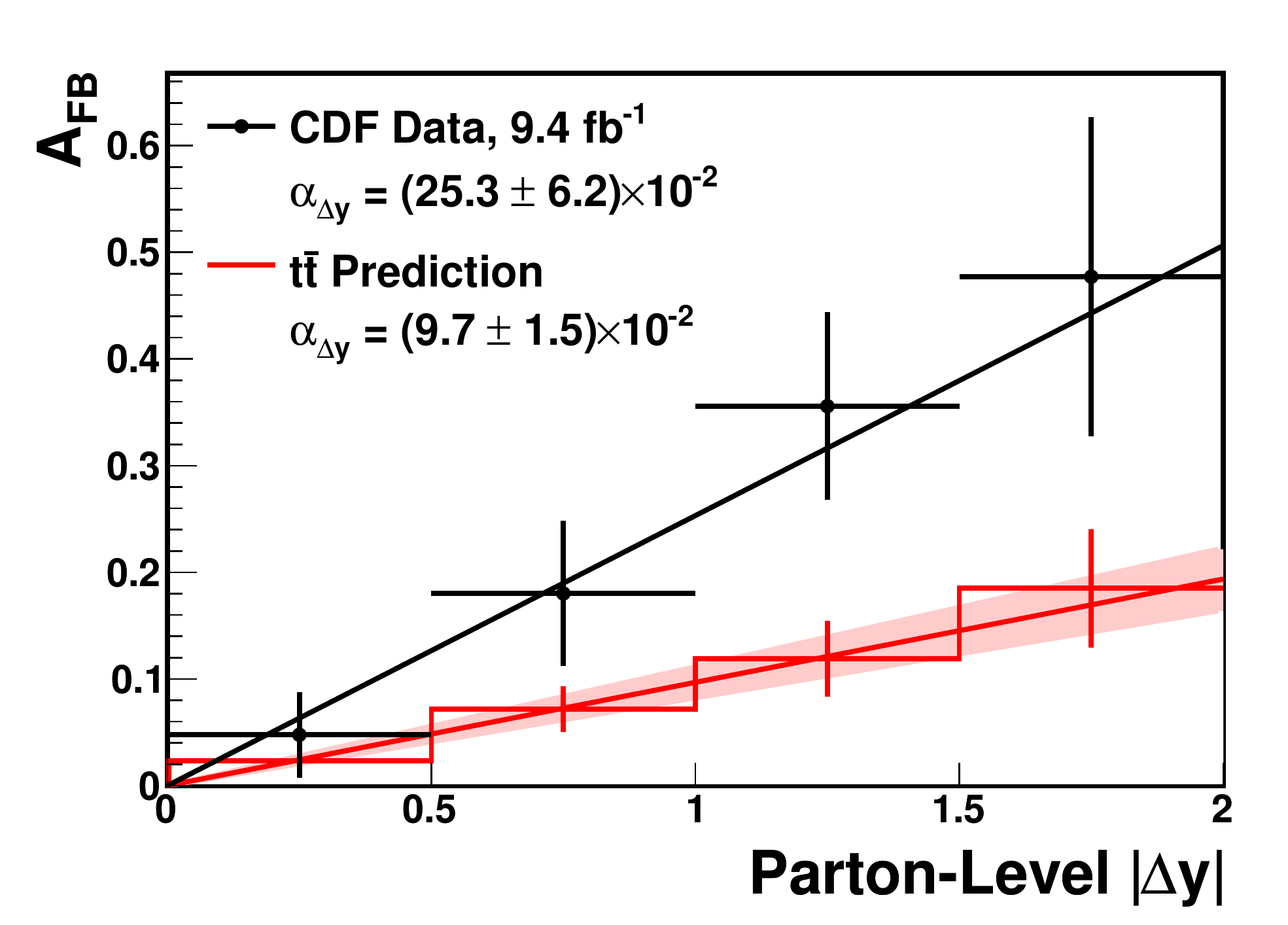}
\caption{{\small The parton-level forward-backward asymmetry as a function of $|\dy|$ with a best-fit line superimposed.  Uncertainties are correlated and include both statistical and systematic contributions.  The shaded region represents the theoretical uncertainty on the slope of the prediction.} \label{fig:afb_v_dely_parton}}
\end{center}
\end{figure}

The behavior of the asymmetry as a function of $|\dy|$ is also measured after the removal of the background contribution as described previously.  Figure~\ref{fig:afb_v_dely_signal} shows the distribution $\afb(|\dy|)$ for the background-subtracted data, with the measured values summarized in Table~\ref{tab:afb_v_dy_signal}.  Systematic uncertainties on the background-subtraction procedure are included in the error bars.  The data measurements and the predictions are well-fitted by the linear assumption, with an observed slope of $\alpha_{\dy} = (15.5 \pm 3.3) \times 10^{-2}$ that exceeds the prediction of $(5.3 \pm 1.0) \times 10^{-2}$ by approximately $3\sigma$.  The observed slope is larger than at the reconstruction level owing to the removal of the background, with the significance of the difference relative to the standard model staying approximately the same.

The $|\dy|$ dependence of the asymmetry at the parton level can be derived from Fig.~\ref{fig:dely_parton} by comparing the forward and backward bins corresponding to a given value of $|\dy|$.  This parton-level $\afb(|\dy|)$ distribution is shown in Fig.~\ref{fig:afb_v_dely_parton}, with the asymmetries in each bin also listed in Table~\ref{tab:afb_dely_parton}.  A linear fit to the parton-level results yields a slope $\alpha_{\dy}$ = $(25.3 \pm 6.2) \times 10^{-2}$, compared to an expected slope of $(9.7 \pm 1.5) \times 10^{-2}$.  We use the full covariance matrix (including both statistical and systematic contributions) for the corrected $\afb$ values when minimizing $\chi^{2}$ in order to account for the correlations between bins in the parton-level distribution.

\begin{table*}[!htb]
\caption{The asymmetry at the parton level as measured in the data, compared to the SM $\ttbar$ expectation, as a function of $|\dy|$.}\label{tab:afb_dely_parton}
\begin{center}
\begin{tabular}{c c c }
\hline
\hline
Parton level        &    Data                             &  SM $\ttbar$        \\
  $|\dy|$      &    $\afb$ $\pm$ stat $\pm$ syst      &   $\afb$       \\
\hline
0.0 - 0.5          &  \phantom{0}0.048 $\pm$ 0.034 $\pm$ 0.025\phantom{0}         &  \phantom{0}0.023 $\pm$ 0.007\phantom{0}    \\
0.5 - 1.0      &  \phantom{0}0.180 $\pm$ 0.057 $\pm$ 0.046\phantom{0}        &  \phantom{0}0.072 $\pm$ 0.022\phantom{0}    \\
1.0 - 1.5      &  \phantom{0}0.356 $\pm$ 0.080 $\pm$ 0.036\phantom{0}         &  \phantom{0}0.119 $\pm$ 0.036\phantom{0}    \\
$\ge 1.5$        &  \phantom{0}0.477 $\pm$ 0.132 $\pm$ 0.074\phantom{0}        &  \phantom{0}0.185 $\pm$ 0.056\phantom{0}    \\
\hline
$< 1.0$          &  \phantom{0}0.101 $\pm$ 0.040 $\pm$ 0.029\phantom{0}         &  \phantom{0}0.043 $\pm$ 0.013\phantom{0}     \\
$\ge 1.0$        &  \phantom{0}0.392 $\pm$ 0.093 $\pm$ 0.043\phantom{0}         &  \phantom{0}0.139 $\pm$ 0.042\phantom{0}     \\
\hline \hline
\end{tabular}
\end{center}
\end{table*}

\section{Dependence of the asymmetry on $\mttb$}\label{sec:afb_v_mtt}

The dependence of $\afb$ on the invariant mass of the $\ttbar$ system was also studied in the $5~\ifb$ analyses~\cite{cdfafb,d0afb} with only two bins.  $\mttb$ is correlated with the rapidity difference $\dy$, but because $\dy$ depends on the top-quark production angle in addition to $\mttb$, a measurement of the $\mttb$ dependence can provide additional information about the underlying asymmetry relative to the $\afb(|\dy|)$ measurement.  In the previous publications~\cite{cdfafb,d0afb}, the CDF and D0 measurements of $\afb$ at small and large $\mttb$ were consistent within statistical uncertainties but had quite different central values, leading to an ambiguity in the comparison of the results and their interpretation.  We use the full CDF data set and the new techniques introduced in this analysis to clarify the dependence of $\afb$ on $\mttb$.

\begin{figure}[!htbp]
\begin{center}
  \includegraphics[width=0.45\textwidth, clip]{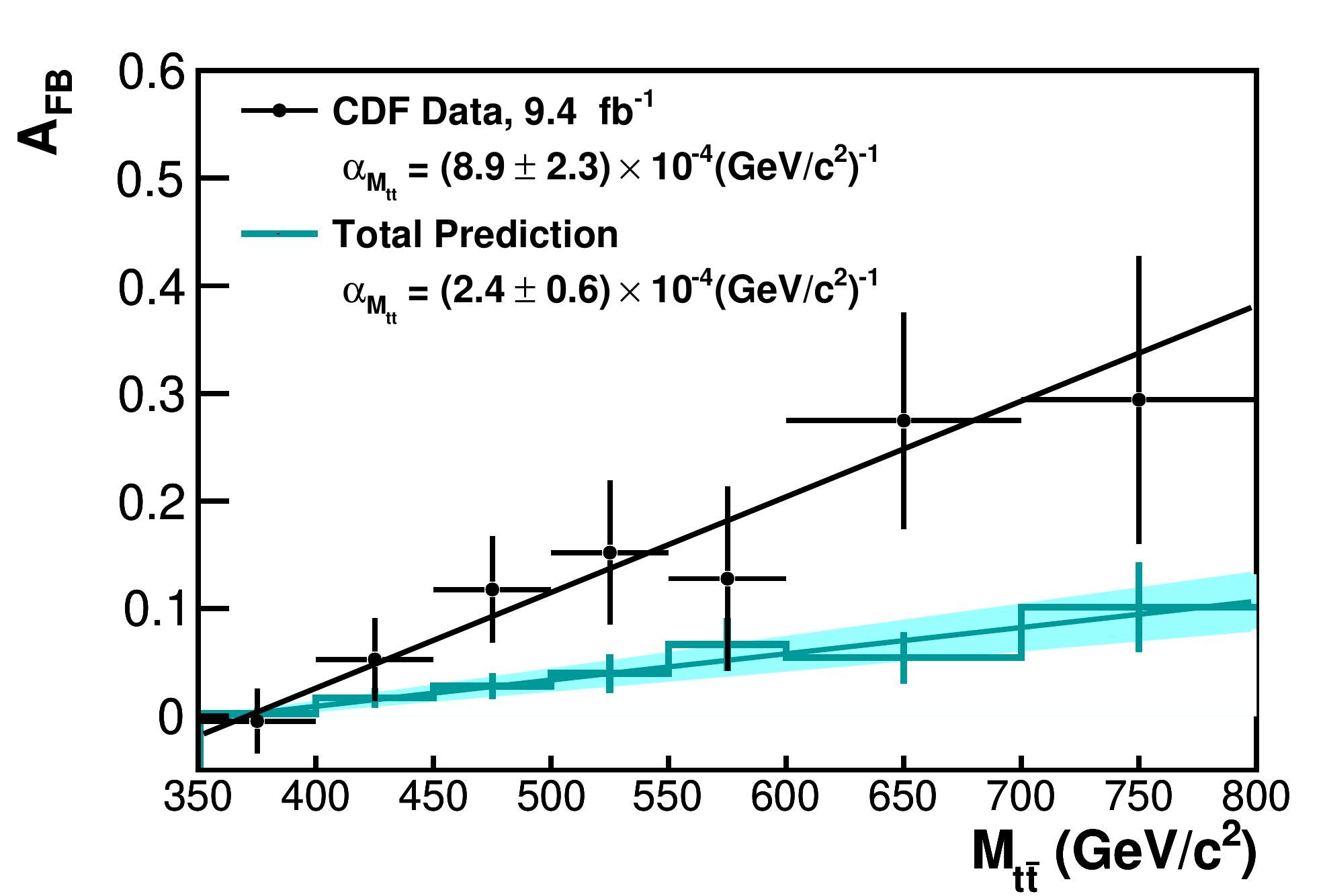}
\caption{{\small The reconstruction-level forward-backward asymmetry as a function of $\mttb$ with a best-fit line superimposed.  The last bin contains overflow events.  The errors on the data are statistical, and the shaded region represents the uncertainty on the slope of the prediction.} \label{fig:afb_v_mtt}}
\end{center}
\end{figure}

\begin{figure}[!htbp]
\begin{center}
\subfigure[]{
\includegraphics[width=0.45\textwidth, clip]{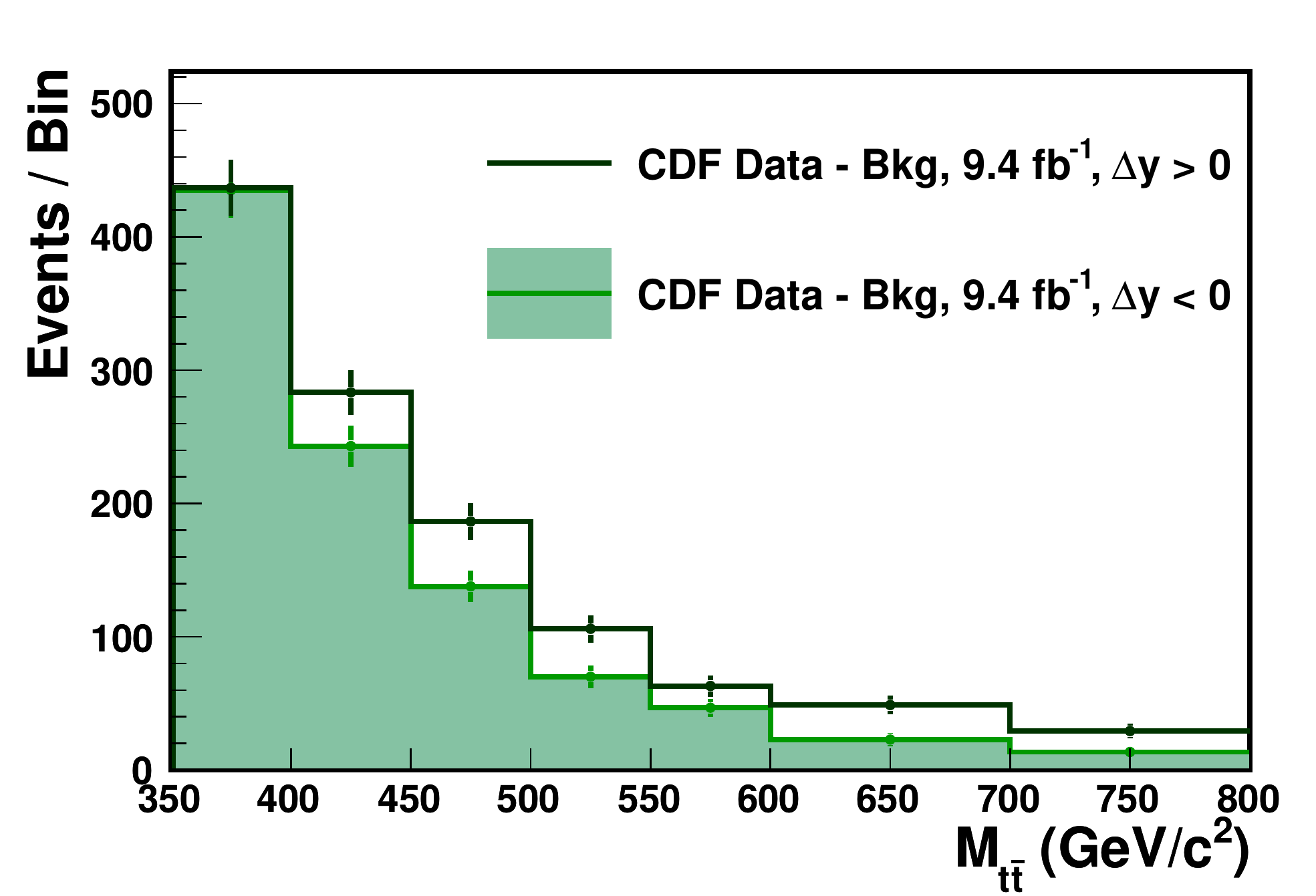}\label{fig:mtt_forbac_sig}}
\vspace*{0.15in}
\subfigure[]{
  \includegraphics[width=0.45\textwidth, clip]{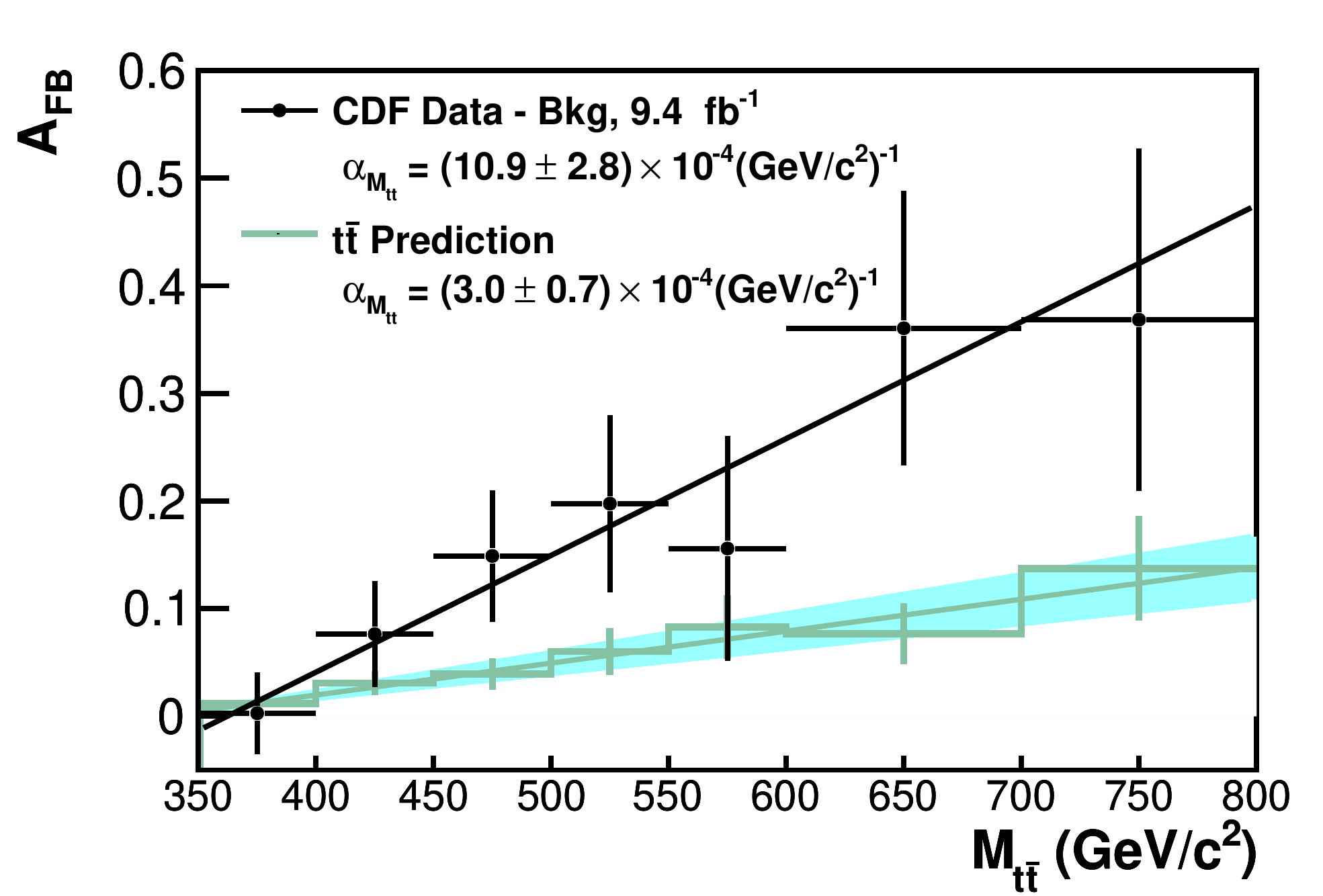}\label{fig:afb_v_mtt_sig}}

\caption{{\small \subref{fig:mtt_forbac_sig} $\mttb$ after background subtraction in events with positive and negative $\dy$ and \subref{fig:afb_v_mtt_sig} background-subtracted $\afb$ as a function of $\mttb$ with a best-fit line superimposed.  The last bin contains overflow events.  Error bars include both statistical and background-related systematic uncertainties.  The shaded region in \subref{fig:afb_v_mtt_sig} represents the theoretical uncertainty on the slope of the prediction.} \label{fig:mtt_fb_signal}}
\end{center}
\end{figure}

\begin{figure*}[!htbp]
\begin{center}
\subfigure[]{
\includegraphics[width=0.45\textwidth, clip]{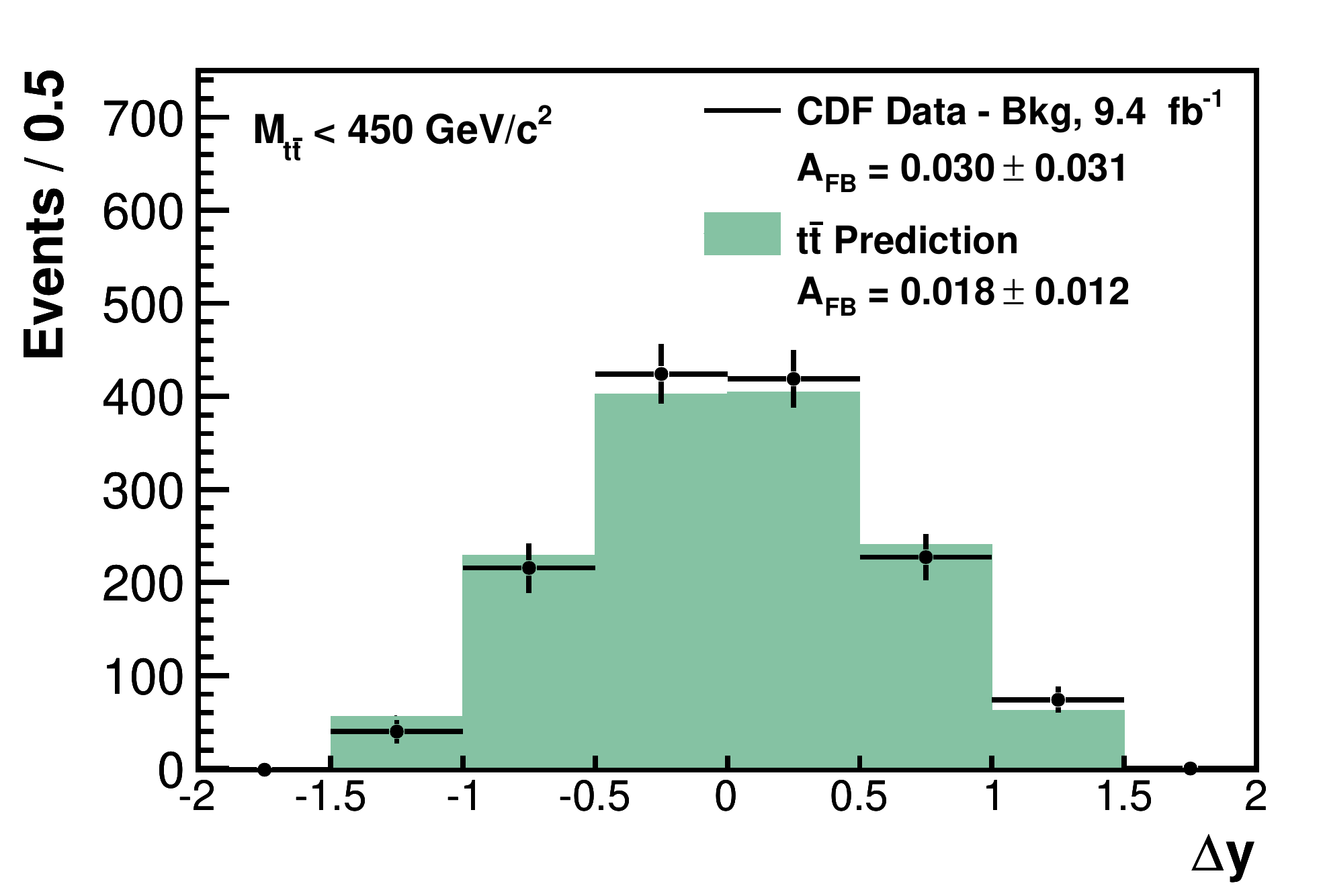}\label{fig:dely_lo_sig}}
\subfigure[]{
  \includegraphics[width=0.45\textwidth, clip]{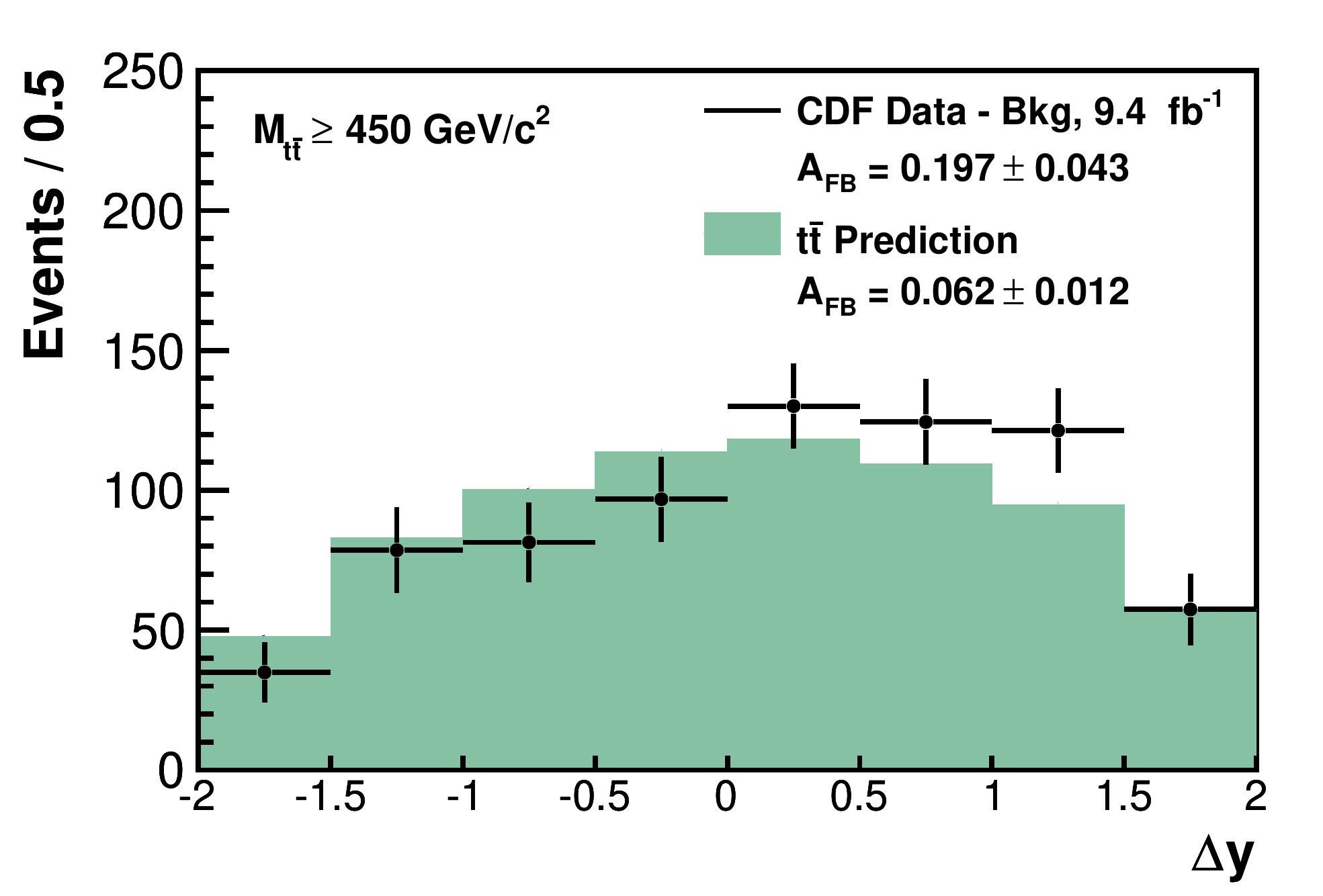}\label{fig:dely_hi_sig}}
\caption{{\small The background-subtracted $\dy$ distributions for events with \subref{fig:dely_lo_sig} $\mttb < 450~\gevcc$ and \subref{fig:dely_hi_sig} $\mttb \ge 450~\gevcc$.  Error bars include both statistical and background-related systematic uncertainties.} \label{fig:dely_hilomass_signal}}
\end{center}
\end{figure*}

We start at the detector level, where we divide the data into several mass bins and determine the number of events with positive ($N_{F}$) and negative ($N_{B}$) $\dy$ in each bin, from which we calculate the asymmetry as a function of $\mttb$ according to

\begin{equation}
\afb(\mttb) = \frac{N_F(\mttb) - N_B(\mttb)}{N_F(\mttb) + N_B(\mttb)}.
\label{afbvm_data}
\end{equation}

\noindent The $\mttb$-dependent asymmetry is compared to the NLO $\ttbar$ plus background prediction in Fig.~\ref{fig:afb_v_mtt} and Table~\ref{tab:afb_v_mtt_reco}. The $\mttb$ spectrum is divided into intervals of $50~\gevcc$ below $600~\gevcc$ and $100~\gevcc$ intervals above $600~\gevcc$, with the final bin containing overflow events. The $\mttb$ resolution across this range varies as a function of mass, being approximately $50~\gevcc$ at the lowest masses and increasing to near $100~\gevcc$ at very high mass.  A linear fit of the observed data has $\chi^{2}/N_{dof} = 1.0/5$ and yields a slope of $\alpha_{\mttb} = (8.9 \pm 2.3) \times 10^{-4}~(\gevcc)^{-1}$, which is non-zero with significance in excess of $3 \sigma$.  The predicted slope at the reconstruction level is $(2.4 \pm 0.6) \times 10^{-4}~(\gevcc)^{-1}$.

\begin{table*}[!htb]
\caption{The asymmetry observed in the reconstructed data, compared to the SM $\ttbar$ plus background expectation, as a function of $\mttb$.}\label{tab:afb_v_mtt_reco}
\begin{center}
\begin{tabular}{c c c }

\hline
\hline
       &    Data       &  SM $\ttbar$ + Bkg.        \\
  $\mttb$ ($\gevcc$)      &     $\afb$ $\pm$ stat      &   $\afb$       \\
\hline
$< 400$       &    $-$0.005 $\pm$ 0.030\phantom{0}                        & \phantom{0}0.002 $\pm$ 0.006\phantom{0}      \\
400 - 450   &    \phantom{$-$}0.053 $\pm$ 0.039\phantom{0}               & \phantom{0}0.017 $\pm$ 0.010\phantom{0}      \\
450 - 500   &    \phantom{$-$}0.118 $\pm$ 0.050\phantom{0}               & \phantom{0}0.028 $\pm$ 0.012\phantom{0}      \\
500 - 550   &    \phantom{$-$}0.152 $\pm$ 0.067\phantom{0}               & \phantom{0}0.040 $\pm$ 0.018\phantom{0}      \\
550 - 600   &    \phantom{$-$}0.128 $\pm$ 0.086\phantom{0}               & \phantom{0}0.067 $\pm$ 0.025\phantom{0}      \\
600 - 700   &    \phantom{$-$}0.275 $\pm$ 0.101\phantom{0}               & \phantom{0}0.054 $\pm$ 0.024\phantom{0}      \\
$\ge 700$     &    \phantom{$-$}0.294 $\pm$ 0.134\phantom{0}               & \phantom{0}0.101 $\pm$ 0.042\phantom{0}      \\
\hline \hline
\end{tabular}
\end{center}
\end{table*}

After removing the background contribution,  Fig.~\ref{fig:mtt_forbac_sig} compares the observed $\mttb$ distributions in forward and backward events, with an excess of forward events in many bins.  These distributions are converted into asymmetries as a function of $\mttb$, as shown in Fig.~\ref{fig:afb_v_mtt_sig} and summarized in Table~\ref{tab:afb_v_mtt_signal}.  The linear fit to the background-subtracted asymmetries yields $\chi^{2}/N_{dof}= 1.1/5$ and a slope of $(10.9 \pm 2.8) \times 10^{-4}~(\gevcc)^{-1}$, with the predicted slope being $(3.0 \pm 0.7) \times 10^{-4}~(\gevcc)^{-1}$.

\begin{table*}[!htb]
\caption{The asymmetry at the background-subtracted level as measured in the data, compared to the SM $\ttbar$ expectation, as a function of $\mttb$.}\label{tab:afb_v_mtt_signal}
\begin{center}
\begin{tabular}{c c c }

\hline
\hline
        &    Data      &  SM $\ttbar$        \\
  $\mttb$ ($\gevcc$)     &     $\afb$ $\pm$ (stat$+$syst)     &   $\afb$       \\
\hline
$< 400$       &    \phantom{$-$}0.003 $\pm$ 0.038\phantom{0}               &  \phantom{0}0.012 $\pm$ 0.006\phantom{0}   \\
400 - 450   &    \phantom{$-$}0.076 $\pm$ 0.049\phantom{0}               &  \phantom{0}0.031 $\pm$ 0.011\phantom{0}    \\
450 - 500   &    \phantom{$-$}0.149 $\pm$ 0.061\phantom{0}               &  \phantom{0}0.039 $\pm$ 0.015\phantom{0}    \\
500 - 550   &    \phantom{$-$}0.198 $\pm$ 0.083\phantom{0}               &  \phantom{0}0.060 $\pm$ 0.022\phantom{0}    \\
550 - 600   &    \phantom{$-$}0.156 $\pm$ 0.104\phantom{0}               &  \phantom{0}0.083 $\pm$ 0.030\phantom{0}    \\
600 - 700   &    \phantom{$-$}0.361 $\pm$ 0.128\phantom{0}               &  \phantom{0}0.077 $\pm$ 0.028\phantom{0}    \\
$\ge 700$     &    \phantom{$-$}0.369 $\pm$ 0.159\phantom{0}               &  \phantom{0}0.137 $\pm$ 0.049\phantom{0}   \\
\hline \hline
\end{tabular}
\end{center}
\end{table*}

\begin{figure}
\begin{center}
\subfigure[]{
\includegraphics[width=0.45\textwidth, clip]{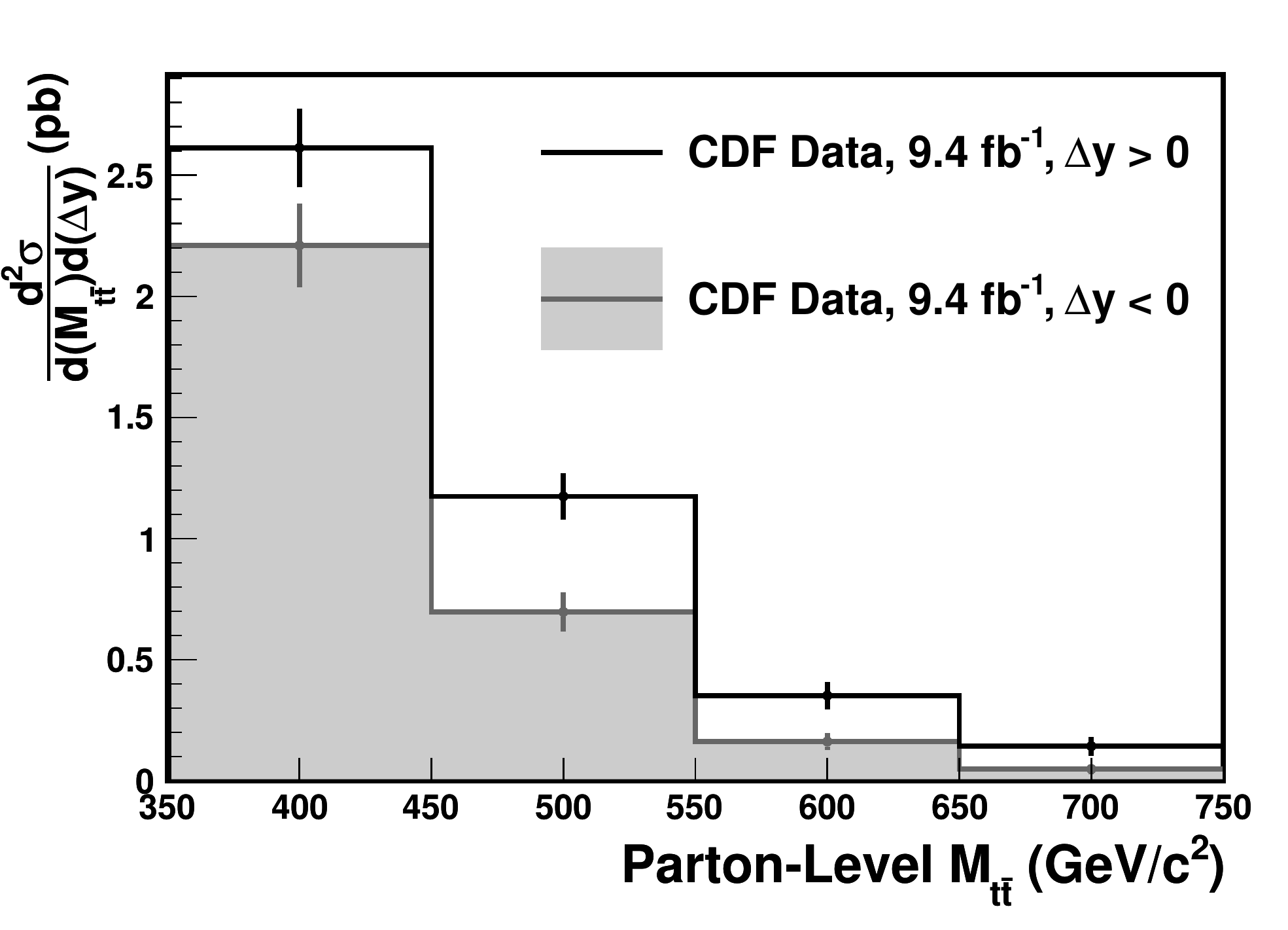}\label{fig:mtt_forbac_parton}}
\vspace*{0.15in}
\subfigure[]{
\includegraphics[width=0.45\textwidth, clip]{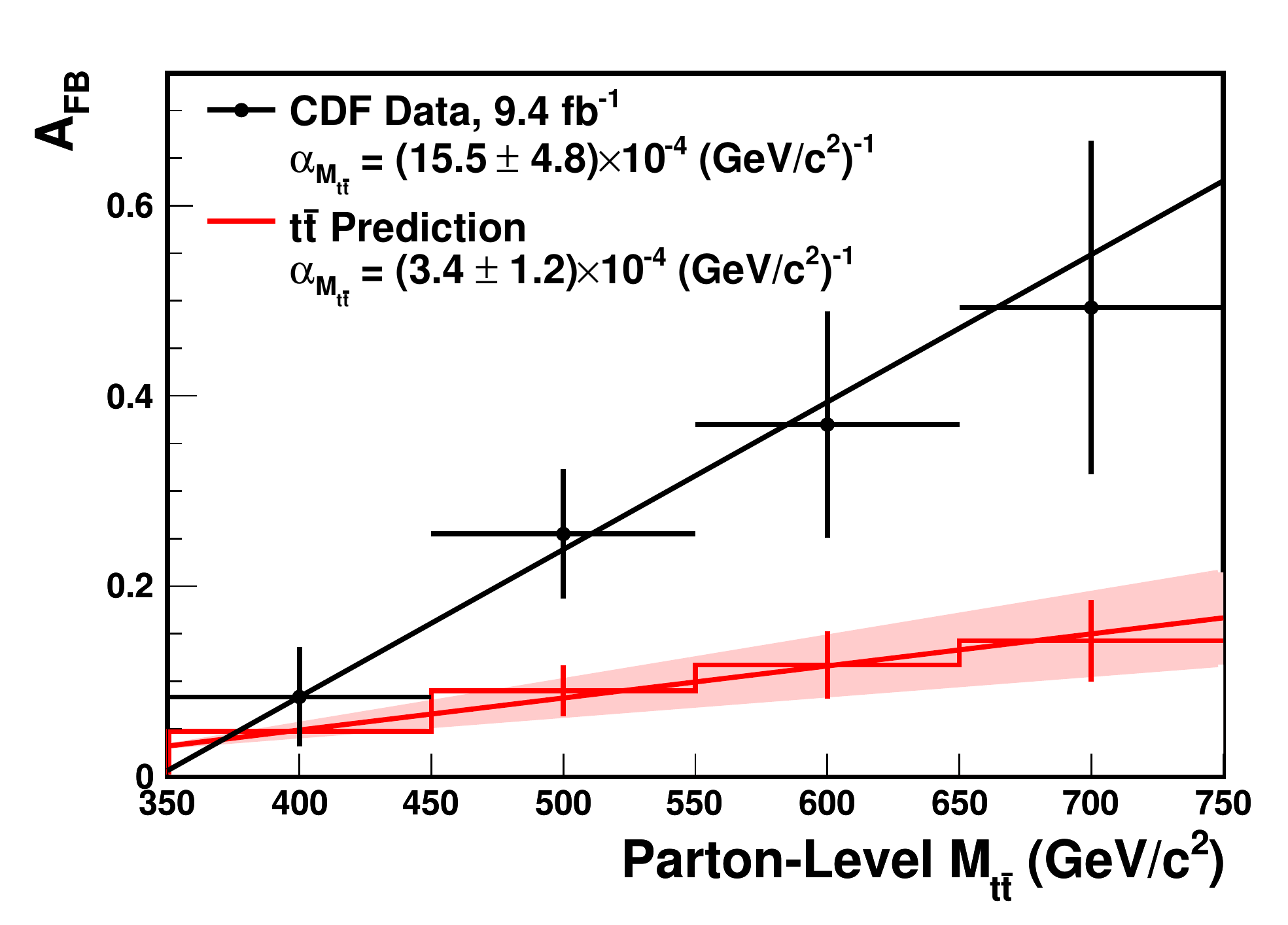}\label{fig:afb_v_mtt_parton}}
\caption{{\small \subref{fig:mtt_forbac_parton} The parton-level $\mttb$ distributions for events with positive and negative $\dy$ and \subref{fig:afb_v_mtt_parton} the parton-level forward-backward asymmetry as a function of $\mttb$ with a best-fit line superimposed.  The last bin contains overflow events.  Uncertainties are correlated and include both statistical and systematic contributions.  The shaded region in \subref{fig:afb_v_mtt_parton} represents the theoretical uncertainty on the slope of the prediction.} \label{fig:mtt_fb_parton}}
\end{center}
\end{figure}

At the background-subtracted level, we divide the data into two regions of $\mttb$ (above and below $ 450~\gevcc$) for direct comparison to the $5.3~\ifb$ CDF analysis~\cite{cdfafb}. The $\dy$ distributions at high and low mass are shown in Fig.~\ref{fig:dely_hilomass_signal},  yielding asymmetries of $0.030 \pm 0.031$ for $\mttb < 450~\gevcc$ and $0.197 \pm 0.043$ for $\mttb\geq 450~\gevcc$, where the uncertainties include statistical and background-related systematic contributions.  These are in good agreement with the values from the $5.3~\ifb$ analysis, which found background-subtracted asymmetries of $-0.022 \pm 0.043$ for $\mttb < 450~\gevcc$ and $0.266 \pm 0.62$ for $\mttb\geq 450~\gevcc$~\cite{cdfafb}.  To check against potential systematic effects, the behavior of the background-subtracted asymmetry at high and low $\mttb$ in various subsets of the data is summarized in Table~\ref{tab:afb_splits_signal}.  The $\mttb$ dependence is consistent across lepton charge and lepton type. It is consistent (within relatively large statistical uncertainties) across single- and double-$b$-tagged events.  The asymmetry is larger in events with exactly four jets than it is in events with at least five jets, an effect that is discussed further in Sec.~\ref{sec:afb_v_pt}.

\begin{table*}[!htb]
\caption{Various measured asymmetries after background subtraction, inclusively and at small and large $\mttb$.}\label{tab:afb_splits_signal}
\begin{center}
\begin{tabular}{l c c c}

\hline
\hline
            &    \multicolumn{3}{c}{$\afb$ $\pm$ (stat$+$syst)}      \\
   Sample             &     \phantom{0}Inclusive\phantom{0}                        & \phantom{0}$\mttb < 450~\gevcc$\phantom{0}             & \phantom{0}$\mttb \ge 450~\gevcc$\phantom{0} \\
\hline
All data      &      \phantom{0}0.087 $\pm$ 0.026\phantom{0}     &  \phantom{0$-$}0.030 $\pm$ 0.031\phantom{0}            & \phantom{0}0.197 $\pm$ 0.043\phantom{0}  \\

Positive leptons   &  \phantom{0}0.094 $\pm$ 0.036\phantom{0}     &  \phantom{0$-$}0.034 $\pm$ 0.044\phantom{0}            & \phantom{0}0.207 $\pm$ 0.060\phantom{0}  \\
Negative leptons   &  \phantom{0}0.080 $\pm$ 0.035\phantom{0}     &  \phantom{0$-$}0.027 $\pm$ 0.043\phantom{0}            & \phantom{0}0.186 $\pm$ 0.057\phantom{0}  \\

Exactly 1 $b$ tags  &\phantom{0}0.100 $\pm$ 0.031\phantom{0}     &  \phantom{0$-$}0.047 $\pm$ 0.036\phantom{0}            & \phantom{0}0.220 $\pm$ 0.049\phantom{0}  \\
At least 2 $b$ tags & \phantom{0}0.037 $\pm$ 0.045\phantom{0}     & \phantom{0}$-$0.018 $\pm$ 0.055\phantom{0}            & \phantom{0}0.134 $\pm$ 0.073\phantom{0}  \\

Electrons    &   \phantom{0}0.079 $\pm$ 0.039\phantom{0}     &  \phantom{0$-$}0.017 $\pm$ 0.047\phantom{0}            & \phantom{0}0.195 $\pm$ 0.062\phantom{0}  \\
Muons        &   \phantom{0}0.094 $\pm$ 0.033\phantom{0}     &  \phantom{0$-$}0.041 $\pm$ 0.040\phantom{0}            & \phantom{0}0.197 $\pm$ 0.055\phantom{0}  \\

Exactly 4 jets    &  \phantom{0}0.110 $\pm$ 0.031\phantom{0}     &  \phantom{0$-$}0.029 $\pm$ 0.037\phantom{0}            & \phantom{0}0.256 $\pm$ 0.049\phantom{0}  \\
At least 5 jets   &   \phantom{0}0.033 $\pm$ 0.044\phantom{0}     & \phantom{0$-$}0.034 $\pm$ 0.053\phantom{0}            & \phantom{0}0.033 $\pm$ 0.077\phantom{0}  \\
\hline \hline
\end{tabular}
\end{center}
\end{table*}

\begin{table*}[!htb]
\caption{The asymmetry at the parton level as measured in the data, compared to the SM $\ttbar$ expectation, as a function of $\mttb$.}\label{tab:afb_mtt_parton}
\begin{center}
\begin{tabular}{c c c }
\hline
\hline
Parton level   &    Data      &  SM $\ttbar$       \\
  $\mttb$ ($\gevcc$)     &     $\afb$ $\pm$ stat $\pm$ syst      &   $\afb$       \\
\hline
$< 450$       & \phantom{0}0.084 $\pm$ 0.046 $\pm$ 0.030\phantom{0}              & \phantom{0}0.047 $\pm$ 0.014\phantom{0}     \\
450 - 550     & \phantom{0}0.255 $\pm$ 0.062 $\pm$ 0.034\phantom{0}              & \phantom{0}0.090 $\pm$ 0.027\phantom{0}     \\
550 - 650     & \phantom{0}0.370 $\pm$ 0.084 $\pm$ 0.087\phantom{0}              & \phantom{0}0.117 $\pm$ 0.035\phantom{0}     \\
$\ge 650$     & \phantom{0}0.493 $\pm$ 0.158 $\pm$ 0.110\phantom{0}              & \phantom{0}0.143 $\pm$ 0.043\phantom{0}     \\
\hline
$< 450$   & \phantom{0}0.084 $\pm$ 0.046 $\pm$ 0.030\phantom{0}                  & \phantom{0}0.047 $\pm$ 0.014\phantom{0}     \\
$\ge 450$ & \phantom{0}0.295 $\pm$ 0.058 $\pm$ 0.033\phantom{0}                  & \phantom{0}0.100 $\pm$ 0.030\phantom{0}     \\
\hline \hline
\end{tabular}
\end{center}
\end{table*}

We determine the parton-level mass dependence of $\afb$ by correcting the $\dy$ and $\mttb$ distributions simultaneously.  To do so, we apply the unfolding procedure to a two-dimensional distribution consisting of two bins in $\dy$ (for forward and backward events) and four bins in $\mttb$.  Since regularization makes use of the second-derivative matrix, which is not well-defined for a two-bin distribution, the regularization constraint is applied only along the $\mttb$ dimension.  The resulting $\mttb$ distributions for forward and backward events are shown in Fig.~\ref{fig:mtt_forbac_parton}.  These distributions are combined to determine the differential asymmetry as a function of $\mttb$ shown in Fig.~\ref{fig:afb_v_mtt_parton} and summarized in Table~\ref{tab:afb_mtt_parton}.  The best-fit line to the measured data asymmetries at parton level has a slope $\alpha_{\mttb}$ = $(15.5 \pm 4.8) \times 10^{-4}~(\gevcc)^{-1}$, compared to the \powheg prediction of $(3.4 \pm 1.2) \times 10^{-4}~(\gevcc)^{-1}$.

\section{Determination of the significance of the dependence of the asymmetry on $|\dy|$ and $\mttb$}\label{sec:sig}

The slopes of the linear dependencies of $\afb$ on $|\dy|$ and $\mttb$ provide a measure of the consistency between the data and the SM prediction.  We quantify this consistency in a more rigorous manner by repeating the measurement on large ensembles of simulated experiments generated according to the SM prediction and determining the probabilities, or p-values, for observing the actual data given the SM assumption.  Each p-value is defined as the fraction of simulated experiments in which the measured slopes are at least as large as those found in the data, $\alpha_{\dy,\mttb}^{\rm simulated} \ge \alpha_{\dy,\mttb}^{\rm data}$.  

We use the background-subtracted sample for measuring these p-values because it provides access to an asymmetry calculation that has been corrected for background but is still independent of the assumptions made when using a regularized unfolding procedure to extract parton-level information.  We start from the predicted distribution at the reconstruction level, created from the standard model predictions of \powheg and the various background contributions proportioned as in Table~\ref{tab:method2}. The population of each bin of this predicted distribution is fluctuated within its uncertainty, which includes the statistical uncertainty on the contents of that bin, the systematic uncertainties on the various background contributions, as described in Sec.~\ref{sec:bkg_sub} above, and the theoretical uncertainty on the \powheg prediction.

Many systematic uncertainties may in principal simultaneously affect both the signal and background models.  However, the theory uncertainty is the dominant uncertainty on the predicted asymmetry (0.011).  As a point of comparison, the uncertainty due to jet energy scales is only 0.0008, and the effects of correlations between uncertainties on the $\ttbar$ prediction and backgrounds are negligible.  For this reason we do not include the effect of correlations between uncertainties on the signal and background models.

For each simulated experiment, the nominal background prediction with the normalizations of Table~\ref{tab:method2} is subtracted, and the slopes of the remaining asymmetries as a function of $|\dy|$ and $\mttb$ are fit.  We find p-values of $2.2 \times 10^{-3}$ for $\afb(|\dy|)$ and $7.4 \times 10^{-3}$ for $\afb(\mttb)$, corresponding to $2.8\sigma$ and $2.4\sigma$ discrepancies respectively (based on a one-sided integration of the normal probability distribution).

\section{Dependence of the asymmetry on the transverse momentum of the $\ttbar$ system}\label{sec:afb_v_pt}

\begin{figure}[!ht]
\begin{center}
\includegraphics[width=0.45\textwidth, clip]{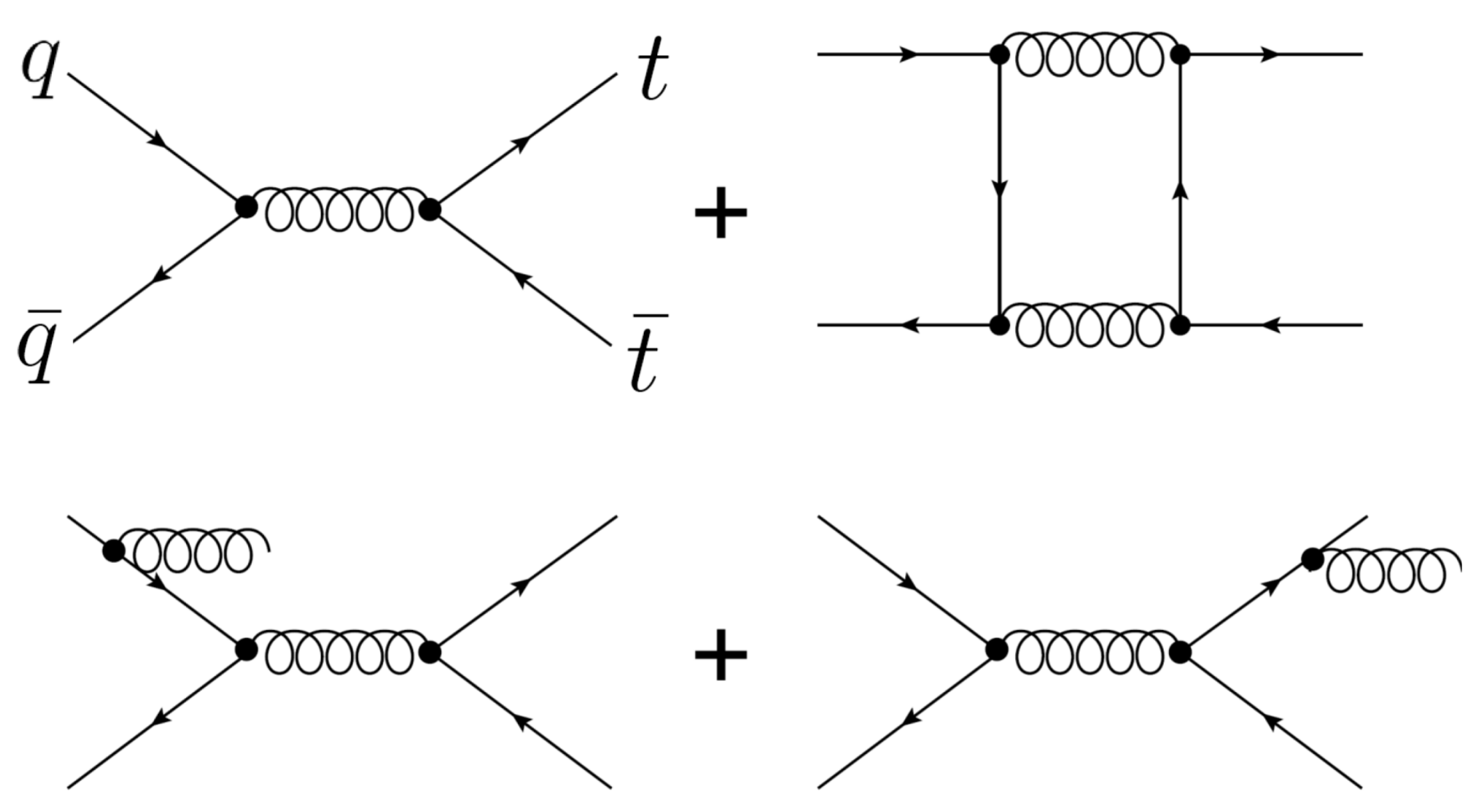}
\caption{{\small Interfering $\qqbar\rightarrow\ttbar$ (top) and $\qqbar\rightarrow\ttbar j$ (bottom) diagrams.} \label{fig:nlo}}
\end{center}
\end{figure}

The QCD asymmetry at NLO arises from the sum of two different effects~\cite{nlotheory}. The interference of the $2 \rightarrow 2$ LO tree-level diagrams (upper left of Fig.~\ref{fig:nlo}) and the NLO box diagrams (upper right) produces a positive asymmetry (``Born-box'' interference), while the interference of $2 \rightarrow 3$ tree-level diagrams with initial-state (lower left) and final-state radiation (lower right) produces a negative asymmetry (``ISR-FSR'' interference). In the latter final state, $\ttbar$ plus an additional jet, the $\ttbar$ system acquires a transverse momentum $\ptran$, while in the former case with an exclusive $\ttbar$ final state, all events have $\ptran = 0$. The resultant SM asymmetry at NLO is therefore the sum of two effects of different sign, with very different $\ptran$ dependence. The virtual effects from Born-box interference are larger, leading to a net positive asymmetry.  Recent work has also emphasized that color coherence during the hadronization process can produce a significant $\ptran$ dependence for the asymmetry in Monte Carlo generators that include hadronization~\cite{d0afb}, with the degree of the $\ptran$ dependence varying greatly depending on the details of the implementation of color coherence~\cite{ttpt}. The verification of the $\ptran$ dependence of the asymmetry is therefore crucial to understanding the reliability of the SM predictions for $\afb$~\cite{d0afb}, as well as testing for possible new effects beyond the SM. 

In this section, we first compare and discuss several predictions for $\afb(\ptran)$. We then compare the data to two of these predictions (the NLO with hadronization prediction from \powheg and the LO with hadronization prediction from \pythia), showing that the asymmetry in the data displays the same trend with respect to $\ptran$ as observed in both \powheg and \pythia, and that the excess inclusive asymmetry in the data is consistent with a $\ptran$-independent component.

We define the $\ptran$ dependence of the asymmetry as 

\begin{equation}
\afb(\ptran) = \frac{N_F(\ptran) - N_B(\ptran)}{N_F(\ptran) + N_B(\ptran)}.
\label{afbvpt_data}
\end{equation}

\begin{figure}[!ht]
\begin{center}
\includegraphics[width=0.45\textwidth, clip]{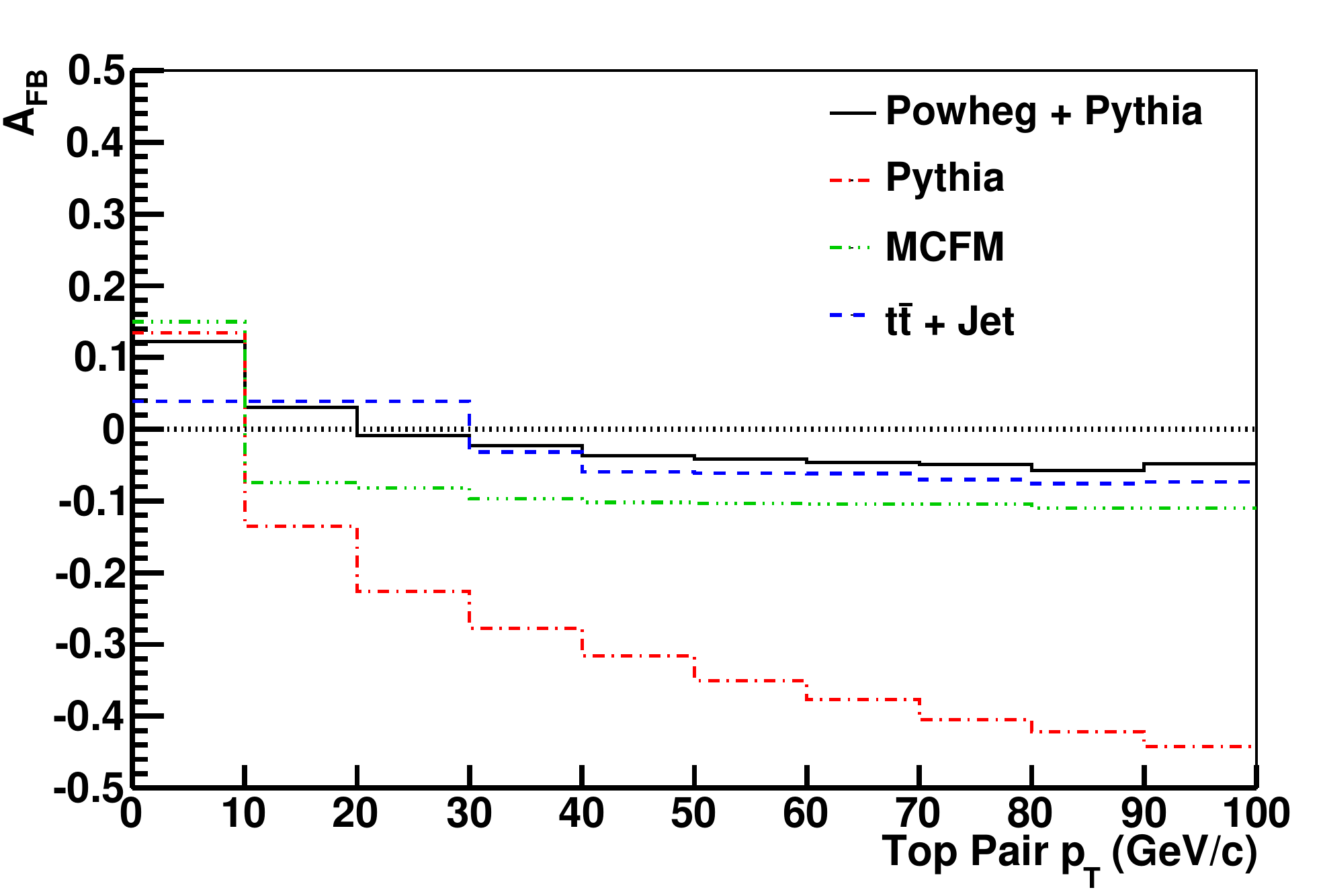}
\caption{{\small Expected $\afb$ as a function of the $\ptran$ of the $\ttbar$ system at the parton level from \mcfm, \powheg, and \pythia, as well as a NLO prediction for events where the top-quark pair is produced in association with an extra energetic jet.   } \label{fig:true_afb_vs_pt}}
\end{center}
\end{figure}

\noindent The expected SM parton-level asymmetry is shown for four predictions in Fig.~\ref{fig:true_afb_vs_pt}. The matrix elements for \pythia are LO for $\ttbar$ production, with some higher-order effects approximated through hadronization. There is essentially no net inclusive asymmetry in \pythia due to the underlying $2 \rightarrow 2$ matrix elements in the hard-scattering process; gluon emission during hadronization results in a negative asymmetry for non-zero $\ptran$ events, leaving a positive asymmetry in the low-$\ptran$ region.
  
The other three curves suggest a different behavior for the $\ptran$ dependence at NLO. The \mcfm calculation uses NLO matrix elements for $\ttbar$ production, and includes both the Born-box and ISR-FSR interference terms, with the result being a parton-level output with two partons ($\ttbar$) or three partons ($\ttbar$ plus a gluon) in the final state. In \mcfm, events produced by the virtual matrix elements with Born-box interference have $\ptran=0$ and a positive asymmetry, while events produced by the real matrix elements describing gluon radiation have non-zero $\ptran$ and a negative asymmetry. \powheg has the same NLO matrix elements as \mcfm, with additional higher-order effects approximated through hadronization performed by \pythia. The additional radiation from the hadronization process results in a migration of events in $\ptran$ and thus a moderation of the otherwise bimodal $\ptran$ behavior observed in \mcfm.   

The \powheg prediction with \pythia hadronization can be partially checked against a recent NLO calculation for $\ttbar$ production in association with an extra energetic jet ($p_T^{\rm jet} > 20~\gevc$ and $|\eta_{\rm jet}| < 2.0$)~\cite{schulze}, shown as ``$\ttbar$+jet''.  This calculation has a Born-level final state with three partons ($\ttbar$ plus a gluon), and thus it is most relevant for comparison to the other predictions at high $\ptran$.  It contains virtual matrix elements for the $\ttbar$+jet final state as well as real corrections from final states with $\ttbar$ and two extra jets.  The negative asymmetry observed in the tree-level prediction for $\ttbar$+jet (as shown in \mcfm at high-$\ptran$) is reduced with the full NLO calculation of this final state. In the high-$\ptran$ region, we see that the \powheg predictions are in good agreement with those from the NLO $\ttbar$+jet calculation. 

\begin{figure}[!htbp]
\begin{center}
\includegraphics[width=0.45\textwidth, clip]{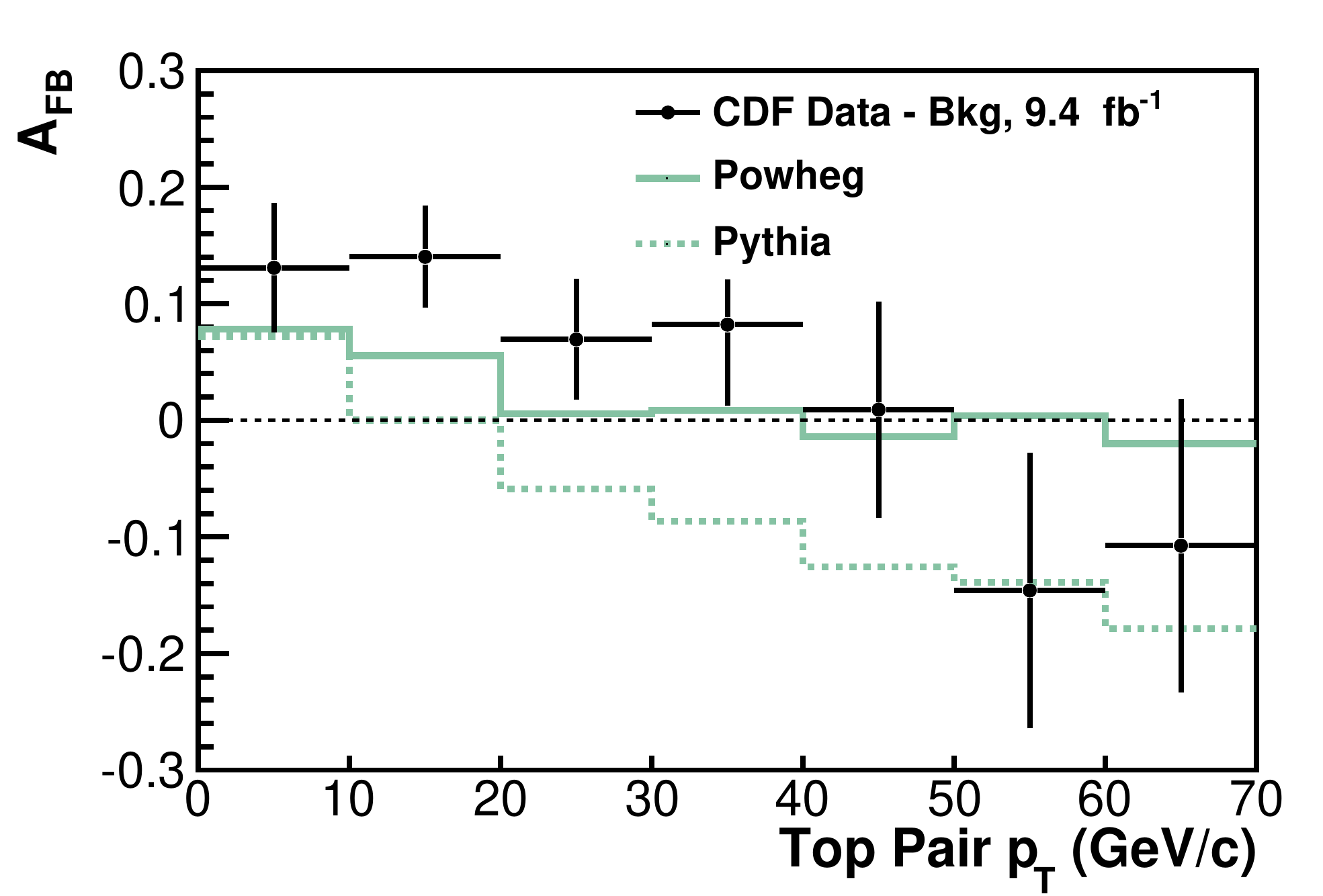}
\caption{{\small The background-subtracted forward-backward asymmetry in the data as a function of the transverse momentum of the $\ttbar$ system, compared to both \powheg and \pythia.  Error bars include both statistical and background-related systematic uncertainties.  The last bin contains overflow events.} \label{fig:afb_v_pt}}
\end{center}
\end{figure}

\begin{figure}[!htbp]
\begin{center}
\includegraphics[width=0.45\textwidth, clip]{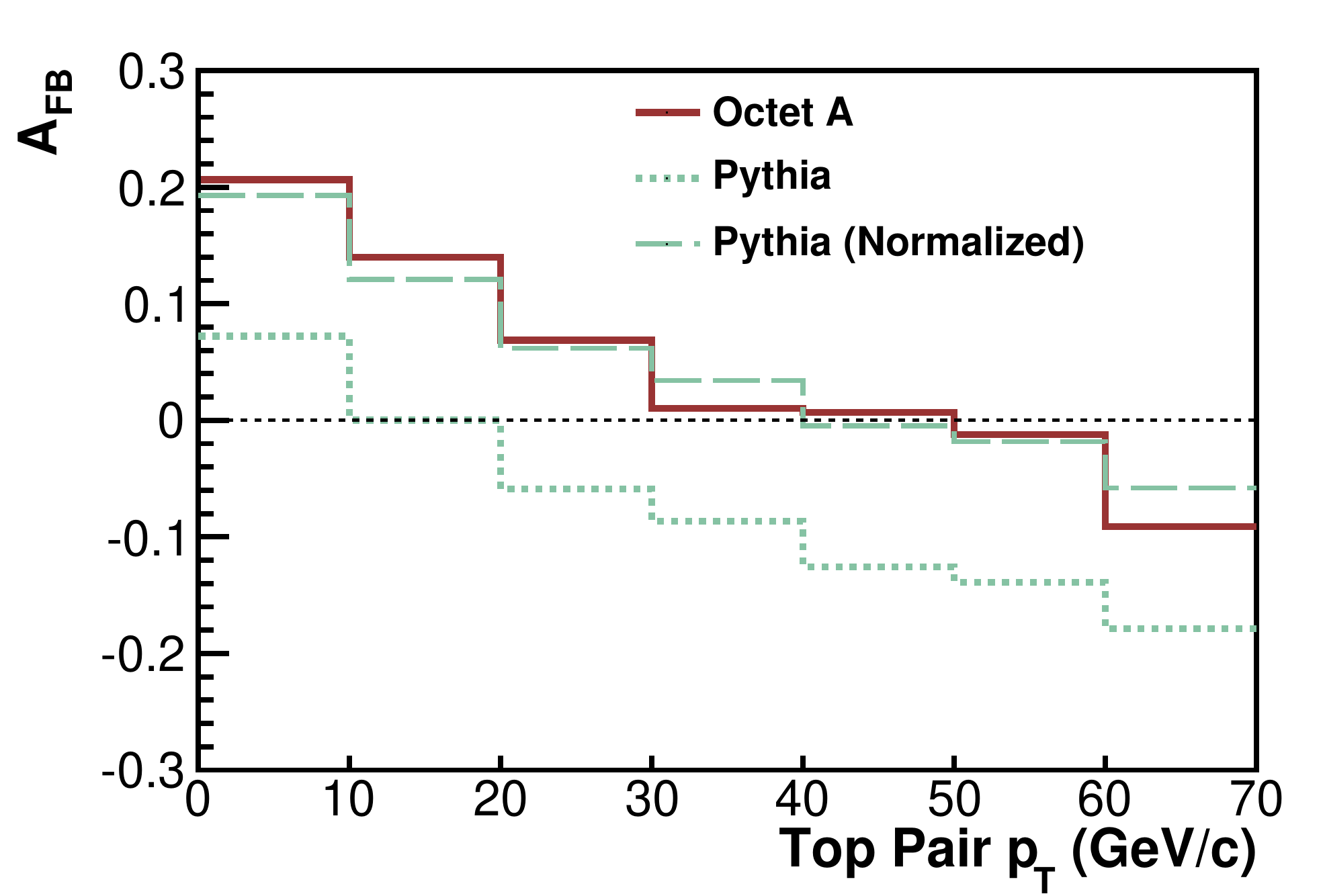}
\caption{{\small The forward-backward asymmetry as a function of the transverse momentum of the $\ttbar$ system for three models at the background-subtracted level: Octet A, SM \pythia, and SM \pythia normalized by the addition of $\Delta A_{\rm Oct}$.  The last bin contains overflow events.} \label{fig:afb_v_pt_pythia_v_octA}}
\end{center}
\end{figure}

\begin{figure}[!htbp]
\begin{center}
\includegraphics[width=0.45\textwidth, clip]{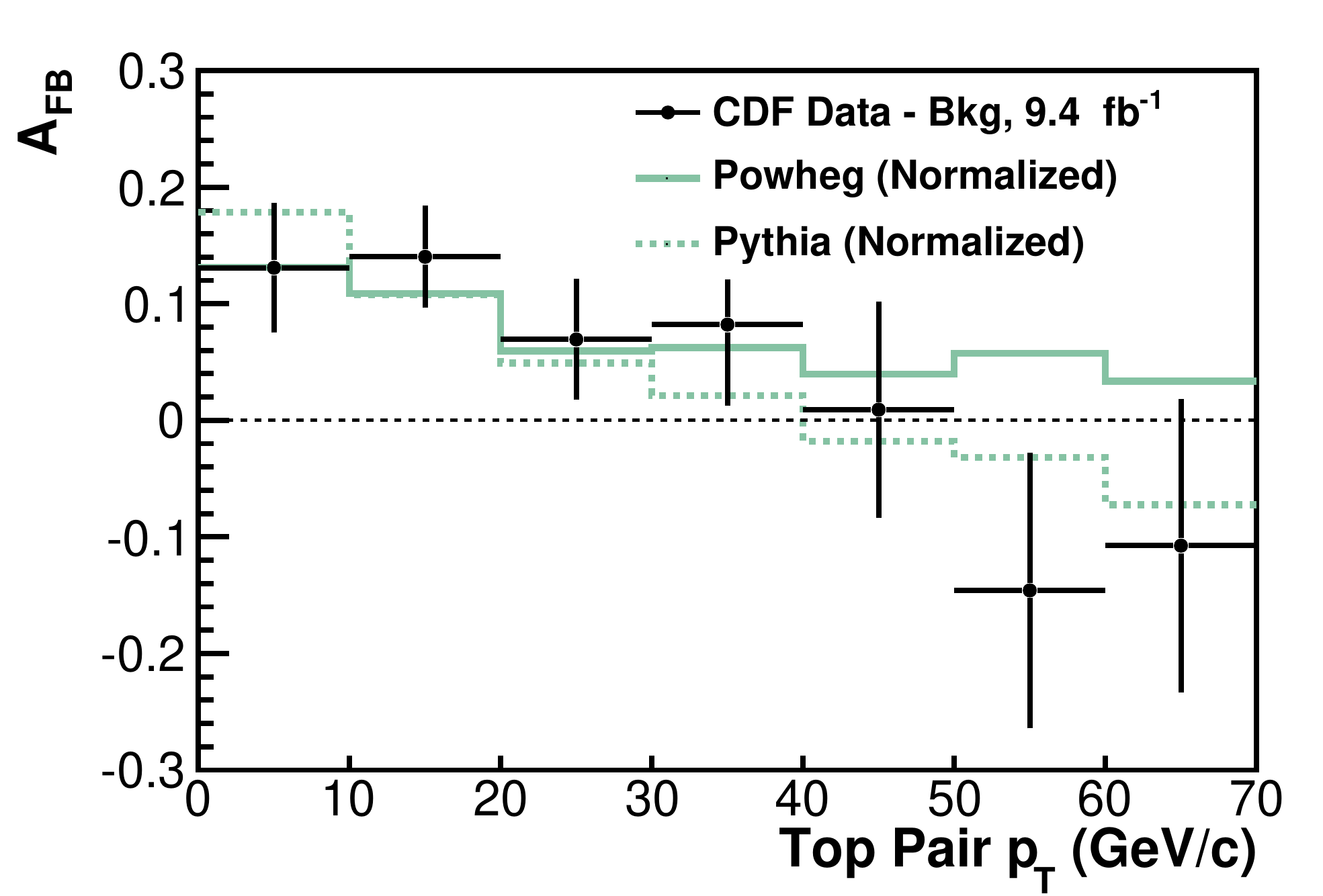}
\caption{{\small The background-subtracted forward-backward asymmetry in the data as a function of the transverse momentum of the $\ttbar$ system, compared to both \powheg and \pythia.  The model predictions have been normalized by the addition of $\Delta A_{\rm NLO}$ to \powheg and $\Delta A_{\rm LO}$ to \pythia as described in the text.  Error bars include both statistical and background-related systematic uncertainties.  The last bin contains overflow events.} \label{fig:afb_v_pt_scaled}}
\end{center}
\end{figure}

In Fig.~\ref{fig:ptsys} we show that the reconstructed $\ptran$ spectrum in the data is well-reproduced by the $\ttbar$ signal and background model simulations. Building on this, we study the $\ptran$ dependence of the asymmetry in the data.  Figure~\ref{fig:afb_v_pt} shows $\afb(\ptran)$ for the data after background subtraction compared to predictions from \powheg (hadronized with \pythia) and from \pythia.  The trends of the parton-level curves in Fig.~\ref{fig:true_afb_vs_pt} are reproduced: the LO prediction has a steady drop, while the NLO prediction tends to zero or slightly below. The data show a clear decrease with $\ptran$, but lie above the models.  We investigate this using the ansatz that the data contain an additional source of asymmetry that is independent of $\ptran$. In this case, because independent asymmetries are additive, it should be possible to normalize the model predictions to the data by adding a constant offset $\Delta A$ that is equal to the excess observed inclusive asymmetry in the data. 
   
We test this ansatz using the color-octet model Octet A (implemented in \textsc{madgraph} with hadronization performed by \pythia) described at the end of Sec.~\ref{sec:mc}. In this LO model, the octet physics induces a $\ptran$-independent inclusive $\ttbar$ asymmetry 0.106 at the background-subtracted level (we neglect very small statistical uncertainties in these large Monte Carlo samples).  We wish to compare the $\ptran$ dependence of this asymmetry to the LO \pythia model, which has a background-subtracted asymmetry of $-0.021$. The inclusive difference is $\Delta A_{\rm Oct} = 0.127$. If the excess asymmetry in Octet A is independent of $\ptran$, we expect that $\afb^{\rm \pythia}(\ptran) + \Delta A_{\rm Oct}$ reproduces satisfactorily $\afb^{\rm Octet~A}(\ptran)$.  Figure~\ref{fig:afb_v_pt_pythia_v_octA} shows this test in the simulated samples, with the $\ptran$-dependent behavior of Octet A being described well by the addition of the constant normalization factor $\Delta A_{\rm Oct}$ to $\afb^{\rm \pythia}(\ptran)$.

\begin{figure*}[!htbp]
\begin{center}
\subfigure[]{
\includegraphics[width=0.45\textwidth, clip]{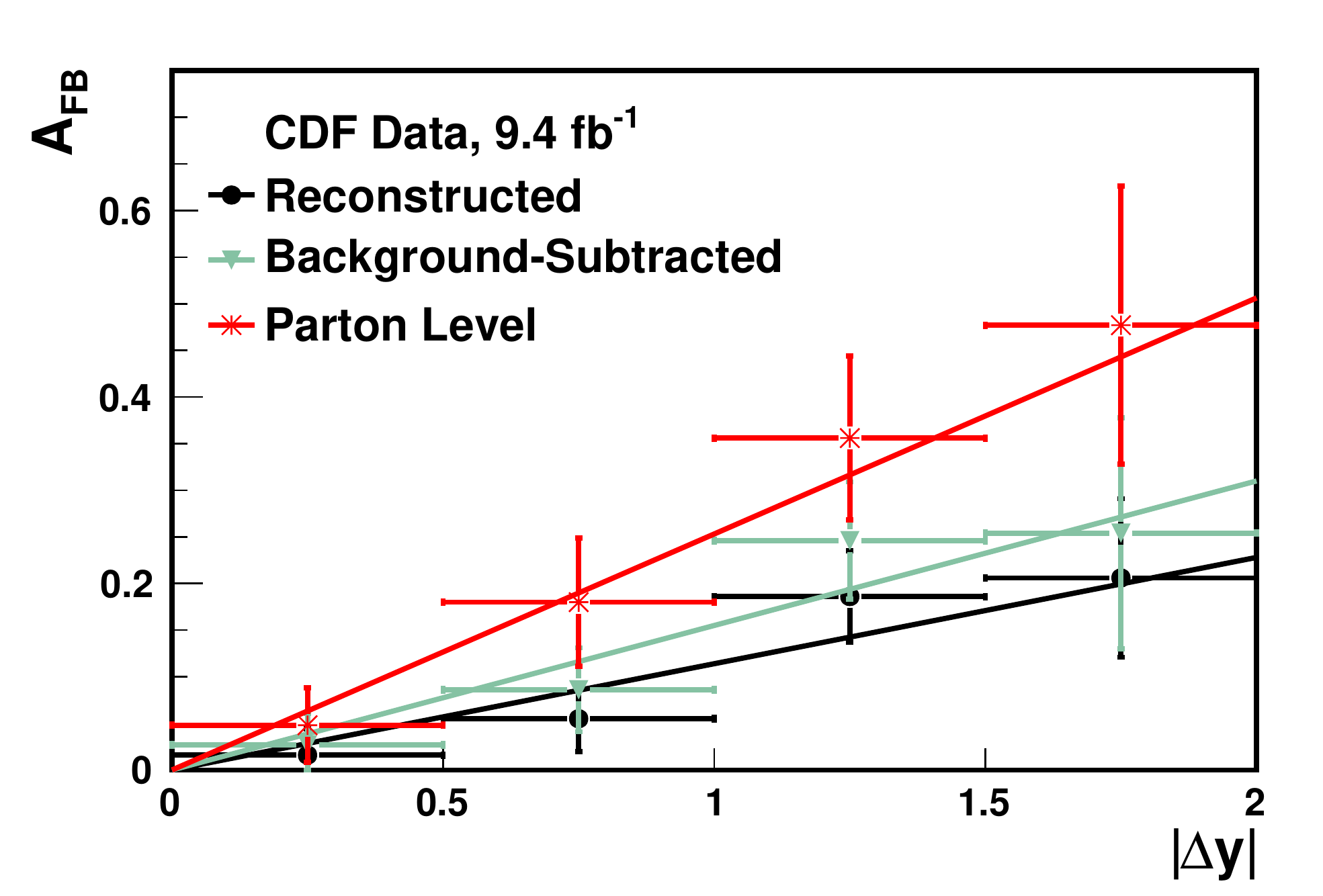}\label{fig:corr_steps_dy}}
\subfigure[]{
  \includegraphics[width=0.45\textwidth, clip]{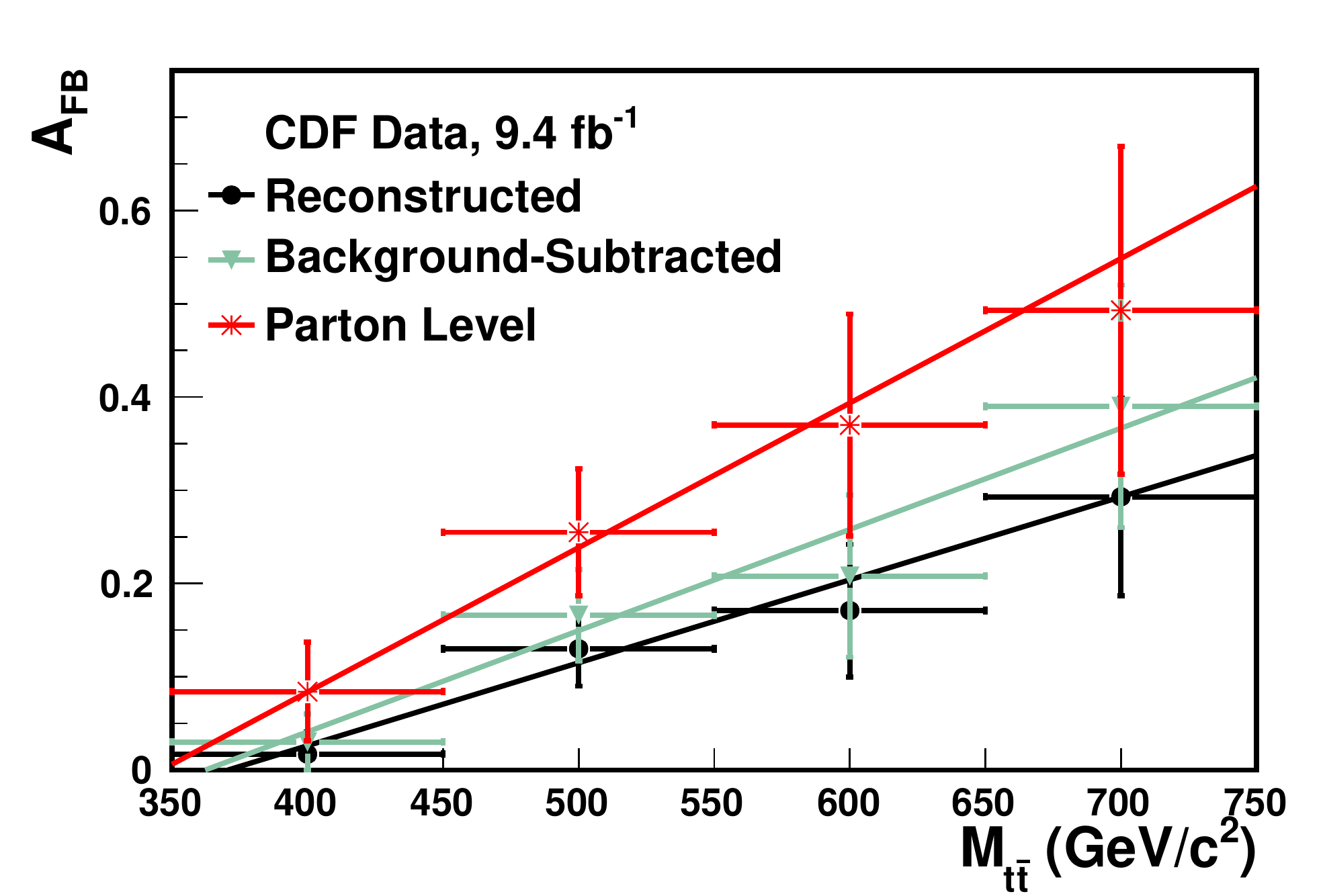}\label{fig:corr_steps_mtt}}
\caption{{\small The measured asymmetries as a function of \subref{fig:corr_steps_dy} $|\dy|$ and \subref{fig:corr_steps_mtt} $\mttb$ in the data at the three different levels of correction.  The error bars include both statistical uncertainties and the appropriate systematic uncertainties for each correction level as described in the text.  The last bin contains overflow events.} \label{fig:corr_steps}}
\end{center}
\end{figure*}

\begin{table*}[!htb]
\caption{The measured inclusive forward-backward asymmetry and the best-fit slopes for $\afb(|\dy|)$ and $\afb(\mttb)$ at the different levels of correction. The uncertainties include the statistical uncertainties and the appropriate systematic uncertainties for each correction level as discussed in the text.}\label{tab:corr_steps}\begin{center}
\begin{tabular}{c c c c}

\hline
\hline
                    &    \phantom{0}Inclusive\phantom{0}        &    \phantom{0}Slope\phantom{0}         &  \phantom{0} Slope\phantom{0}  \\
  Correction level      &   \phantom{0}$\afb$\phantom{0}      &   \phantom{0}$\alpha_{\dy}$\phantom{0}   &   \phantom{0}$\alpha_{\mttb}\phantom{0}$ \\
\hline
Reconstruction         &   \phantom{0}0.063 $\pm$ 0.019\phantom{0}              &   \phantom{0}$(11.4  \pm 2.5) \times 10^{-2}$\phantom{0}  & \phantom{0} $(8.9 \pm 2.3) \times 10^{-4}~(\gevcc)^{-1}$\phantom{0}   \\
Background-subtracted  &   \phantom{0}0.087 $\pm$ 0.026\phantom{0}              &   \phantom{0}$(15.5  \pm 3.3) \times 10^{-2}$\phantom{0}  &  \phantom{0}$(10.9 \pm 2.8) \times 10^{-4}~(\gevcc)^{-1}$\phantom{0}   \\
Parton                 &    \phantom{0}0.164 $\pm$ 0.047\phantom{0}              &  \phantom{0}$(25.3  \pm 6.2) \times 10^{-2}$\phantom{0}  & \phantom{0}$(15.5 \pm 4.8) \times 10^{-4}~(\gevcc)^{-1}$\phantom{0}     \\
\hline \hline
\end{tabular}
\end{center}
\end{table*}

We use this procedure to normalize the $\afb(\ptran)$ models of \powheg and \pythia to the total inclusive asymmetry observed in the data. Since this artificial procedure adjusts the mean values such that they are exactly equal, we do not assign uncertainties to the offsets. The asymmetry after background subtraction is $0.087$ in the data, $0.033$ in NLO \powheg (Table~\ref{tab:inc_asyms_posneg}), and $-0.021$ in LO \pythia, resulting in offset terms $\Delta A_{\rm NLO} = 0.054$ and $\Delta A_{\rm LO} = 0.108$. 

The normalized $\afb(\ptran)$ models are compared to the data in Fig.~\ref{fig:afb_v_pt_scaled}. Within the experimental uncertainties, the $\afb(\ptran)$ behavior of the data is described well by both models. We conclude that the excess asymmetry in the data is consistent with being independent of $\ptran$.

Finally we note the connection between $\ptran$ and jet multiplicity. In events with one or more extra energetic jets, we expect the $\ttbar$ system to have large $\ptran$ due to recoil against these additional jets. In Table~\ref{tab:afb_splits_signal} a difference was noted in the asymmetry measurements at the background-subtracted level between events with exactly four jets and at least five jets. Rephrasing this in terms of $\ptran$, we find that the mean $\ptran$ in five-jet events is $34.4 \pm 0.6~\gevc$ compared to $18.6 \pm 0.3~\gevc$ in events with only four jets. The smaller asymmetry in events with extra jets is seen to be consistent with the observed $\afb(\ptran)$ behavior.

\section{Conclusions}

We study the forward-backward asymmetry $\afb$ in top-quark pair production using the full CDF Run~II data set. Using the reconstructed $\ttbar$ rapidity difference in the detector frame, after removal of backgrounds, we observe an inclusive asymmetry of $0.063 \pm 0.019$(stat) compared to $0.020 \pm 0.012$ expected from the NLO standard model (with both QCD and electroweak contributions). Looking differentially, the asymmetry is found to have approximately linear dependence on both $|\dy|$ and $\mttb$, as expected for the NLO charge asymmetry, although with larger slopes then the NLO prediction. The probabilities to observe the measured values or larger for the detector-level dependencies are $2.8\sigma$ and $2.4\sigma$ for $|\dy|$ and $\mttb$ respectively. 

The results are corrected to the parton level to find the differential cross section $d\sigma/d(\dy)$, where we measure an inclusive parton-level asymmetry of $0.164 \pm 0.047$(stat$+$syst).The asymmetries and their functional dependencies at the three stages of the analysis procedure are summarized in Fig.~\ref{fig:corr_steps} and Table~\ref{tab:corr_steps}.

We also study the dependence of $\afb$ on the transverse momentum of the $\ttbar$ system. We find a significant momentum dependence that is consistent with either of the LO or NLO predictions, and evidence that the excess asymmetry is independent of the momentum.

This new measurement of the top quark production asymmetry serves as a means to better understand higher-order corrections to the standard model or potential effects from non-standard model processes.

\section*{Acknowledgments}
We thank the Fermilab staff and the technical staffs of the participating institutions for their vital contributions. This work was supported by the U.S. Department of Energy and National Science Foundation; the Italian Istituto Nazionale di Fisica Nucleare; the Ministry of Education, Culture, Sports, Science and Technology of Japan; the Natural Sciences and Engineering Research Council of Canada; the National Science Council of the Republic of China; the Swiss National Science Foundation; the A.P. Sloan Foundation; the Bundesministerium f\"ur Bildung und Forschung, Germany; the Korean World Class University Program, the National Research Foundation of Korea; the Science and Technology Facilities Council and the Royal Society, UK; the Russian Foundation for Basic Research; the Ministerio de Ciencia e Innovaci\'{o}n, and Programa Consolider-Ingenio 2010, Spain; the Slovak R\&D Agency; the Academy of Finland; and the Australian Research Council (ARC). 

\appendix
\setcounter{secnumdepth}{0}
\section{Appendix I: Validation of charge asymmetry measurements with the CDF~II detector}\label{sec:det_asym}

With a $\ppbar$ initial state and a $\ttbar$ final state that are symmetric under charge conjugation, the forward-backward asymmetry is equivalent to a charge asymmetry. The asymmetry measurements rely crucially on measurement of the lepton charge to determine the charges of all reconstructed particles in the $\ttbar$ final state. This is particularly important when the lepton $\peetee$ is large, as large $\mttb$ is correlated with large-lepton-$\peetee$ events, in which the determination of the lepton charge is more challenging. It is therefore important to verify that the lepton charge determination is modeled correctly over the lepton $\peetee$ and $\eta$ ranges pertinent to the $\ttbar$ measurements. 

\begin{figure}
\begin{center}
\includegraphics[width=0.45\textwidth, clip]{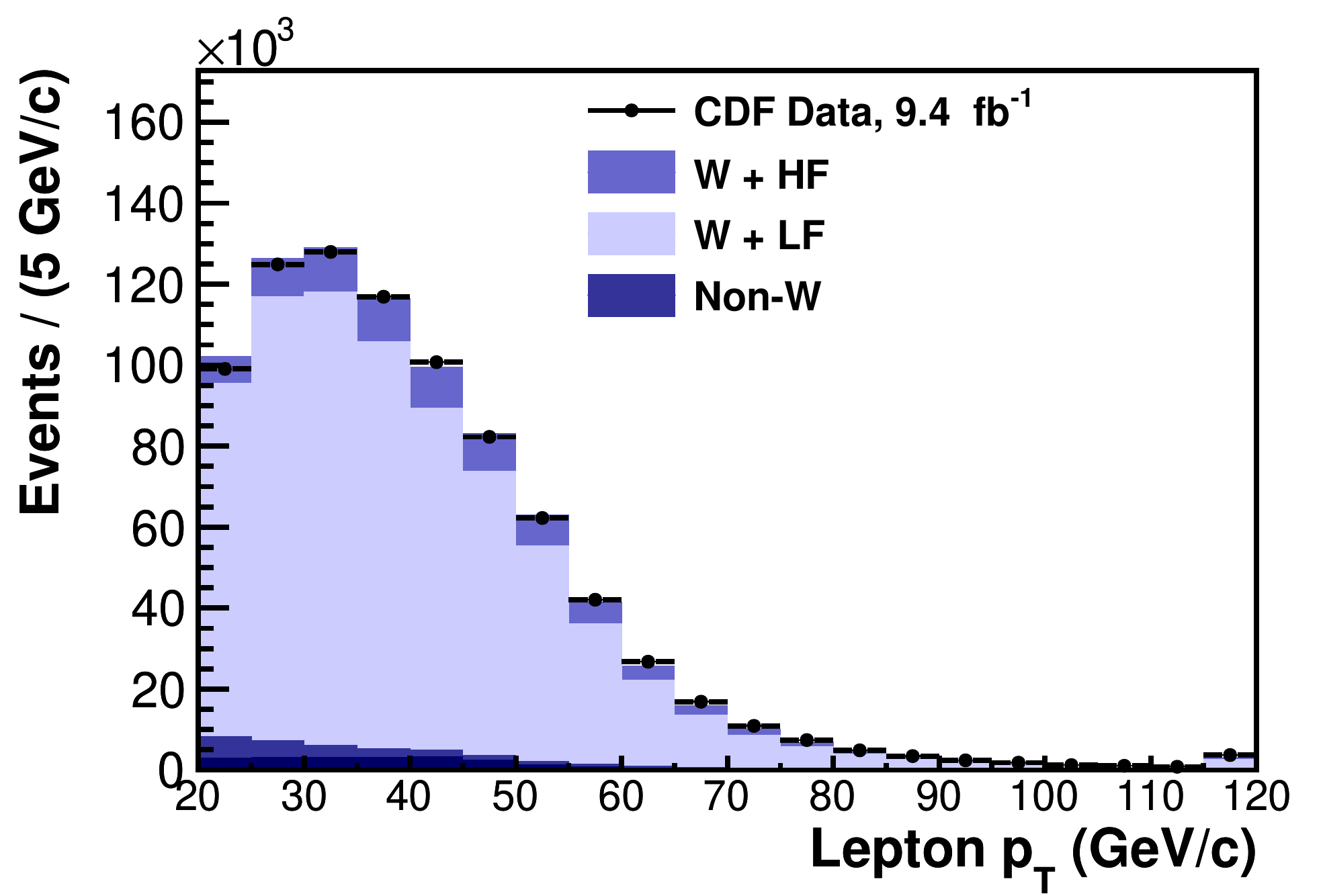}
\caption{{\small The lepton $\peetee$ distribution in events with a $W$ boson and only one observed jet. The last bin contains overflow events.} \label{fig:1jet_leppt}}
\end{center}
\end{figure}

\begin{figure*}[!htbp]
\begin{center}
\subfigure[]{
\includegraphics[width=0.45\textwidth, clip]{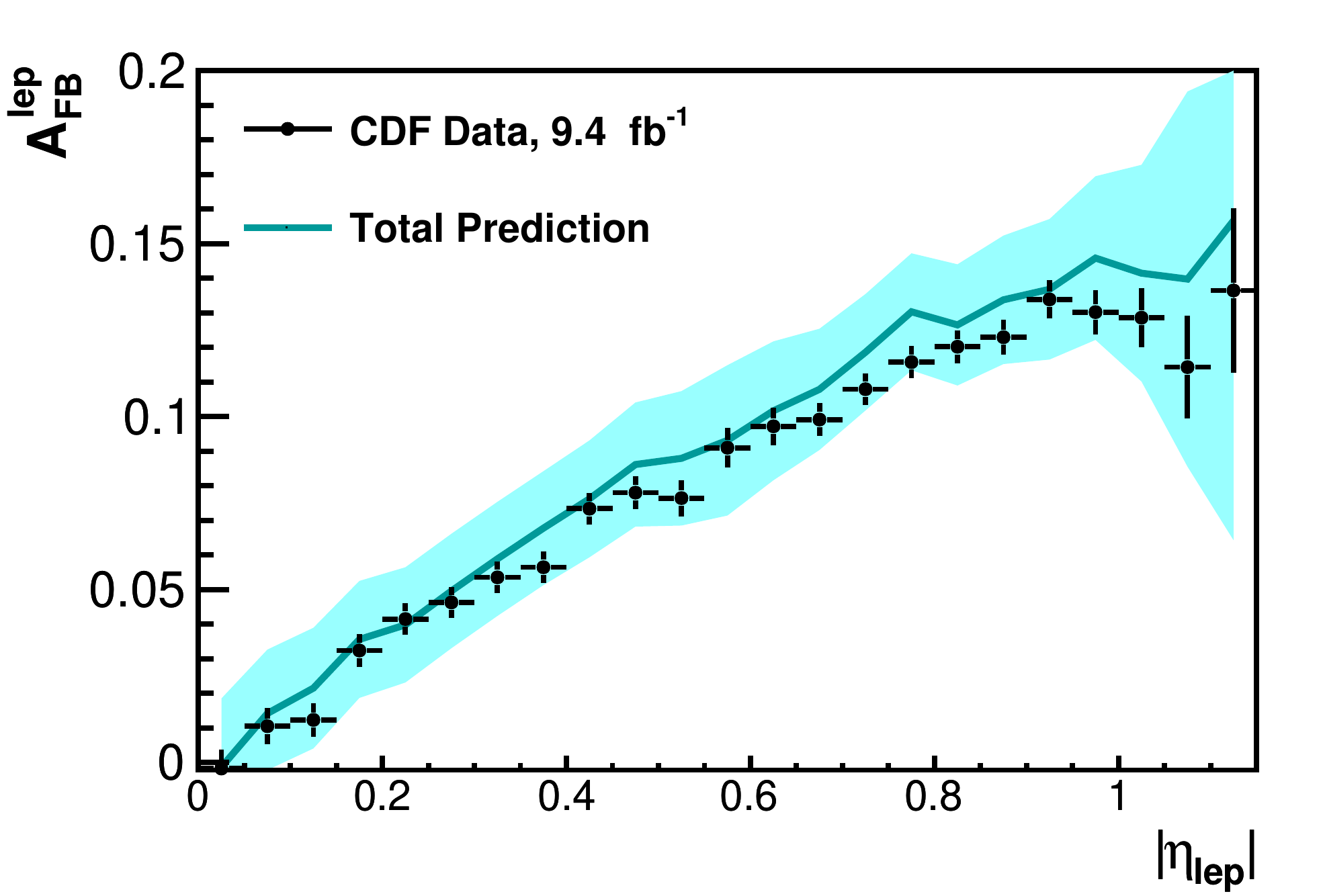}\label{fig:1jet_afb_eta}}
\vspace*{0.15in}
\subfigure[]{
\includegraphics[width=0.45\textwidth, clip]{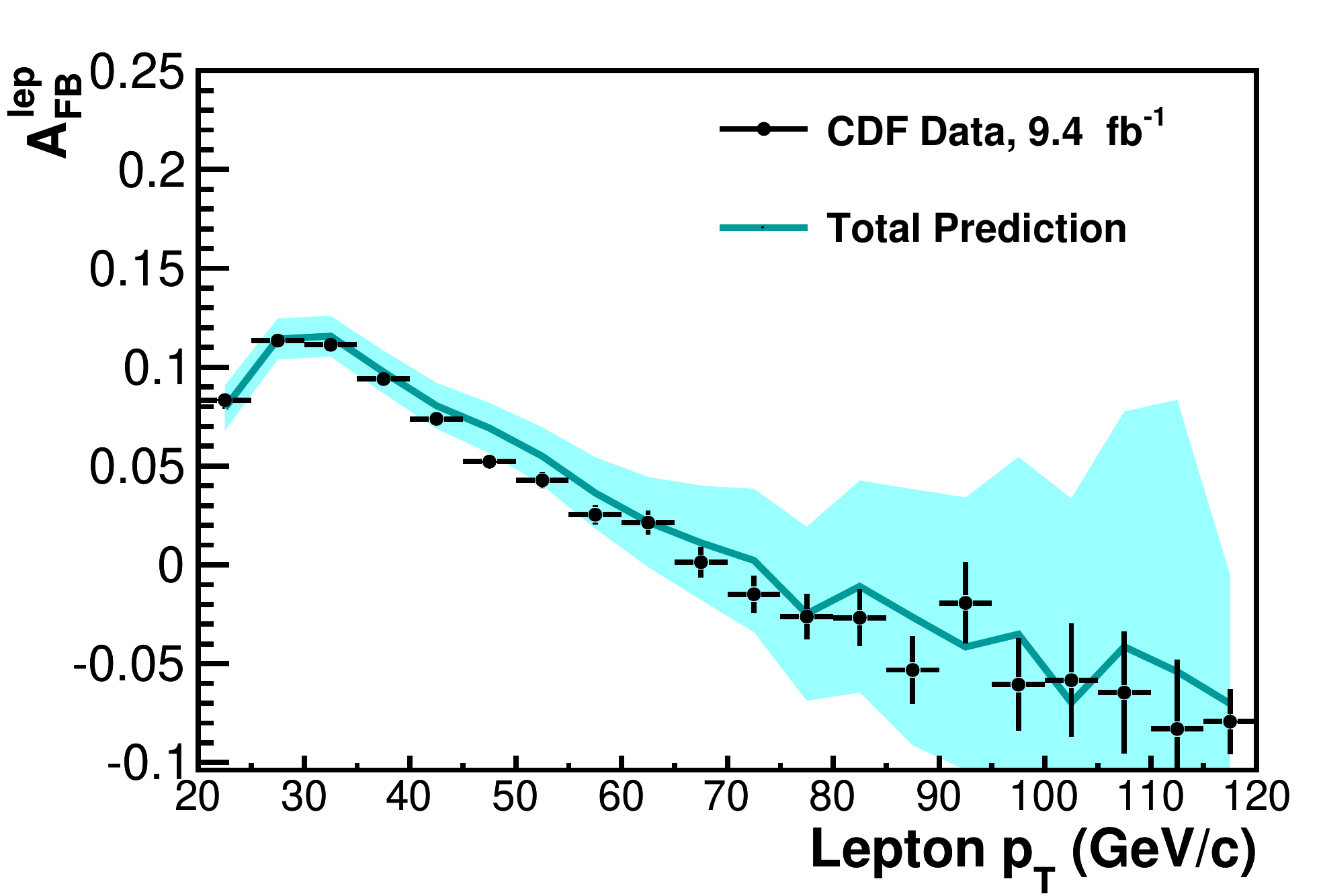}\label{fig:1jet_afb_pt}}
\caption{{\small The forward-backward asymmetry in $q\cdot\eta_{\rm lep}$ at the detector level as a function of \subref{fig:1jet_afb_eta} $|\eta_{\rm lep}|$ and \subref{fig:1jet_afb_pt} lepton $\peetee$ in events with a $W$ boson and only one observed jet.  The error bars on the data include only a statistical contribution, with the uncertainty on the SM prediction shown as a band around the predicted asymmetry. The last bin contains overflow events.} \label{fig:1jet_afb_v_leppt}}
\end{center}
\end{figure*}

\begin{table*}[!htb]
\caption{The $q\cdot\eta_{\rm lep}$ asymmetry in the $W$+1 jet sample, compared to SM expectations, for small and large lepton $\peetee$.}\label{tab:1jet_afb_v_pt}
\begin{center}
\begin{tabular}{l c c }

\hline
\hline
        &      \phantom{00}Lepton $\peetee < 60~\gevc$\phantom{00}     &    \phantom{00}Lepton $\peetee \ge 60~\gevc$\phantom{00}  \\
\hline
Observed data            &     \phantom{$-$}0.083 $\pm$ 0.001               &  $-$0.009 $\pm$ 0.004  \\
SM prediction	         &     \phantom{$-$}0.089 $\pm$ 0.004               &  $-$0.001 $\pm$ 0.013  \\
Data minus prediction    &               $-$0.006 $\pm$ 0.004               &  $-$0.008 $\pm$ 0.014  \\
\hline \hline
\end{tabular}
\end{center}
\end{table*}

\begin{table*}[!htb]
\caption{The $q\cdot\eta_{\rm lep}$ asymmetry in the $W$+1 jet sample, compared to SM expectations, for small and large $|\eta_{\rm lep}|$.}\label{tab:1jet_afb_v_eta}
\begin{center}
\begin{tabular}{l c c }

\hline
\hline
        &      \phantom{00}$|\eta_{\rm lep}| < 0.75$ \phantom{00}     &    \phantom{00}$|\eta_{\rm lep}| \ge 0.75$ \phantom{00}  \\
\hline
Observed data            &     \phantom{$-$}0.059 $\pm$ 0.001               &  \phantom{$-$}0.124 $\pm$ 0.002  \\
SM prediction	         &     \phantom{$-$}0.063 $\pm$ 0.005               &  \phantom{$-$}0.134 $\pm$ 0.008  \\
Data minus prediction    &               $-$0.004 $\pm$ 0.005               &            $-$0.010 $\pm$ 0.008  \\
\hline \hline
\end{tabular}
\end{center}
\end{table*}

We do this in the large sample of CDF events containing a $W$ boson and only one observed hadronic jet.  In addition to an abundant, low-background signal, this sample features a well-understood, lepton-$\peetee$-dependent asymmetry in the direction of motion of the lepton from the $W$-boson decay, which is used to gauge the charge measurement.  We measure the asymmetry in the observable $q\cdot\eta_{\rm lep}$, where $q$ is the lepton charge and $\eta_{\rm lep}$ is the pseudorapidity of the lepton.  At low lepton $\peetee$, the asymmetry is positive and dominated by asymmetric contributions to the proton parton-distribution function from $u$ and $d$ quarks, while at large lepton $\peetee$, the asymmetry is negative and dominated by effects from the electroweak decay of the $W$ boson. We compare the data and prediction for the leptonic asymmetry over the relevant ranges of lepton $\peetee$ and $\eta_{\rm lep}$ to test whether we reproduce the known SM asymmetries in this important control region.  As in the analysis of the $\ttbar$ signal sample, SM $W$+jet production is modeled using the {\sc alpgen} generator~\cite{alpgen}.

The lepton selection in this sample is the same as that for $\ttbar$ candidate events. We require there to be only one observed hadronic jet, for which no $b$-tag requirement is applied.  We also release the $H_T$ requirement for this sample.  Finally, we introduce a new variable, the minimum $W$-boson mass.  We add the four-momenta of the identified lepton and the ``neutrino'', which is defined to be a massless particle with the $x$- and $y$-components of momentum given by the $\vec{\met}$, and the $z$-component chosen to minimize the total mass of the lepton+neutrino system.  We require this mass to exceed 20 $\gevcc$, removing most of the non-$W$ contribution to this data sample.  After applying this selection, we have approximately 800~000 total data events.

The lepton $\peetee$ in the $W$+1 jet sample is given in Fig.~\ref{fig:1jet_leppt}. Good agreement with the prediction is seen over the entire spectrum. Compared to the lepton $\peetee$ in $\ttbar$ decays shown in Fig.~\ref{fig:jetet_leppt}, this distribution is softer, but it still provides sufficient precision in the high-lepton-$\peetee$ region relevant to the $\ttbar$ sample. Figure~\ref{fig:1jet_afb_v_leppt} shows the asymmetries in $q\cdot\eta_{\rm lep}$ as a function of $|\eta_{\rm lep}|$ and lepton $\peetee$.  Across the entire spectrum, the asymmetry is measured with good accuracy and is in excellent agreement with the SM prediction. The biggest difficulty with the comparison is the uncertainty due to model sampling at very large $|\eta_{\rm lep}|$ and lepton $\peetee$. 

In $\ttbar$ events, lepton $\peetee$ is correlated with $\mttb$, with higher mass $\ttbar$ pairs leading to larger lepton $\peetee$.  Therefore, in the context of $\afb(\mttb)$, where a large asymmetry is observed at high mass, we are particularly interested in events with high lepton $\peetee$. The measured asymmetries in two bins of lepton $\peetee$ are given in Table~\ref{tab:1jet_afb_v_pt} for direct comparison to the SM prediction. In the context of $\afb(|\dy|)$, we also list the asymmetries for two bins of $|\eta_{\rm lep}|$ in Table~\ref{tab:1jet_afb_v_eta}. Excellent agreement, to within $1\%$, is found between the data and the prediction in all regions of lepton $\peetee$ and $\eta_{\rm lep}$ using this high-precision control sample, supporting confidence in the understanding and modeling of the detector's lepton charge reconstruction.

\appendix
\setcounter{secnumdepth}{0}
\section{Appendix II: Covariance Matrices for the Parton-Level Results}\label{sec:covariances}

The unfolding procedure used to determine the parton-level results presented in this paper corrects for migrations of events between bins.  In doing so, it introduces correlations between bins in the measured results, as each ``detector-level event'' affects multiple bins at the parton level.  Proper error treatment requires the use of a covariance matrix to describe these correlations.  This is particularly important when measuring quantities that involve multiple bins, such as the linear fits discussed in the main body of the paper.  In Table~\ref{tab:xsec_covariance}, we provide the eigenvalues and eigenvectors of the covariance matrix for the parton-level differential cross section measurement.  Tables~\ref{tab:dy_covariance} and \ref{tab:mtt_covariance} display the same information for the parton-level measurements of $\afb(|dy|)$ and $\afb(\mttb)$ respectively.  The bins are the same as those described in Tables~\ref{tab:xsecmeas}, \ref{tab:afb_dely_parton}, and \ref{tab:afb_mtt_parton}.  The covariance matrices include both statistical and systematic contributions, with the systematics uncertainties assumed to be 100\% correlated across bins.

\begin{table*}[!htb]
\caption{The eigenvalues and eigenvectors of the covariance matrix for the parton-level measurement of $d\sigma/d(\dy)$.  A single vertical column contains first an eigenvalue, then the error eigenvector that corresponds to that eigenvalue.}\label{tab:xsec_covariance}
\begin{center}
\begin{tabular}{l c c c c c c c c }

\hline
\hline
Eigenvalue $\lambda$                  &   \phantom{0}0.0380\phantom{0} &  \phantom{0}0.0204\phantom{0} & \phantom{0}0.00782\phantom{0} & \phantom{0}0.00559\phantom{0} &  \phantom{0}0.00346\phantom{0} &  \phantom{0}0.00245\phantom{0} & \phantom{0}0.000134\phantom{0} & \phantom{0}0.000414\phantom{0}   \\
\hline
$\dy \le -1.5$             &   -0.185 &  -0.014 &    -0.260 &    0.051 &   -0.439 &    0.286 &    0.788 &    0.017   \\    
$-1.5 \le \dy < -1.0$      &   -0.249 &  -0.171 &    -0.366 &    0.264 &   -0.515 &    0.281 &   -0.585 &   -0.141   \\   
$-1.0 \le \dy < -0.5$      &    0.111 &  -0.563 &    -0.139 &    0.679 &    0.196 &   -0.312 &    0.145 &    0.185   \\   
$-0.5 \le \dy < 0.0$       &    0.649 &  -0.261 &     0.489 &    0.078 &   -0.275 &    0.426 &   -0.002 &   -0.089   \\   
$0.0 \le \dy < 0.5$        &    0.490 &   0.571 &    -0.215 &    0.255 &   -0.370 &   -0.430 &   -0.012 &    0.022   \\              
$0.5 \le \dy < 1.0$	   &   -0.140 &   0.504 &     0.146 &    0.556 &    0.321 &    0.501 &   -0.019 &    0.204   \\ 
$1.0 \le \dy < 1.5$        &   -0.364 &   0.076 &     0.524 &    0.292 &   -0.170 &   -0.292 &    0.095 &   -0.615   \\ 
$\dy \ge 1.5$              &   -0.280 &  -0.016 &     0.444 &   -0.030 &   -0.400 &   -0.198 &   -0.084 &    0.719   \\
\hline \hline
\end{tabular}
\end{center}
\end{table*}

\begin{table*}[!htb]
\caption{The eigenvalues and eigenvectors of the covariance matrix for the parton-level measurement of $\afb(|\dy|)$.  A single vertical column contains first an eigenvalue, then the error eigenvector that corresponds to that eigenvalue.}\label{tab:dy_covariance}
\begin{center}
\begin{tabular}{l c c c c }

\hline
\hline
Eigenvalue $\lambda$                  &   \phantom{0}0.0293\phantom{0} &  \phantom{0}0.00734\phantom{0} & \phantom{0}0.000721\phantom{0} & \phantom{0}0.000497\phantom{0}   \\
\hline
$|\dy| < 0.5$              &    -0.062 &    0.376 &   -0.921 &   -0.080   \\              
$0.5 \le |\dy| < 1.0$	   &     0.033 &    0.838 &    0.300 &    0.455   \\ 
$1.0 \le |\dy| < 1.5$      &     0.471 &    0.347 &    0.179 &   -0.791   \\ 
$|\dy| \ge 1.5$            &     0.880 &   -0.191 &   -0.171 &    0.401   \\
\hline \hline
\end{tabular}
\end{center}
\end{table*}

\begin{table*}[!htb]
\caption{The eigenvalues and eigentvectors of the covariance matrix for the parton-level measurement of $\afb(\mttb)$.  A single vertical column contains first an eigenvalue, then the error eigenvector that corresponds to that eigenvalue.}\label{tab:mtt_covariance}
\begin{center}
\begin{tabular}{l c c c c }

\hline
\hline
Eigenvalue $\lambda$                               &   \phantom{0}0.0431\phantom{0} &  \phantom{0}0.00158\phantom{0} &  \phantom{0}0.00441\phantom{0} &   \phantom{0}0.0105\phantom{0}    \\
\hline
$\mttb < 450 \gevcc$                    &   -0.019  &   -0.753 &    0.641 &   -0.151    \\
$450 \gevcc \le \mttb < 550 \gevcc$	&   -0.009  &    0.612 &    0.597 &   -0.519    \\      
$550 \gevcc \le \mttb < 650 \gevcc$     &    0.419  &   -0.223 &   -0.431 &   -0.767   \\     
$\mttb \ge 650 \gevcc$                  &    0.908  &    0.094 &    0.218 &    0.346   \\
\hline \hline
\end{tabular}
\end{center}
\end{table*}


\begin{thebibliography}{99}
\bibitem{cc} The charge-conjugate states are implied.
\bibitem{nlo_v_lo} Arguments have been made that, because the $\afb$ numerator, $N(\dy > 0) - N(\dy< 0)$, is effectively a LO calculation of the asymmetry, the LO cross section should be used in the denominator, $N(\dy > 0) + N(\dy< 0) \propto \sigma_{\ttbar}$, when defining $\afb$. For the studies presented here, we use the clearly defined and self-consistent prediction that results from \powheg, which, as an NLO generator, includes the NLO cross section for the denominator.
\bibitem{tait} We are indebted to T.~Tait for the {\sc madgraph} implementation of the color-octet models.
\bibitem{coords} The polar angle is $\theta$ and the azimuthal angle is $\phi$. With total energy $E$ and momentum $p$, the transverse energy is defined as $\etran = E\sin\theta$ and the transverse momentum is $\peetee = p\sin\theta$. The missing transverse energy ($\met$) is the magnitude of $\vec{\met} = -\Sigma_{i}\etran^{i}\hat{n}_i$ where $\hat{n}_i$ is a unit vector perpendicular to the beam axis and pointing to the $i$th calorimeter tower. The pseudorapidity is $\eta = -\ln(\tan(\theta/2))$.
\end{thebibliography}
\end{document}